\documentclass[thmsa]{article}
\usepackage{amsmath}
\usepackage{amssymb}
\usepackage{sw20lart}

\input{tcilatex}
\begin{document}

\author{J. G. Cardoso \\
%EndAName
Department of Mathematics\\
Centre for Technological Sciences-UDESC\\
Joinville 89223-100 SC\\
Brazil.\\
e-mail: jgcardoso@brturbo.com.br
}
\title{The Classical World and Spinor Formalisms of General Relativity}
\date{}
\maketitle

\begin{abstract}
A review of some facts concerning classical spacetime geometry is presented
together with a description of the most elementary aspects of the
two-component spinor formalisms of Infeld and van der Waerden. Special
attention is concentrated upon the gauge characterization of the basic
geometric objects borne by the formalisms. It is pointed out that
spin-affine configurations may be naively defined by carrying out parallel
displacements of null world vectors within the framework of the $\gamma $%
-formalism. The standard result that assigns a covariant gauge behaviour to
the symmetric parts of any admissible spin connexions is deduced out of
building up a generalized version of spin transformation laws. A fairly
complete algebraic description of curvature splittings is carried out on the
basis of the construction of a set of spinor commutators for each formalism.
The pertinent computations take up the utilization of some covariant
differential prescriptions which facilitate specifying the action of the
commutators on arbitrary spin tensors and densities. It turns out that the
implementation of such commutators under certain circumstances gives rise to
a system of wave equations for gravitons and Infeld-van der Waerden photons
which possess in either formalism a gauge-invariance property associated
with appropriate spinor-index configurations. The situation regarding the
accomplishment of the couplings between Dirac fields and electromagnetic
curvatures is entertained to a considerable extent.
\end{abstract}

\tableofcontents

\section{INTRODUCTION}

In the realm of the theory of general relativity, typical physical
environments are viewed as curved four-real-dimensional spacetime continua
equipped with torsionless covariant derivative operators along with
symmetric metric tensors having either of the pseudo-Riemannian signatures $%
(+---)$ and $(-+++)$. Any covariant differentials in generally relativistic
spacetimes are uniquely associated to symmetric affine connexions which fix
linear displacements whose implementation leaves arc lengths invariant under
the action of manifold mapping groups [1]. Each metric tensor is thus
locally subject to a compatibility condition which just amounts to a
covariant constancy property. One of the most important features of this
theoretical framework is the fact that a generally relativistic spacetime
admits spinor structures locally [2, 3].

The first two-component spinor approach for general relativity was proposed
by Infeld [4] much earlier than the achievement of the definitive conditions
for a curved spacetime to admit spinor structures. In this context, the
independent entry of the representative matrix for a characteristic metric
spinor is taken as a nowhere-vanishing differentiable real-valued function
defined on a generally relativistic spacetime. A relationship between this
function and the functional determinant of a spacetime metric tensor, as
well as a system of equivalent expressions for the corresponding Ricci
scalar and cosmological constant, were then derived from the utilization of
simple spinor computational devices. These techniques took up the
combination of the coordinate-derivative operator with some constant
connecting objects, and thence made it feasible to write down for the first
time a curved-space version of Dirac's theory. Soon after the presentation
of this approach, a geometric generalization of it was exhibited by Infeld
and van der Waerden [5], with a couple of different two-component formalisms
having arisen from this generalization. The formalisms of Infeld and van der
Waerden constitute the classical spinor framework for general relativity,
and are traditionally designated as the $\gamma \varepsilon $-formalisms. In
accordance with either of them, two conjugate spin spaces are set up at any
non-singular point of a curved spacetime, but the special Lorentzian role
played by the unimodular linear group $SL(2,C)$ had unavoidably to be taken
over by a group of gauge transformations whose determinants amount to
complex numbers that depend essentially upon a real parameter. Actually, it
had been pointed out in conjunction with the formulation of a generalized
principle of gauge invariance [6] that such transformations could be
naturally implemented within the context of general relativity.

The $\gamma $-formalism version of the basic geometric objects is prescribed
in such a way that a smooth complex-valued function of some spacetime
coordinates is utilized in place of the real-valued metric function borne by
the Infeld formulation. All metric spinors for the $\gamma $-formalism bear
an invariant character as regards the action of manifold mapping groups, and
additionally behave themselves as spin tensors under the action of the gauge
group. Any connecting object for the $\gamma $-formalism thus appears to
bear a combination of a spin-tensor character with either a covariant or a
contravariant world-vector character. The metric spinors and connecting
objects for the $\varepsilon $-formalism are considered as entities that
carry the same world characters as the ones for the $\gamma $-formalism.
However, a spin-density character is ascribed to each of them, whence
geometric quantities generally enter into the $\varepsilon $-formalism as
spin densities. Incidentally, the theory of spin densities had already been
introduced [7, 8] at the time of the advent of the $\gamma \varepsilon $%
-formalisms.

Within the $\gamma \varepsilon $-framework, the specification of
spin-affinity patterns rests upon both the geometric properties of the usual
world-affine connexions and the implementation of a strong requirement which
amounts to taking any Hermitian connecting objects as covariantly constant
entities. Hence, a spinor version of the world metric compatibility
condition comes about, thereby stating that covariant differentials of any
outer product that consists of the coupling of two conjugate metric spinors
for either formalism must be taken to vanish. The procedures for building up
any suitable spin connexion yield a pair of conjugate contracted spin-affine
structures which carry two world-covariant quantities having different spin
characters. One of these quantities appears as a world vector that undergoes
a local gauge transformation in a spin space. It is identified with a
geometric electromagnetic potential that satisfies the gauge principle, and
likewise provides the imaginary parts of the contracted structures. Its
physical significance depends only upon the selection of covariant
derivatives for the individual $\gamma $-metric spinors [5]. The other
quantity emerges as the common real part of the contracted structures. In
the $\gamma $-formalism, it must be expressed as the partial derivative of
the logarithm of a covariantly constant real spin-scalar density that bears
a spacetime-metric character. There are some particular cases where it
becomes reexpressible in terms of a gauge-invariant world density that
formally allows the recovery of the world covariance of the pertinent affine
structures. It can be shown [9] that the treatment of such cases brings
forth world-spin affine connexions that are involved in the geometric
structure of a well-known class of conformally flat spacetimes.
Nevertheless, no spacetime relationship carrying the real part of a
contracted spin-affine structure for the $\varepsilon $-formalism does
really arise. The metric spinors for the $\varepsilon $-formalism are chosen
at the outset as covariantly constant objects in both the formalisms. In
fact, this choice comes into play without affecting at all the physical
specification of any affine electromagnetic potentials. Combining it with
the covariant constancy of the spin density which enters the real part of a
contracted $\gamma $-affinity, implies that all the $\varepsilon $%
-connecting objects must bear covariant constancy in either formalism [10].
The rules for computing covariant derivatives of spin densities in either
formalism are fixed in terms of spin-affine prescriptions which arise
directly from invoking the covariant constancy of the $\varepsilon $-metric
spinors. Such computational devices are thus constituted by world-vector
configurations which effectively emerge from contracted spin affinities.

The construction of spin-curvature structures is modelled upon the
traditional procedure that includes taking commutators between covariant
derivative operators. As originally formulated [5], the covariant constancy
of any Hermitian connecting objects gives rise to curvature splittings which
involve only the sum of purely gravitational and electromagnetic
contributions. Nonetheless, the computational tools that had been put into
practice thereabout could not cope with the spinor splittings of the
bivector configurations borne by the commutators utilized. Consequently, the
complete algebraic description of curvatures was not accomplished at that
time. Indeed, what seems to be the most striking physical feature of the $%
\gamma \varepsilon $-formalisms is the possible occurrence of wave functions
for gravitons and photons in the curvature structures of generally
relativistic spacetimes [10, 11]. This insight stems from the achievement of
some of the most significant developments of the spinor calculational
techniques, which are related to the construction of sets of algebraic
expansions and formal valence-reduction devices [12]. An important property
of such techniques is that they may be applicable equally well to specially
and generally relativistic situations because of their intrinsic symbolic
character. Loosely speaking, wave functions for photons amount to contracted
spin-curvature pieces borne by spinor decompositions of Maxwell bivectors.
The presence of electromagnetic fields in spin curvatures is bound up with
the imposition of a single gauge-covariant condition upon the metric spinors
for the $\gamma $-formalism, which is just the same as that associated with
the physical significance of affine electromagnetic potentials. Wave
functions for gravitons are defined as totally symmetric curvature pieces
that occur in spinor representations of Weyl tensors [13], but the algebraic
characterization of gravitational contributions has always to be made up by
underlying world configurations. Gravitational wave functions are
geometrically expressed in the same way as for the cases of covariantly
constant $\gamma $-metric spinors, while wave functions for photons are in
any such case automatically made into useless vanishing quantities. In
spacetimes which admit nowhere-vanishing electromagnetic and gravitational
wave functions, background photons eventually interact with underlying
gravitons, with the occurrent couplings turning out to be in both formalisms
exclusively borne by the equations that control the electromagnetic
propagation [14]. The gravitational contributions for the $\varepsilon $%
-formalism were utilized in Refs. [12, 13] to support a spinor translation
of Einstein's equations. It had been established somewhat earlier [15] that
any of them should show up as a spinor pair which must be associated to the
irreducible decomposition of a Riemann tensor. Only recently, however, has
the $\gamma \varepsilon $-description of the propagation of spin curvatures
in spacetime been fully exhibited [16].

In the presence of geometric electromagnetic fields, the affine
computational devices for the $\varepsilon $-formalism can be obtained from
the ones for the $\gamma $-formalism by allowing for a limiting case that
involves an independent $\gamma $-metric component. The imaginary part of
any former device, which actually carries an electromagnetic potential for
the $\gamma $-formalism, remains essentially the same when the limiting
process is carried through in some gauge frame whilst the respective former
real part, which does in fact bear a spacetime-metric character, gets
replaced with a physically meaningless quantity. Putting such a limit into
effect in the absence of electromagnetic fields, yields contracted
spin-affine expressions that vanish in a gauge frame. Under these
circumstances, any affine potentials are expressed as useless gradients, and
the $\varepsilon $-formalism turns out to bear a weaker meaning.

The Infeld-van der Waerden formalisms have been largely utilized over the
years for various purposes by several authors in many different ways
[17-28], particularly to construct alternative spinor patterns for classical
world structures and to carry out a spinor transcription of the famous
Petrov classification schemes for world-curvature tensors [29, 30]. An
apparently appropriate spinor technique for treating Einstein's equations
has also been proposed [31]. It has been claimed by some authors that the
relevance of the $\varepsilon $-formalism as far as classification schemes
are concerned relies upon the occurrence of a technical simplification over
the Petrov schemes [30]. Somewhat surprisingly, both the utmost importance
of spin densities and the gauge structure inherently borne by the formalisms
were entirely ruled out by several of the works we have referred to.
Notwithstanding the fact that the construction of curvature spinors is
implicitly carried by the $\gamma \varepsilon $-formalisms, the spin
curvatures that occur in the classification schemes and some of the spinor
structures mentioned above were obtained in an artificial way by carrying
out straightforward spinor translations of Riemann and Weyl tensors. More
recently, it has been suggested [32] that a description of some of the
physical properties of the cosmic microwave background may be achieved by
looking at the propagation in Friedmann-like conformally flat spacetimes of
Infeld-van der Waerden photons. A full description of the interaction
couplings that take place in the formulation of Dirac's theory in curved
spacetimes, has likewise been given [33].

The present work is primarily aimed at emphasizing that the $\gamma
\varepsilon $-formalisms should be thought of as constituting the definitive
framework for describing general spacetime properties. Attention will
therefore be concentrated upon many of the elementary aspects of the
formalisms, whence we will certainly describe the key structures associated
to the fundamental role played by spin densities in the $\varepsilon $%
-formalism. A correspondence principle associated with the limiting process
will be established in a self-consistent way by looking into two systems of
eigenvalue equations for the $\gamma $-metric spinors. A heuristic procedure
for controlling the presence or absence of geometric electromagnetic fields
is trivially realized from these equations [5, 21]. We will show that the
metric information supposedly carried by the real part of a contracted
spin-affine structure for the $\gamma $-formalism can be totally extracted
from some of the eigenvalues. The standard result [10] which states that the
symmetric parts of any admissible spin connexions behave covariantly under
the action of the gauge group, is deduced from the introduction of a set of
generalized spin transformation laws. A fairly complete description of
curvature splittings is indeed obtained out of the construction of a set of
covariant spinor commutators for both formalisms. The pertinent computations
take up the utilization of some differential prescriptions which facilitate
visualizing the action of the commutators on arbitrary spin tensors and
densities. It will be seen that the implementation of these commutators in
the presence of electromagnetic fields produces a system of wave equations
for gravitons and photons which possess in either formalism a
gauge-invariance property associated with appropriate spinor-index
configurations. We will also exhibit the patterns that describe the standard
couplings between Dirac fields and electromagnetic curvatures.

Our work has been divided into five Sections. For the sake of consistency,
we have included as Section 2 a concise review of some facts concerning
spacetime geometry which will not only impart an organizational character to
our presentation, but will also enhance many formal world-spin analogies.
The outlines of Sections 2 through 4 will be given in due course. In Section
5, we make some remarks on the formalisms. We have decided from the
beginning to adopt the following conventions. Greek and Latin letters are
broadly used as kernel letters for world and spin quantities. Kernel letters
for world densities will especially appear as Gothic letters. Components of
world and spin quantities are labelled by lower-case and upper-case Latin
letters, respectively. The primed-unprimed index notation of Ref. [12] will
be applied to the case of conjugate spinor components. World indices all
range over the four values $0,1,2,3$ whereas spinor indices take either the
values $0,1$ or $0^{\prime },1^{\prime }$. We will utilize the convention
[12, 34] according to which the effect on any index block of the actions of
the symmetry and antisymmetry operators is indicated by surrounding the
relevant indices with round and square brackets, respectively. Vertical bars
surrounding an index block will mean that the indices singled out are not to
partake of a symmetry operation. Any world quantity having $p$ upper and $q$
lower indices will sometimes be referred to as a quantity of valence $%
\{p,q\} $. Similarly, a spinor carrying $a$ upper and $b$ lower unprimed
indices together with $c$ upper and $d$ lower primed indices will be termed
as a spinor of valence $\{a,b;c,d\}$. Use will be made of the natural system
of units in which $c=\hbar =1$. We will continue using the words \textit{%
object} and \textit{quantity} without making any conceptual specifications.
Further conventions will be explained occasionally.

\section{CLASSICAL SPACETIME GEOMETRY}

As was pointed out before, we will begin by reviewing some facts related to
classical world geometry. We allow for a spacetime $\mathfrak{M}$ endowed
with a torsionless covariant derivative operator $\nabla _{a}$ and a
covariant metric tensor $g_{ab}$ whose components amount to smooth real
functions on $\mathfrak{M}$. Throughout the work, it will be assumed that
the signature of $g_{ab}$ is $(+---)$. Unprimed and primed kernel letters
will be used\footnote{%
Later on, we will unambiguously make use of this kernel-letter convention
also in the case of gauge transformations.} to refer to outcomes of
allowable (invertible) world coordinate transformations $x^{a}\mapsto $%
{\tiny \ }$x^{\prime a}(x)$, with the $``x"$ in parentheses generally
meaning functional dependence on some spacetime coordinates $x^{0},$ $x^{1},$
$x^{2}$ and $x^{3}$ on $\mathfrak{M}$. The partial derivative operators $%
\partial /\partial x^{a}$ and $\partial /\partial x^{\prime a}$ will be
written as $\partial _{a}$ and $\partial _{a}^{\prime }$. Only holonomic
coordinate systems should be utilized in spite of the fact that some of our
expressions would still remain valid in case anholonomic coordinates were
implemented. In Subsection 2.1, we introduce the definitions of parallel
displacements and covariant differentials in $\mathfrak{M}$ along with the
definition and properties of world curvature objects. Subsection 2.2 deals
with the construction of covariant derivatives of world densities. In
Subsection 2.3, we will touch upon the formulation of the conventional
least-action principle for Einstein's equations without exhibiting the
details of the pertinent calculations [35-39]. There, a particular procedure
for introducing the cosmological term into the field equations will be
described. The metric traces of any spacetime quantities of valences $%
\{0,2\} $ and $\{2,0\}$ will be denoted by the kernel letters used to write
the aforesaid quantities. It will be convenient to define a pair of
operators whose actions entail picking up the traceless parts and reversing
the signs of the traceful pieces of any of those two-world-index quantities.
Such operators were utilized in Ref. [12] to obtain a set of covariant
relations involving the gravitational tensors. Upon building up covariant
differentials, we will assume that all the geometric objects being dealt
with bear a purely world character.

\subsection{Elementary Structures}

Usually, the tensor $g_{ab}$ provides the length $ds$ of an arbitrary linear
displacement $dx^{a}$ in $\mathfrak{M}$ according to the formula%
\begin{equation}
ds={\huge \mid }\sqrt{g_{ab}dx^{a}dx^{b}}{\huge \mid }.  \tag{2.1}
\end{equation}%
The symmetry of $g_{ab}$ implies that there exists a contravariant metric
tensor $g^{ab}$ which satisfies the relation 
\begin{equation}
g_{ah}g^{hb}=\delta _{a}{}^{b},  \tag{2.2}
\end{equation}%
with $\delta _{a}{}^{b}$ being the (invariant) world Kronecker delta. Such
metric tensors may be particularly used for lowering and raising world
indices of any spacetime quantities. Tangent spaces of $\mathfrak{M}$ are
locally identified with independent Minkowski spaces to such a degree that a
single tangent space may be set up at each non-singular point of $\mathfrak{M%
}$. The independence borne by this world settlement just means that
Minkowski spaces at different points of $\mathfrak{M}$ have no point in
common. A signature-preserving coordinate transformation can thus be
performed which carries the metric tensors of $\mathfrak{M}$ into those of
special relativity, there being likewise a one-to-one correspondence between
local directions in $\mathfrak{M}$ and Minkowskian directions.

In contradistinction to ordinary $d$-differentials, covariant differentials
are often brought in $\mathfrak{M}$ by carrying out affine displacements of
world tensors from one tangent space to another. These displacements provide
an invariant way of connecting geometric objects defined at neighbouring
points, and it is from the choice of such a displacement that the local
geometric characterization of $\mathfrak{M}$ partially arises. A covariant
differential at $x^{a}$ of a world-tensor quantity amounts to the difference
between the value of the given quantity at $x^{a}+dx^{a}$ and the value of
the new quantity that results from the implementation of an affine
displacement, with the covariant differential itself appearing as the outer
product of $dx^{a}$ with another world tensor. This latter tensor is called
the covariant derivative at $x^{a}$ of the displaced quantity. The
traditional procedure [1] for prescribing covariant differentials in $%
\mathfrak{M}$ requires linearity and homogeneity in $dx^{a}$ as well as
applicability of the Leibniz-rule property. In addition, any covariant
differentials are taken to coincide with ordinary ones whenever the
displaced quantities bear a scalar character. For such a quantity $f$ on $%
\mathfrak{M}$, we then have\footnote{%
Many authors use a semi-colon to denote a covariant derivative. We have
adopted the better notation of Ref. [12].}%
\begin{equation}
Df=df\Rightarrow \nabla _{a}f=\partial _{a}f,  \tag{2.3}
\end{equation}%
where $D\doteqdot dx^{a}\nabla _{a}$ stands for the covariant differential
for $\nabla _{a}$. The prescriptions for vectors $u^{a}$ and $w_{a}$ are set
as 
\begin{equation}
Du^{a}\doteqdot du^{a}+\Gamma _{bc}{}^{a}u^{c}dx^{b}\Rightarrow \nabla
_{a}u^{b}=\partial _{a}u^{b}+\Gamma _{ac}{}^{b}u^{c}  \tag{2.4}
\end{equation}%
and%
\begin{equation}
Dw_{a}=dw_{a}-\Gamma _{ba}{}^{c}w_{c}dx^{b}\Rightarrow \nabla
_{a}w_{b}=\partial _{a}w_{b}-\Gamma _{ab}{}^{c}w_{c}.  \tag{2.5}
\end{equation}%
In the case of either vector, the quantity $\Gamma _{ab}{}^{c}$ effectively
specifies the displacement allowed for, and constitutes the world-affine
connexion associated to $\nabla _{a}$. We must emphasize that the individual
pieces carried by the right-hand sides of the expansions (2.4) and (2.5) do
not bear a tensor character, but each of the overall expansions does. It
should be obvious that these expansions can be obtained from one another by
using the Leibniz rule along with the property (2.3) for $f=u^{h}w_{h}$.
Covariant derivative prescriptions for world tensors of any valences may be
readily built up by invoking the result that generic tensors can always be
given as linear combinations of outer products between vectors, and likewise
performing Leibniz expansions. For instance, for a world tensor of valence $%
\{2,1\}$, we have the expansion 
\begin{equation}
\nabla _{a}H_{b}{}^{cd}=\partial _{a}H_{b}{}^{cd}-\Gamma
_{ab}{}^{h}H_{h}{}^{cd}+\Gamma _{ah}{}^{c}H_{b}{}^{hd}+\Gamma
_{ah}{}^{d}H_{b}{}^{ch}.  \tag{2.6}
\end{equation}%
It is useful to notice that covariant derivatives can be thought of as
symbolically involving index-displacement rules.

An important property of the geometric structure of $\mathfrak{M}$\ is
related to the fact that we can determine out of $g_{ab}$\ a unique
symmetric affine connexion $\Gamma _{ab}{}^{c}$\ which fixes displacements
whose implementation leaves $ds$\ invariant. This affine symmetry may be
immediately brought forth by accounting for the torsionlessness of $\nabla
_{a}$, namely 
\begin{equation}
\nabla _{\lbrack a}\nabla _{b]}f=-\Gamma _{\lbrack ab]}{}^{c}\nabla _{c}f=0,
\tag{2.7}
\end{equation}%
which accordingly gives rise to the property%
\begin{equation}
\Gamma _{ab}{}^{c}=\Gamma _{(ab)}{}^{c}.  \tag{2.8}
\end{equation}%
The displacement invariance of $ds$ implies that $g_{ab}$ must be taken
covariantly constant with respect to $\nabla _{a}$, whence we have the
metric compatibility condition 
\begin{equation}
Dg_{bc}=0\Leftrightarrow \nabla _{a}g_{bc}=0,  \tag{2.9a}
\end{equation}%
which means that\footnote{%
The symbol $\Gamma _{ab}{}^{c}$ possesses $4^{3}=64$ components in all, but
the symmetry occurring in (2.8) implies taking $64-24=4\times 10=40$ as the
number of its independent components.} 
\begin{equation}
\partial _{a}g_{bc}=2\Gamma _{a(bc)},\text{ }\Gamma _{abc}\doteqdot \Gamma
_{ab}{}^{h}g_{hc}.  \tag{2.9b}
\end{equation}%
Obviously, the condition (2.9a) also yields the covariant constancy of both $%
g^{bc}$ and $\delta _{a}{}^{b}$ whence the action of $\nabla _{a}$ is taken
to commute with the lowering and raising of world indices. The metric
expression for $\Gamma _{ab}{}^{c}$ thus reads 
\begin{equation}
\Gamma _{ab}{}^{c}=\frac{1}{2}g^{ch}(2\partial _{(a}g_{b)h}-\partial
_{h}g_{ab}),  \tag{2.10}
\end{equation}%
which defines a Christofell connexion in $\mathfrak{M}$. We observe that the
expression (2.10) is invariant under constant rescalings of $g_{ab}$.

The basic curvature structure of $\mathfrak{M}$ arises when we carry out
affine displacements along infinitesimal loops. In essence, this structure
appears as an invariant difference between two generally distinct displaced
world tensors that are obtained from some given tensorial object by
displacing it along two different paths of a loop which have the same
starting and end points. It turns out that the information carried by the
overall tensor difference can be extracted from either of the commutator
configurations 
\begin{equation}
2\nabla _{\lbrack a}\nabla _{b]}u^{c}=[\nabla _{a},\nabla
_{b}]u^{c}=R_{abh}{}^{c}u^{h}  \tag{2.11a}
\end{equation}%
and%
\begin{equation}
2\nabla _{\lbrack a}\nabla _{b]}w_{c}=[\nabla _{a},\nabla
_{b}]w_{c}=-R_{abc}{}^{h}w_{h},  \tag{2.11b}
\end{equation}%
with $R_{abc}{}^{d}$ being the curvature tensor of $\nabla _{a}$. Since the
Leibniz rule is applicable to $\nabla _{\lbrack a}\nabla _{b]}$, we can
carry out a commutator expansion for an arbitrary world tensor by using the
prescriptions supplied by (2.11). For the tensor borne by Eq. (2.6), for
instance, we have 
\begin{equation}
\lbrack \nabla _{a},\nabla
_{b}]H_{c}{}^{ks}=R_{abm}{}^{k}H_{c}{}^{ms}+R_{abm}{}^{s}H_{c}{}^{km}-R_{abc}{}^{m}H_{m}{}^{ks}.
\tag{2.12}
\end{equation}%
Hence, applying (2.12) to $g_{cd}$ and invoking (2.9a), yields the relation $%
R_{ab(cd)}=0$, whence%
\begin{equation}
R_{abcd}=R_{[ab][cd]}.  \tag{2.13a}
\end{equation}%
The torsionless character of $\nabla _{a}$ as expressed by (2.7) is most
transparently passed on to $R_{abcd}$ through the cyclic property%
\begin{equation}
R_{[abc]d}=0,  \tag{2.13b}
\end{equation}%
as can be seen by utilizing (2.11b) with $w_{c}=\nabla _{c}f$ and performing
a skew symmetrization over the indices $a$, $b$ and $c$. We can see, in
addition, that the combination of (2.13a) and (2.13b) produces the
index-pair symmetry\footnote{%
Owing to the symmetries of $R_{abcd}$, the number of its independent
components equals $\frac{4}{3}(16-1)=20$.}%
\begin{equation}
R_{abcd}=R_{cdab},  \tag{2.13c}
\end{equation}%
which also reflects the torsionlessness of $\nabla _{a}$. By allowing for
either of Eqs. (2.11), we deduce the Riemann-Christoffel expression 
\begin{equation}
R_{abc}{}^{d}=2(\partial _{\lbrack a}\Gamma _{b]c}{}^{d}+\Gamma
_{m[a}{}^{d}\Gamma _{b]c}{}^{m}).  \tag{2.14}
\end{equation}%
\ 

The torsion-freeness of $\nabla _{a}$ likewise gives rise to the covariant
differential identity%
\begin{equation}
\nabla _{\lbrack a}R_{bc]dh}=0,  \tag{2.15a}
\end{equation}%
which may be easily derived [12] by utilizing (2.11a) and (2.12) for
simultaneously working out the configurations%
\begin{equation}
2\nabla _{\lbrack a}\nabla _{b}\nabla _{c]}u^{d}=2\nabla _{\lbrack \lbrack
a}\nabla _{b]}(\nabla _{c]}u^{d})  \tag{2.15b}
\end{equation}%
and%
\begin{equation}
2\nabla _{\lbrack a}\nabla _{b}\nabla _{c]}u^{d}=2\nabla _{\lbrack a}(\nabla
_{\lbrack b}\nabla _{c]]}u^{d}).  \tag{2.15c}
\end{equation}%
Equation (2.15a) is traditionally known in the literature as the
gravitational Bianchi identity. By carrying out the skew expansion involved
in it, and making suitable index contractions, we promptly get the relation%
\begin{equation}
2\nabla ^{b}R_{ab}{}-\nabla _{a}R=0{},  \tag{2.15d}
\end{equation}%
where $R_{ab}{}$ and $R$ are the Ricci tensor and scalar of $R_{abcd}$,
which are defined by\footnote{%
Our sign convention for the Ricci tensor is the same as that adopted in Ref.
[12].} 
\begin{equation}
R_{ab}{}=R_{ahb}{}^{h}=R_{(ab)}{},\text{ }R=g^{ab}R_{ab}{}=R_{ab}{}^{ab}. 
\tag{2.15e}
\end{equation}%
We notice that the symmetry of $R_{ab}$ comes about because of the property
(2.13c). It is worthwhile to observe that the Ricci tensor occurs in either
of the contracted commutators 
\begin{equation}
\lbrack \nabla _{a},\nabla _{b}]u^{b}=R_{ab}{}u^{b},\text{ }[\nabla
^{a},\nabla ^{b}]w_{b}=R^{ab}w_{b},  \tag{2.16}
\end{equation}%
with $\nabla ^{a}\doteqdot g^{ab}\nabla _{b}$. The explicit expression for $%
R $ in terms of $\Gamma _{ab}{}^{c}$ is thus written as%
\begin{equation}
R=2g^{ab}(\partial _{\lbrack a}\Gamma _{h]b}{}^{h}+\Gamma _{m[a}{}^{h}\Gamma
_{h]b}{}^{m}).  \tag{2.17}
\end{equation}

\subsection{World Densities}

The concept of the simplest world densities emerges from the observation
that the action of the manifold mapping group of $\mathfrak{M}$ on any
totally antisymmetric tensors of valences $\{0,4\}$ and $\{4,0\}$, implies
that each of the relevant components undergoes a transformation law of the
same type as that for a world scalar, but with either the Jacobian
functional determinant 
\begin{equation}
\delta _{W}=\frac{\partial (x^{\prime 0},x^{\prime 1},x^{\prime 2},x^{\prime
3})}{\partial (x^{0},x^{1},x^{2},x^{3})},  \tag{2.18}
\end{equation}%
or its inverse, being effectively taken up as a factor by the corresponding
outcome. In the covariant case, 
\begin{equation}
W_{abcd}=W_{[abcd]},  \tag{2.19}
\end{equation}%
we have the prescription 
\begin{equation}
W_{0123}^{\prime }=4!(\partial _{\lbrack 0}^{\prime }x^{0})(\partial
_{1}^{\prime }x^{1})(\partial _{2}^{\prime }x^{2})(\partial _{3]}^{\prime
}x^{3})W_{0123}=\Delta _{W}W_{0123},  \tag{2.20}
\end{equation}%
with $\Delta _{W}\doteqdot (\delta _{W})^{-1}$. The determinants $\Delta
_{W} $ and $\delta _{W}$ are formally obtained from one another by
interchanging the roles of the unprimed and primed world frames.
Consequently, we can write the tensor law 
\begin{equation}
W_{abcd}^{\prime }=\Delta _{W}W_{abcd}.  \tag{2.21}
\end{equation}%
In the contravariant case, we similarly obtain 
\begin{equation}
U^{\prime 0123}=(\Delta _{W})^{-1}U^{0123}\Rightarrow U^{\prime
abcd}=(\Delta _{W})^{-1}U^{abcd}.  \tag{2.22}
\end{equation}%
Any numerical quantities that undergo the same laws as the components
occurring in (2.20) and (2.22) are called world-scalar densities of weights $%
+1$ and $-1$, respectively. A world-scalar density $\mathfrak{D}$ of weight $%
w$ in $\mathfrak{M}$ is thus defined as a quantity which transforms as 
\begin{equation}
\mathfrak{{D}^{\prime }}=(\Delta _{W})^{w}\mathfrak{D}.  \tag{2.23}
\end{equation}%
The value of the weight of any world density remains conventionally
unaffected under the interchange $\Delta _{W}\leftrightarrow \delta _{W}$.
Thus, the right-hand side of Eq. (2.23) may be rewritten as $(\delta
_{W})^{-w}\mathfrak{D}$. An important world-scalar density of weight $+2$ is
the determinant $\mathfrak{g}$ of $g_{ab}$. In effect, we have 
\begin{equation}
\mathfrak{g}^{\prime }\doteqdot 4!g_{0[0}^{\prime }g_{\mid 1\mid 1}^{\prime
}g_{\mid 2\mid 2}^{\prime }g_{\mid 3\mid 3]}^{\prime }=(\Delta _{W})^{2}%
\mathfrak{g}.  \tag{2.24}
\end{equation}

Arbitrary world-tensor densities are defined as outer products between
tensors and scalar densities. The valences of such densities are specified
in terms of those borne by the objects which enter into the products in much
the same way as for the case of world tensors. Any product between
world-tensor densities carries a weight which equals the sum of the weights
of the factors involved. Evidently, a tensor can be particularly viewed as a
world-tensor density whose weight equals zero. Hence, for a world-tensor
density $\mathfrak{Y}$ of valence $\{p,q\}$ and weight $w$ in $\mathfrak{M}$%
, we have the homogeneous transformation law 
\begin{equation}
\mathfrak{Y}_{a...b}^{\prime }{}^{k...s}=(\Delta _{W})^{w}(\partial
_{a}^{\prime }x^{h})...(\partial _{b}^{\prime }x^{j})(\partial _{m}x^{\prime
k})...(\partial _{n}x^{\prime s})\mathfrak{Y}_{h...j}{}^{m...n}.  \tag{2.25}
\end{equation}%
World tensors can be naively constructed by performing outer products
between suitable world densities. A very useful example is afforded by 
\begin{equation}
e_{abcd}=(-\mathfrak{g})^{1/2}\mathfrak{E}_{abcd},\text{ }e^{abcd}=(-%
\mathfrak{g})^{-1/2}\mathfrak{E}^{abcd},  \tag{2.26}
\end{equation}%
where the $\mathfrak{E}$-objects are the alternating Levi-Civita world
densities in $\mathfrak{M}$. We thus have the laws 
\begin{equation}
e_{abcd}^{\prime }=\Delta _{W}e_{abcd},\text{ }e^{\prime abcd}=(\Delta
_{W})^{-1}e^{abcd},  \tag{2.27a}
\end{equation}%
together with the invariance properties 
\begin{equation}
\mathfrak{E}_{abcd}^{\prime }=(\Delta _{W})^{-1}(\partial _{a}^{\prime
}x^{h})(\partial _{b}^{\prime }x^{j})(\partial _{c}^{\prime }x^{k})(\partial
_{d}^{\prime }x^{s})\mathfrak{E}_{hjks}=\mathfrak{E}_{abcd}  \tag{2.27b}
\end{equation}%
and%
\begin{equation}
\mathfrak{E}^{\prime abcd}=\Delta _{W}(\partial _{h}x^{\prime a})(\partial
_{j}x^{\prime b})(\partial _{k}x^{\prime c})(\partial _{s}x^{\prime d})%
\mathfrak{E}^{hjks}=\mathfrak{E}^{abcd},  \tag{2.27c}
\end{equation}%
whence either of the $\mathfrak{E}$-densities possesses only one independent
world-scalar component which is usually taken as a constant.\footnote{%
The usual $\mathfrak{E}$-densities satisfy the invariant relation $\mathfrak{%
E}_{abcd}\mathfrak{E}^{hjks}=4!\delta _{a}{}^{[h}\delta _{b}{}^{j}\delta
_{c}{}^{k}\delta _{d}{}^{s]}$.} These densities are frequently utilized to
define dual world tensors and write formal expressions for determinants. For
example, 
\begin{equation}
^{\ast }R_{abcd}=\frac{1}{2}(-\mathfrak{g})^{1/2}\mathfrak{E}%
_{abks}R^{ks}{}_{cd}  \tag{2.28a}
\end{equation}%
and%
\begin{equation}
\Delta _{W}=\frac{1}{4!}\mathfrak{E}^{abcd}(\partial _{a}^{\prime
}x^{h})(\partial _{b}^{\prime }x^{j})(\partial _{c}^{\prime }x^{k})(\partial
_{d}^{\prime }x^{s})\mathfrak{E}_{hjks},  \tag{2.28b}
\end{equation}%
where $^{\ast }R_{abcd}$ is the so-called first-left dual of $R_{abcd}$. Of
course, the value of $\Delta _{W}$ as given by (2.28b) is invariant under
the kernel-letter replacement $\mathfrak{E}$ $\mapsto e$. By making use of
the dualization schemes exhibited in Ref. [12], one can reexpress the
properties (2.13b) and (2.15a) as 
\begin{equation}
^{\ast }R_{ab}{}^{bc}=0,\text{ }\nabla ^{a\ast }R_{abcd}{}=0.  \tag{2.29}
\end{equation}

The construction of covariant derivatives of world densities in $\mathfrak{M}
$ is also based upon the patterns of covariant differentials of totally
antisymmetric world tensors of valences $\{0,4\}$ and $\{4,0\}$. In order to
achieve the relevant configurations, it suffices to consider the case of
either valence. The crucial point as regards this construction is that the
expansion (2.6) for either of the tensors (2.21) and (2.22) turns out to be
simplified when we implement the total skewness. For $W_{abcd}$, say, we get 
\begin{equation}
\nabla _{a}W_{bcdh}=\partial _{a}W_{bcdh}-4\Gamma _{a[h}{}^{m}W_{bcd]m}, 
\tag{2.30a}
\end{equation}%
which, after some index manipulations, yields 
\begin{equation}
\nabla _{a}W_{bcdh}=\partial _{a}W_{bcdh}-\Gamma _{a}W_{bcdh},  \tag{2.30b}
\end{equation}%
with 
\begin{equation}
\Gamma _{a}\doteqdot \Gamma _{ab}{}^{b}=\partial _{a}\log (-\mathfrak{g}%
)^{1/2},  \tag{2.30c}
\end{equation}%
and the relations (2.9) having been accounted for. Working out the covariant
derivative of $U^{abcd}$ leads to a structure which can be built up from
Eqs. (2.30) by rearranging indices and substituting $(-\mathfrak{g})^{-1/2}$
for $(-\mathfrak{g})^{1/2}$, namely 
\begin{equation}
\nabla _{a}U^{bcdh}=\partial _{a}U^{bcdh}+\Gamma _{a}U^{bcdh}.  \tag{2.31}
\end{equation}%
When written out explicitly in terms of $W_{0123}$ and $U^{0123}$, the
expansions (2.30b) and (2.31) provide us with the $\nabla $-patterns for
world-scalar densities of weights $+1$ and $-1$, respectively. For the
density (2.23), we thus define%
\begin{equation}
\nabla _{a}\mathfrak{D}=\partial _{a}\mathfrak{D}-w\Gamma _{a}\mathfrak{D}, 
\tag{2.32}
\end{equation}%
whence allowing for a tensor density like 
\begin{equation}
\mathfrak{Y}_{a...b}{}^{k...s}=\mathfrak{D}Y_{a...b}{}^{k...s},  \tag{2.33}
\end{equation}%
and utilizing the Leibniz rule, gives the covariant expansion for the case
of weight $w$ and arbitrary valence. For instance, 
\begin{equation}
\nabla _{a}\mathfrak{Y}_{b}{}^{c}=\partial _{a}\mathfrak{Y}_{b}{}^{c}-\Gamma
_{ab}{}^{h}\mathfrak{Y}_{h}{}^{c}+\Gamma _{ah}{}^{c}\mathfrak{Y}%
_{b}{}^{h}-w\Gamma _{a}\mathfrak{Y}_{b}{}^{c},  \tag{2.34a}
\end{equation}%
which thus conforms to the generalized law 
\begin{equation}
\nabla _{a}^{\prime }\mathfrak{Y}_{b...c}^{\prime }{}^{k...s}=(\Delta
_{W})^{w}(\partial _{a}^{\prime }x^{h})(\partial _{b}^{\prime
}x^{j})...(\partial _{c}^{\prime }x^{r})(\partial _{m}x^{\prime
k})...(\partial _{n}x^{\prime s})\nabla _{h}\mathfrak{Y}_{j...r}{}^{m...n}. 
\tag{2.34b}
\end{equation}%
We can now recall (2.30c) to obtain the integrability condition 
\begin{equation}
\lbrack \nabla _{a},\nabla _{b}]\mathfrak{D}=2\mathfrak{D}\nabla _{\lbrack
a}(\mathfrak{D}^{-1}\nabla _{b]}\mathfrak{D})=(-2w\mathfrak{D})\partial
_{\lbrack a}\Gamma _{b]}=(-w\mathfrak{D})R_{abh}{}^{h}\equiv 0,  \tag{2.35}
\end{equation}%
whence the commutator expansions for world-tensor densities are formally the
same as the ones for world tensors. As an example, we have 
\begin{equation}
\lbrack \nabla _{a},\nabla _{b}]\mathfrak{Y}_{c}{}^{d}=R_{abm}{}^{d}%
\mathfrak{Y}_{c}{}^{m}-R_{abc}{}^{m}\mathfrak{Y}_{m}{}^{d}.  \tag{2.36}
\end{equation}

An immediate consequence of Eq. (2.32) is the covariant constancy of the
density $(-\mathfrak{g})^{N}$, with $N$ being any real number. It should be
clear that this result comes out of the applicability of (2.30c). We have,
in effect, 
\begin{equation}
\nabla _{a}(-\mathfrak{g})^{N}=0.  \tag{2.37}
\end{equation}%
It follows that, by invoking (2.9a) and (2.37), we obtain the combined
formulae 
\begin{equation}
\nabla _{a}[(-\mathfrak{g})g_{bc})=0  \tag{2.38a}
\end{equation}%
and 
\begin{equation}
g^{bc}\partial _{a}[(-\mathfrak{g})^{-1/4}g_{bc}]=0,\text{ }g_{bc}\partial
_{a}[(-\mathfrak{g})^{1/4}g^{bc}]=0.  \tag{2.38b}
\end{equation}%
Equation (2.37) can likewise be employed for establishing the useful
contracted relation 
\begin{equation}
\nabla _{a}[(-\mathfrak{g})^{1/2}U^{abcd}]=\partial _{a}[(-\mathfrak{g}%
)^{1/2}U^{abcd}],  \tag{2.39a}
\end{equation}%
which produces the famous divergence formula 
\begin{equation}
\nabla _{a}u^{a}=(-\mathfrak{g})^{-1/2}\partial _{a}[(-\mathfrak{g}%
)^{1/2}u^{a}].  \tag{2.39b}
\end{equation}%
Hence, both of $\mathfrak{E}_{abcd}$ and $e_{abcd}$ bear covariant
constancy, that is to say, 
\begin{equation}
\nabla _{a}\mathfrak{E}_{bcdh}=0\Leftrightarrow \nabla _{a}e_{bcdh}=0. 
\tag{2.40}
\end{equation}

\subsection{Einstein's Equations}

In vacuum, Einstein's equations without cosmological terms emerge out of the
variational principle [35-39] 
\begin{equation}
\delta \int_{\Omega }(-\mathfrak{g})^{1/2}Rd^{4}x\hspace{0.15cm}=0, 
\tag{2.41a}
\end{equation}%
where $\Omega $ stands for a bounded region in $\mathfrak{M}$ whose closure
is compact, and 
\begin{equation}
d^{4}x=\frac{1}{4!}\mathfrak{E}_{abcd}dx^{a}\wedge dx^{b}\wedge dx^{c}\wedge
dx^{d}  \tag{2.41b}
\end{equation}%
defines an elementary volume-density in $\Omega .$ Presumably, the metric
variation $\delta g_{ab}$ is taken as an arbitrary quantity that vanishes on
the boundary of $\Omega $. The components of the functional derivative of $(-%
\mathfrak{g})^{1/2}R$ thus appear as functions of $g_{ab}$, $\partial
_{a}g_{bc}$ and $\partial _{a}\partial _{b}g_{cd}$, with the derivative
itself being given by the gravitational density 
\begin{equation}
\mathfrak{G}_{ab}=(-\mathfrak{g})^{1/2}G_{ab},  \tag{2.42}
\end{equation}%
where $G_{ab}$ is the Einstein tensor. This tensor can be obtained by
operating on $R_{ab}$ with the trace-reversal operator $\hat{\tau}$ which is
defined by\footnote{%
The operator $\hat{\tau}$ reverses the signs of traces, but it preserves
symmetry. It is linear and possesses the involutory property $\hat{\tau}%
^{2}= $ identity. In particular, $G=-R$.} 
\begin{equation}
G_{ab}=\hat{\tau}R_{ab}\doteqdot R_{ab}-\frac{1}{2}Rg_{ab}.  \tag{2.43}
\end{equation}%
Equations (2.15d) and (2.37) tell us that both $\mathfrak{G}_{ab}$ and $%
G_{ab}$ possess the divergencelessness property 
\begin{equation}
\nabla ^{a}\mathfrak{G}_{ab}=0,\text{ }\nabla ^{a}G_{ab}=0.  \tag{2.44}
\end{equation}%
Accordingly, the field equations associated to the statement (2.41a) are
written as 
\begin{equation}
R_{ab}=0\Leftrightarrow G_{ab}=0,  \tag{2.45}
\end{equation}%
in which case both $R_{ab}$ and $G_{ab}$ bear tracelessness.

A notable procedure [12] for introducing a cosmological term into Eqs.
(2.45) involves the utilization of the splitting relation%
\begin{equation}
R_{ab}=(-2)\Xi _{ab}+6\varkappa g_{ab},  \tag{2.46}
\end{equation}%
with the quantity $(-2)\Xi _{ab}$ being identified with the trace-free part
of $R_{ab}$ such that $R\doteqdot 24\varkappa $ (see Eq. (2.48) below).
Thus, Eqs. (2.45) must be replaced with either of the equivalent statements 
\begin{equation}
R_{ab}-6\varkappa g_{ab}=0  \tag{2.47a}
\end{equation}%
and 
\begin{equation}
(-2)\Xi _{ab}=G_{ab}+6\varkappa g_{ab}=0.  \tag{2.47b}
\end{equation}%
The relation (2.46) can be reexpressed in a somewhat formal way by defining
an operator $\hat{s}$ as\footnote{%
The operator $\hat{s}$ picks out the trace-free part of any world
configurations of valences $\{0,2\}$ and $\{2,0\}$. It is linear and
commutes with $\hat{\tau}$. It also satisfies $\hat{s}\hat{\tau}+\hat{\tau}%
\hat{s}=2\hat{s}$ and $\hat{s}^{n}=\hat{s}$ for any integer $n$.} 
\begin{equation}
\hat{s}R_{ab}\doteqdot R_{ab}-\frac{1}{4}Rg_{ab}\Rightarrow \hat{s}%
R_{ab}\doteqdot (-2)\Xi _{ab}.  \tag{2.48}
\end{equation}%
Therefore, applying $\hat{s}$ to $G_{ab}$ yields the symbolic relationships 
\begin{equation}
\hat{s}G_{ab}=\hat{s}\hat{\tau}R_{ab}=\hat{\tau}\hat{s}R_{ab}=\hat{s}R_{ab},
\tag{2.49}
\end{equation}%
while Eqs. (2.47) become 
\begin{equation}
R_{ab}-\lambda g_{ab}=0\Leftrightarrow G_{ab}+\lambda g_{ab}=0,  \tag{2.50}
\end{equation}%
where $\lambda =6\varkappa $ is the cosmological constant.

In the presence of sources, Eqs. (2.50) have to be modified to 
\begin{equation}
R_{ab}-(12\varkappa -\lambda )g_{ab}=-\kappa T_{ab}  \tag{2.51}
\end{equation}%
and 
\begin{equation}
G_{ab}+\lambda g_{ab}=-\kappa T_{ab},  \tag{2.52}
\end{equation}%
where $T_{ab}$ amounts to the world version of the energy-momentum tensor of
the sources, and $\kappa $ is the Einstein gravitational constant. Hence,
transvecting with $g^{ab}$ either of Eqs. (2.51) and (2.52), yields the
extended trace relation 
\begin{equation}
\varkappa =\frac{1}{6}\lambda +\frac{1}{24}\kappa T,  \tag{2.53}
\end{equation}%
which particularly means that the suppression of the cosmological context
must just be ruled by the vanishing of $\lambda $. It turns out that the
full field equations are written as%
\begin{equation}
2\Xi _{ab}=\kappa (T_{ab}-\frac{1}{4}Tg_{ab}),  \tag{2.54a}
\end{equation}%
which amount to the same thing as%
\begin{equation}
\Xi _{ab}=\frac{\kappa }{2}\hat{s}T_{ab}.  \tag{2.54b}
\end{equation}%
In the case of a trace-free $T_{ab}$, Eqs. (2.54) get simplified to 
\begin{equation}
\Xi _{ab}=\frac{\kappa }{2}T_{ab}.  \tag{2.55}
\end{equation}

\section{SPIN-AFFINE GEOMETRY}

A natural procedure for bringing spinor covariant differentials in $%
\mathfrak{M}$ consists in carrying out affine displacements from one spin
space to another, which absorb the same geometric definition as the one for
the world situation. It appears that a set of world-spin affine correlations
may be most easily attainable by combining the strongly required covariant
constancy of the Hermitian connecting objects for the $\gamma $-formalism
and the covariant Leibniz expansion of an appropriate spin-tensor outer
product associated to a null world vector. It was originally realized [5]
that contracted spin affinities carrying nowhere-vanishing imaginary parts
should be introduced into the $\gamma \varepsilon $-framework because of the
necessity of balancing the overall numbers of independent world-spin affine
components. The expression for a spin affinity of either formalism can
consequently be obtained by first performing an appropriate index-splitting
of $\Gamma _{abc}$, and then calling for the corresponding world covariant
derivative patterns. An allowable spin-affine connexion is thus made out of
the spinor versions of both $\Gamma _{a[bc]}$ and the traceful part of $%
\Gamma _{a(bc)}$. In either formalism, the former $\Gamma $-contribution
supplies the symmetric part of a general two-piece spinor splitting which
has to be added to a non-Hermitian partial derivative. In the $\gamma $%
-formalism, the latter $\Gamma $-contribution makes up the scalar-density
prescription for the absolute value of a spin-metric function as brought up
in Section 1. A recovery of the real part of a contracted spin affinity for
the $\gamma $-formalism can be accomplished from such configurations, but
the feasibility of such a recovery ceases happening when the metric limiting
situation that yields the affine computational devices for the $\varepsilon $%
-formalism is implemented. The spacetime information carried by the metric
spinors of the $\gamma $-formalism is usually extracted from their partial
derivatives and brought out by a set of world-covariant vectors. One then
becomes able to derive in an elegant way a classical relationship involving
the metric quantities of the $\gamma $-formalism and the parts of the
respective contracted spin-affine structures. It is worth stressing that the
completion of this derivation does not depend upon the choice of any
expression for the torsion freeness of $\nabla _{a}$. The absolute value and
polar argument of the complex-valued function that defines a basic $\gamma $%
-metric component accordingly appear as world scalars, the absolute value
being formally given as the product of two world-scalar densities. It is
shown in Ref. [10] that the information on one of these densities is totally
contained in a suitably contracted partial derivative of an Hermitian
connecting object for the $\gamma $-formalism, whereas the information on
the other is carried by $\mathfrak{g}$, with the former density having to be
thought of as bearing a double world-spin character.

Before completing the geometric specifications of the metric spinors and
connecting objects for the $\varepsilon $-formalism, we will have to call
upon the result that any non-vanishing totally antisymmetric spin quantity
is proportional in the case of either formalism to one of the respective
metric spinors. Such specifications come all from a description of the gauge
transformation laws for the metric spinors of both formalisms. The usual
definition of spin densities [7, 8] is shaped upon the one which is adopted
in the world framework. It turns out that all metric and spin-affine
prescriptions have ultimately to be combined together with the world
invariance of the metric spinors. The full geometric characterization of the
systems of eigenvalue equations mentioned in Section 1, emerges from the
combination of the covariant constancy of $g_{ab}$ with the standard
relationships between the metric and connecting objects for the $\gamma $%
-formalism. We will place emphasis on the fact that the eigenvalues carried
by these equations may supply a technique for controlling the gauge
behaviours of the quantities involved in the limiting process. The
procedures concerning the specification of the gauge behaviours of
spin-affine connexions afford certain differential devices which enable one
to mix up and keep track of gauge frames when computing covariant
derivatives in the $\gamma $-formalism.

Subsection 3.1 exhibits the definitions of the metric spinors and connecting
objects for both formalisms. In Subsection 3.2, the gauge behaviours of the
basic objects for the $\varepsilon $-formalism are specified in conjunction
with the definitions of spin tensors and densities. We shall have to include
the definition of densities that bear a combined world-spin character
because of the occurrence of such an object in the expression for a typical $%
\gamma $-metric component. The definition of spin affinities along with the
relevant covariant derivative patterns and computational devices are shown
in Subsection 3.3. All eigenvalue equations and metric expressions are
deduced in Subsection 3.4. The gauge transformation laws for spin-affine
connexions as well as the introduction of the correspondence principle and a
detailed description of the limiting process are considered together in
Subsection 3.5. Gothic letters will also be used to denote weights of spin
densities. Without any risk of confusion, we will utilize the same symbol as
the one for the world-covariant differentials of Section 2 upon dealing with
covariant derivatives in both formalisms. It will be understood from now on
that world-spin characters are intrinsic geometric attributes which must not
as such depend upon the implementation of any $\nabla $-differentiation. A
horizontal bar lying over some kernel letter will denote the operation of
complex conjugation.

\subsection{Metric Spinors and Connecting Objects}

One of the fundamental metric spinors of the $\gamma $-formalism is taken as
a spin tensor of valence $\{0,2;0,0\}$, which bears skewness and invariance
under world-coordinate transformations. We have, in effect, 
\begin{equation}
{\large (}\gamma _{AB}{\large )}=\left( 
\begin{array}{ll}
0 & \gamma \\ 
-\gamma & 0%
\end{array}%
\right) ,\text{ }\gamma =\mid \gamma \mid \exp (i\Phi ).  \tag{3.1}
\end{equation}%
Either entry of the pair $(\mid \gamma \mid ,\Phi )$ is a smooth real-valued
function on $\mathfrak{M}$, and $\mid \gamma \mid \neq 0$ everywhere. The
inverse of $(\gamma _{AB})$ appears as a world-invariant spin tensor of
valence $\{2,0;0,0\}$, which is set as 
\begin{equation}
{\large (}\gamma ^{AB}{\large )}=\left( 
\begin{array}{ll}
0 & \gamma ^{-1} \\ 
-\gamma ^{-1} & 0%
\end{array}%
\right) .  \tag{3.2}
\end{equation}%
We have the component relationships 
\begin{equation}
\gamma _{AB}=\gamma \varepsilon _{AB},\text{ }\gamma ^{AB}=\gamma
^{-1}\varepsilon ^{AB},  \tag{3.3}
\end{equation}%
with 
\begin{equation}
(\varepsilon _{AB})=(\varepsilon _{A^{\prime }B^{\prime }})=\left( 
\begin{array}{ll}
0 & 1 \\ 
-1 & 0%
\end{array}%
\right) =(\varepsilon ^{AB})=(\varepsilon ^{A^{\prime }B^{\prime }}), 
\tag{3.4}
\end{equation}%
being the metric spinors for the $\varepsilon $-formalism, which are
likewise taken to bear world invariance. Hence, the independent component $%
\gamma $ of $\gamma _{AB}$ is a world scalar.\footnote{%
The gauge specification of $\gamma $ will be given in Subsection 3.2.} It
will be shown in Subsection 3.2 that the $\varepsilon $-metric spinors bear
a natural gauge-invariance property, whence we can say that Eq. (3.4)
defines the only metric spinors that occur in the $\varepsilon $-formalism.
We have the useful relations 
\begin{equation}
M^{CB}M_{CA}=M_{A}{}^{B}=-M^{B}{}_{A},  \tag{3.5a}
\end{equation}%
where the kernel letter $M$ stands here as elsewhere for either $\gamma $ or 
$\varepsilon $, and 
\begin{equation}
{\large (}M_{A}{}^{B}{\large )\doteqdot (}\delta _{A}{}^{B}{\large )}=\left( 
\begin{array}{ll}
1 & 0 \\ 
0 & 1%
\end{array}%
\right) .  \tag{3.5b}
\end{equation}

The metric spinors and their complex conjugates serve particularly for
lowering and raising indices of arbitrary spinor and world-spin quantities.
For some elementary spinor $\nu ^{A}$, for instance, we have the
upper-lower-index prescriptions 
\begin{equation}
\nu ^{A}=\gamma ^{AB}\nu _{B},\text{ }\nu _{A}=\nu ^{B}\gamma
_{BA}\Leftrightarrow \nu _{0}=-\gamma \nu ^{1},\text{ }\nu _{1}=\gamma \nu
^{0}  \tag{3.6a}
\end{equation}%
and 
\begin{equation}
\nu ^{A}=\varepsilon ^{AB}\nu _{B},\text{ }\nu _{A}=\nu ^{B}\varepsilon
_{BA}\Leftrightarrow \nu _{0}=-\nu ^{1},\text{ }\nu _{1}=\nu ^{0}. 
\tag{3.6b}
\end{equation}%
The processes of lowering and raising spinor indices in the $\gamma $%
-formalism always preserve intrinsic spin characters because of the
spin-tensor character of the metric configurations (3.1) and (3.2). It will
be emphasized in Subsection 3.2 that the action of the $\varepsilon $-metric
spinors does not generally retain the spin characters of the former objects.
However, in view of the world invariance of the structures (3.1)-(3.4), the
world characters of any spin objects will remain unchanged when we implement
the action of the metric spinors for either formalism.

The connecting objects of the $\gamma $-formalism are defined as 
\begin{equation}
2\sigma _{AA^{\prime }(a}\sigma _{b)}^{BA^{\prime }}=\gamma _{A}{}^{B}g_{ab},
\tag{3.7a}
\end{equation}%
or, alternatively, as the complex conjugate of (3.7a). Similarly, for the $%
\varepsilon $-formalism, we have%
\begin{equation}
2\Sigma _{AA^{\prime }(a}\Sigma _{b)}^{BA^{\prime }}=\varepsilon
_{A}{}^{B}g_{ab}.  \tag{3.7b}
\end{equation}%
All the entries of the set\footnote{%
Henceforth, the kernel letter $S$ will denote either $\sigma $ or $\Sigma $.}
\begin{equation}
\mathbf{H}=\{S_{aAA^{\prime }},S_{AA^{\prime }}^{a},S_{a}^{AA^{\prime
}},S^{aAA^{\prime }}\}{\large ,}  \tag{3.8}
\end{equation}%
are components of Hermitian $(2\times 2)$-matrices that depend smoothly upon 
$x^{a}$. The ordering of the indices carried by any $S$-symbol is immaterial
as unprimed and primed spinor indices supposedly take algebraically
independent values here. We should notice that the Hermiticity of the
elements of the set (3.8) is lost when we let their spinor indices share out
both stairs. Hence, manipulating the spinor indices of Eqs. (3.7) suitably,
and symmetrizing both sides over $AB$, yields the property 
\begin{equation}
S_{aA^{\prime }}^{(A}S_{b}^{B)A^{\prime }}=S_{A^{\prime
}[a}^{(A}S_{b]}^{B)A^{\prime }}=S_{A^{\prime }[a}^{A}S_{b]}^{BA^{\prime }}, 
\tag{3.9a}
\end{equation}%
and, consequently, we can also write 
\begin{equation}
S_{AA^{\prime }[a}S_{b]}^{AA^{\prime }}=0\Leftrightarrow S_{aAA^{\prime
}}S_{b}^{AA^{\prime }}=S_{AA^{\prime }(a}S_{b)}^{AA^{\prime }}.  \tag{3.9b}
\end{equation}%
The index configurations of (3.9) can be worked out so as to give the
contracted commutator%
\begin{equation}
\lbrack S_{aA^{\prime }}^{A},S_{b}^{BA^{\prime }}]=0,  \tag{3.10a}
\end{equation}%
which leads to the relations 
\begin{equation}
S_{A^{\prime }}^{a(A}S_{a}^{B)A^{\prime }}=0\Leftrightarrow S_{A^{\prime
}}^{aA}S_{a}^{BA^{\prime }}=S_{A^{\prime }}^{a[A}S_{a}^{B]A^{\prime }}. 
\tag{3.10b}
\end{equation}

In either formalism, the pertinent $S$-objects provide a one-to-one
correspondence between world and spin objects, which is written in terms of
adequate outer products.\footnote{%
This correspondence does not apply to $x^{a}$, but it naturally applies to $%
dx^{a}$.} Some notable examples are the following: 
\begin{equation}
g_{ab}{}=S_{a}^{AA^{\prime }}S_{b}^{BB^{\prime }}M_{AB}M_{A^{\prime
}B^{\prime }},\text{ }M_{AB}M_{A^{\prime }B^{\prime }}=S_{AA^{\prime
}}^{a}S_{BB^{\prime }}^{b}g_{ab}{}  \tag{3.11a}
\end{equation}%
and 
\begin{equation}
\partial _{a}=S_{a}^{AA^{\prime }}\partial _{AA^{\prime }}.  \tag{3.11b}
\end{equation}%
Thus, the spinor structure that corresponds to Eq. (2.26) is expressed by
[12] 
\begin{equation}
e_{AA^{\prime }BB^{\prime }CC^{\prime }DD^{\prime
}}=i(M_{AC}M_{BD}M_{A^{\prime }D^{\prime }}M_{B^{\prime }C^{\prime
}}-M_{AD}M_{BC}M_{A^{\prime }C^{\prime }}M_{B^{\prime }D^{\prime }}), 
\tag{3.12a}
\end{equation}%
which agrees with the trivial identities 
\begin{equation}
M_{[AB}M_{C]D}\equiv 0  \tag{3.12b}
\end{equation}%
and 
\begin{equation}
M_{A(B}M_{C)D}=M_{(A\mid B}M_{C\mid D)}=M_{B(A}M_{D)C}.  \tag{3.12c}
\end{equation}%
The combination of (3.3) and (3.11) produces the Hermitian associations 
\begin{equation}
\sigma _{AA^{\prime }}^{a}=\mid \gamma \mid \Sigma _{AA^{\prime }}^{a},\text{
}\sigma ^{aAA^{\prime }}=\mid \gamma \mid ^{-1}\Sigma ^{aAA^{\prime }}, 
\tag{3.13a}
\end{equation}%
along with the lower-world-index ones. An example of a $\sigma \Sigma $%
-association in the non-Hermitian case is given by 
\begin{equation}
\sigma _{aA}^{A^{\prime }}=\exp (i\Phi )\Sigma _{aA}^{A^{\prime }}. 
\tag{3.13b}
\end{equation}

It was said in Section 1 that any connecting object for either formalism is
thought of as a vector as regards world-coordinate transformations, whence
any outer products of $S$-objects must bear a world-tensor character. It
follows that any spinor associated to a world tensor will behave as a scalar
if only transformations belonging to the mapping group of $\mathfrak{M}$ are
performed. Likewise, since all the connecting objects for the $\gamma $%
-formalism are also considered as spin tensors, any couplings of $\sigma $%
-objects with purely world quantities will surely yield spin tensors, but
this generally fails to hold for the case of the $\varepsilon $-formalism.

\subsection{Spin Tensors and Densities}

The generalized gauge group [5, 6] consists of the set of all non-singular
complex $(2\times 2)$-matrices $(\Lambda _{A}{}^{B})$ whose components are
prescribed as\footnote{%
The symbol $\delta _{A}${}$^{B}$ denotes the spinor Kronecker delta such as
in (3.5b).} 
\begin{equation}
\Lambda _{A}{}^{B}=\sqrt{\rho }\exp (i\theta )\delta _{A}{}^{B}.  \tag{3.14a}
\end{equation}%
In Eq. (3.14a), $\rho $ is a positive-definite differentiable real-valued
function of $x^{a}$ and $\theta $ amounts to the gauge parameter of the
group, which is usually taken as an arbitrary differentiable real-valued
function on $\mathfrak{M}$. This group operates locally on the spin spaces
of $\mathfrak{M}$, independently of the effective action of the spacetime
mapping group. For the determinant of $(\Lambda _{A}{}^{B})$, we have the
expression 
\begin{equation}
\det (\Lambda _{A}{}^{B})\doteqdot \Delta _{{\small \Lambda }}=\rho \exp
(2i\theta ),  \tag{3.14b}
\end{equation}%
whence 
\begin{equation}
\Lambda _{A}{}^{B}\Lambda _{C}{}^{D}=\Delta _{{\small \Lambda }}\delta
_{A}{}^{B}\delta _{C}{}^{D},  \tag{3.14c}
\end{equation}%
and $\rho \doteqdot \mid \Delta _{\Lambda }\mid $. Any spin scalar is
defined as a numerical quantity that is invariant under gauge
transformations. By definition, one of the simplest indexed spin tensors is
an unprimed covariant spin vector which undergoes the transformation law 
\begin{equation}
\xi _{A}^{\prime }=\Lambda _{A}{}^{B}\xi _{B}.  \tag{3.15a}
\end{equation}%
Hence, requiring the inner product $\zeta ^{A}\xi _{A}$ to be gauge
invariant, yields the basic unprimed contravariant law 
\begin{equation}
\zeta ^{\prime A}=\zeta ^{B}\Lambda _{B}^{-1}{}^{A}.  \tag{3.15b}
\end{equation}%
Obviously, the transformation laws for primed spin vectors take up either
the complex conjugate matrix $(\Lambda _{A^{\prime }}{}^{B^{\prime }})$ or
its inverse.

The defining transformation laws for spin tensors of arbitrary valences are
usually obtained by performing outer products between spin vectors and
applying appropriately the prescriptions (3.15). Thus, the spin-tensor
character of the metric and connecting objects for the $\gamma $-formalism
is brought out by the covariant and contravariant configurations 
\begin{equation}
\gamma _{AB}^{\prime }=\Lambda _{A}{}^{C}\Lambda _{B}{}^{D}\gamma _{CD},%
\text{ }\gamma ^{\prime AB}=\gamma ^{CD}\Lambda _{C}^{-1}{}^{A}\Lambda
_{D}^{-1}{}^{B}  \tag{3.16}
\end{equation}%
and 
\begin{equation}
\sigma _{AA^{\prime }}^{\prime a}=\Lambda _{A}{}^{B}\Lambda _{A^{\prime
}}{}^{B^{\prime }}\sigma _{BB^{\prime }}^{a},\text{ }\sigma ^{\prime
aAA^{\prime }}=\sigma ^{aBB^{\prime }}\Lambda _{B}^{-1}{}^{A}\Lambda
_{B^{\prime }}^{-1}{}^{A^{\prime }},  \tag{3.17}
\end{equation}%
along with their complex conjugates and the lower-world-index versions. By
virtue of (3.14c), the laws (3.16) and (3.17) can be rewritten as 
\begin{equation}
\gamma _{AB}^{\prime }=\Delta _{{\small \Lambda }}\gamma _{AB},\text{ }%
\gamma ^{\prime AB}=\delta _{{\small \Lambda }}\gamma ^{AB}  \tag{3.18}
\end{equation}%
and\footnote{%
Our choices of world and spin frames are somehow reversed.} 
\begin{equation}
\sigma _{AA^{\prime }}^{\prime a}=\mid \Delta _{{\small \Lambda }}\mid
\sigma _{AA^{\prime }}^{a},\text{ }\sigma ^{\prime aAA^{\prime }}=\mid
\Delta _{{\small \Lambda }}\mid ^{-1}\sigma ^{aAA^{\prime }},  \tag{3.19}
\end{equation}%
with $\delta _{\Lambda }\doteqdot (\Delta _{\Lambda })^{-1}$. For the
non-Hermitian $\sigma $-objects, we have, for instance, 
\begin{equation}
\sigma _{aA^{\prime }}^{\prime B}=\Lambda _{A^{\prime }}{}^{B^{\prime
}}\sigma _{aB^{\prime }}^{C}\Lambda _{C}^{-1}{}^{B}\Leftrightarrow \sigma
_{aA^{\prime }}^{\prime B}=\exp (-2i\theta )\sigma _{aA^{\prime }}^{B}. 
\tag{3.20}
\end{equation}

Inasmuch as the spin spaces of $\mathfrak{M}$ are all two-dimensional, the
only useful totally antisymmetric spin objects bear two indices of the same
type. In the spin-tensor case, such an object $\eta _{AB}$ has the form 
\begin{equation}
\eta _{AB}=\eta _{\lbrack AB]}=\frac{1}{2}\eta \gamma _{AB},  \tag{3.21}
\end{equation}%
with $\eta =\eta _{C}{}^{C}$ thus being a spin scalar. Traditionally [1, 5],
the definitions of complex spin-scalar densities of weights $+1$ and $-1$
were obtained from the combination of the transformation laws (3.18) with
the prescription (3.21) and its contravariant version. Such entities thus
undergo the same gauge transformation laws as the individual independent
components of $\gamma _{AB}$ or $\gamma ^{AB}$, respectively. For a complex
spin-scalar density $\alpha $ of weight $\mathfrak{w}$, we have the extended
definition 
\begin{equation}
\alpha ^{\prime }=(\Delta _{{\small \Lambda }})^{\mathfrak{w}}\alpha . 
\tag{3.22}
\end{equation}%
It is clear that the action of the operation of complex conjugation on
spin-scalar densities can be defined as an interchange involving the
non-vanishing unprimed and primed $\gamma $-metric components. The complex
conjugate of $\alpha $ is sometimes called [5] a spin-scalar density of
antiweight $\mathfrak{w}$. Performing outer products between these densities
produces other spin-scalar densities whose weights and antiweights equal the
sums of the corresponding attributes carried by the couplings. Therefore, a
spin-scalar density $\beta $ of weight $\mathfrak{a}$ and antiweight $%
\mathfrak{b}$ transforms as 
\begin{equation}
\beta ^{\prime }=(\Delta _{{\small \Lambda }})^{\mathfrak{a}}(\bar{\Delta}_{%
{\small \Lambda }})^{\mathfrak{b}}\beta .  \tag{3.23a}
\end{equation}%
When $\mathfrak{a}=\mathfrak{b}$, the density $\beta $ is said to bear an
absolute weight $2\mathfrak{a}$, whence it would behave under gauge
transformations as 
\begin{equation}
\beta ^{\prime }=\mid \Delta _{{\small \Lambda }}\mid ^{2\mathfrak{a}}\beta .
\tag{3.23b}
\end{equation}%
Then, spin-scalar densities of absolute weights $\pm 1$ are subject to the
same transformation laws as the components of the connecting objects
involved in (3.19). The pattern (3.23a) may be specialized still further in
case Hermiticity is required to be preserved under gauge transformations.
Consequently, any real spin-scalar density must bear an absolute weight. It
is of some interest to take into consideration spin-scalar densities that
simultaneously bear weights, antiweights as well as absolute weights. For
such a composite density $\breve{A}$, we have the prescription 
\begin{equation}
\breve{A}^{\prime }=(\Delta _{{\small \Lambda }})^{\mathfrak{a}}(\bar{\Delta}%
_{{\small \Lambda }})^{\mathfrak{b}}\mid \Delta _{{\small \Lambda }}\mid ^{%
\mathfrak{c}}\breve{A}.  \tag{3.24}
\end{equation}

Arbitrary spin-tensor densities were originally defined [7, 8] as outer
products between spin tensors and scalar densities, in formal analogy with
the world situation. Conventionally, the entries of the arrays that specify
the valences of outer products between any spin-tensor densities are taken
as the sums of the corresponding entries of the valences borne by the
involved coupled tensors, while the overall weights and antiweights are
prescribed in the same way as for coupled spin-scalar densities. In
particular, any Hermitian spin-tensor density must be viewed as the product
of an Hermitian tensor with a real spin-scalar density. Of course, we can
occasionally build up spin tensors by performing products that carry
suitable spin scalar and tensor densities. Configurations that possess a
mixed world-spin density character can also be constructed by performing
outer products between world and spin densities. Particularly interesting
world-spin scalar densities have the form $(-\mathfrak{g})^{N}\alpha $.

The easiest procedure for bringing forward the gauge characters of the $%
\varepsilon $-metric spinors involves the combination of Eqs. (3.3) and
(3.18). In effect, we have the laws 
\begin{equation}
\varepsilon _{AB}^{\prime }=(\Delta _{{\small \Lambda }})^{-1}\Lambda
_{A}{}^{C}\Lambda _{B}{}^{D}\varepsilon _{CD}=\varepsilon _{AB}  \tag{3.25a}
\end{equation}%
and 
\begin{equation}
\varepsilon ^{\prime AB}=\Delta _{{\small \Lambda }}\varepsilon ^{CD}\Lambda
_{C}^{-1}{}^{A}\Lambda _{D}^{-1}{}^{B}=\varepsilon ^{AB},  \tag{3.25b}
\end{equation}%
along with their complex conjugates. It follows that we can write down the
conjugate schemes 
\begin{equation*}
\begin{array}{lll}
\varepsilon _{AB} & \rightarrow & \text{invariant spin-tensor density of
weight }-1 \\ 
\varepsilon ^{AB} & \rightarrow & \text{invariant spin-tensor density of
weight }+1%
\end{array}%
\end{equation*}%
and 
\begin{equation*}
\begin{array}{lll}
\varepsilon _{A^{\prime }B^{\prime }} & \rightarrow & \text{invariant
spin-tensor density of antiweight }-1 \\ 
\varepsilon ^{A^{\prime }B^{\prime }} & \rightarrow & \text{invariant
spin-tensor density of antiweight }+1.%
\end{array}%
\end{equation*}%
Any metric spinor for the $\varepsilon $-formalism can then be considered as
a spin Levi-Civita symbol. It should be stressed by this point that $\Delta
_{\Lambda }$ is formally expressed in both formalisms as\footnote{%
The expression (3.26) is analogous to (2.28b).} 
\begin{equation}
\Delta _{{\small \Lambda }}=\frac{1}{2}M{}^{AB}\Lambda _{A}{}^{C}\Lambda
_{B}{}^{D}M_{CD}.  \tag{3.26}
\end{equation}%
Whereas the metric components $(\gamma ,$ $\gamma ^{-1})$ and $(\overline{%
\gamma },$ $\overline{\gamma }^{-1})$ thus have to be regarded as
spin-scalar densities of weights $(+1,-1)$ and antiweights $(+1,-1)$, the
absolute values $(\mid \gamma \mid ,\mid \gamma \mid ^{-1})$ must be looked
upon as real spin-scalar densities of absolute weights $(+1,-1)$,
respectively. In addition, the polar piece $\exp (i\Phi )$ of $\gamma $ must
behave as a composite spin-scalar density, namely 
\begin{equation}
\exp (i\Phi ^{\prime })=\Delta _{{\small \Lambda }}\mid \Delta _{{\small %
\Lambda }}\mid ^{-1}\exp (i\Phi ).  \tag{3.27}
\end{equation}%
Accordingly, Eqs. (3.13a) yield at once the Hermitian prescriptions 
\begin{equation*}
\begin{array}{lll}
\Sigma _{aAA^{\prime }} & \rightarrow & \text{invariant spin-tensor density
of absolute weight }-1 \\ 
\Sigma _{a}^{AA^{\prime }} & \rightarrow & \text{invariant spin-tensor
density of absolute weight }+1.%
\end{array}%
\end{equation*}%
More explicitly, we have 
\begin{equation}
\Sigma _{aAA^{\prime }}^{\prime }=\mid \Delta _{{\small \Lambda }}\mid
^{-1}\Lambda _{A}{}^{B}\Lambda _{A^{\prime }}{}^{B^{\prime }}\Sigma
_{aBB^{\prime }}=\Sigma _{aAA^{\prime }}  \tag{3.28a}
\end{equation}%
and 
\begin{equation}
\Sigma _{a}^{\prime AA^{\prime }}=\mid \Delta _{{\small \Lambda }}\mid
\Sigma _{a}^{BB^{\prime }}\Lambda _{B}^{-1}{}^{A}\Lambda _{B^{\prime
}}^{-1}{}^{A^{\prime }}=\Sigma _{a}^{AA^{\prime }}.  \tag{3.28b}
\end{equation}

We can now see that the implementation of (3.16) ensures the preservation of
spin characters when the processes of lowering and raising spinor indices
take place in the $\gamma $-formalism. In turn, Eqs. (3.25) and their
complex conjugates show us that the change in the $\varepsilon $-formalism
of the spinor-index configuration of an arbitrary spin object generally
produces a modification of the values of the pertinent weights and
antiweights. Hence, correspondences between world and spin objects in the $%
\varepsilon $-formalism do not generally involve spin tensors.

\subsection{Spin Affinities and Covariant Derivatives}

Following the work of Ref. [10], we consider two neighbouring spin spaces of 
$\mathfrak{M}$ which are set up at $x^{a}$ and $x^{a}+dx^{a}$. A covariant
differential of some contravariant spin vector $\zeta ^{A}$ is defined as
the local difference between the value of $\zeta ^{A}$ at $x^{a}+dx^{a}$ and
the value at $x^{a}$ of the spin vector that results from an affine
displacement of $\zeta ^{A}$. The patterns of spin displacements were
originally chosen [5] so as to resemble closely the form borne by the ones
which occur in the purely world framework. In either formalism, a typical
covariant differential configuration looks like%
\begin{equation}
D\zeta ^{A}=d\zeta ^{A}+\vartheta _{aB}{}^{A}\zeta ^{B}dx^{a},  \tag{3.29}
\end{equation}%
with $\vartheta _{aB}{}^{A}$ amounting to the unprimed-index spin-affine
connexion associated to the displacement eventually carried out. For the
corresponding covariant derivative, we have 
\begin{equation}
\nabla _{a}\zeta ^{A}=\partial _{a}\zeta ^{A}+\vartheta _{aB}{}^{A}\zeta
^{B}.  \tag{3.30}
\end{equation}%
Either $D$-differential of a covariant spin vector $\xi _{A}$ can be rapidly
obtained from (3.29) by taking for granted the Leibniz rule and demanding
that 
\begin{equation}
D(\zeta ^{A}\xi _{A})=d(\zeta ^{A}\xi _{A}),  \tag{3.31}
\end{equation}%
whence we also have 
\begin{equation}
\nabla _{a}\xi _{A}=\partial _{a}\xi _{A}-\vartheta _{aA}{}^{B}\xi _{B}, 
\tag{3.32}
\end{equation}%
together with the complex conjugates of the prescriptions (3.30) and (3.32).
We stress that each of the pieces which occur on the right-hand sides of
Eqs. (3.30) and (3.32) must behave covariantly under the action of the
manifold mapping group of $\mathfrak{M}$, in contrast with the world
situation. As for the world case, covariant derivatives of spin tensors of
arbitrary valences can be obtained by allowing for outer products between
spin vectors and carrying out Leibniz expansions thereof.

World and spin displacements in $\mathfrak{M}$ turn out to be induced by
each other when the covariant constancy requirement 
\begin{equation}
DS_{AA^{\prime }}^{a}=0  \tag{3.33}
\end{equation}%
is implemented. Whenever a tensor quantity that carries both world and spin
indices is differentiated covariantly in both formalisms, we will thus have
to incorporate into the pertinent expansions the affine contributions
associated with all the indices borne by the quantity being considered. Any
such mixed expansion must be regarded as a result of the implementation of
combined world-spin displacements in $\mathfrak{M}$. In fact, the simplest
procedure for establishing this geometric property of the formalisms just
accounts for affine displacements of the following $\gamma $-formalism
configuration: 
\begin{equation}
n^{a}=\sigma _{AA^{\prime }}^{a}\zeta ^{A}\zeta ^{A^{\prime }},\text{ }%
n^{a}n_{a}=0.  \tag{3.34a}
\end{equation}%
Hence, writing 
\begin{equation}
Dn^{a}=\sigma _{AA^{\prime }}^{a}D(\zeta ^{A}\zeta ^{A^{\prime }}), 
\tag{3.34b}
\end{equation}%
and performing a Leibniz expansion, yields the correlation [10] 
\begin{equation}
\Gamma _{bc}{}^{a}n^{b}dx^{c}=\sigma _{AA^{\prime }}^{a}(\gamma
_{bB}{}^{A}\zeta ^{B}\zeta ^{A^{\prime }}+\gamma _{bB^{\prime
}}{}^{A^{\prime }}\zeta ^{A}\zeta ^{B^{\prime }})dx^{b}-\zeta ^{A}\zeta
^{A^{\prime }}d\sigma _{AA^{\prime }}^{a},  \tag{3.34c}
\end{equation}%
with $\gamma _{aA}{}^{B}$ standing for the $\gamma $-formalism version of $%
\vartheta _{aA}{}^{B}$. It becomes clear that (3.33) has now to be written
out as the vanishing derivative 
\begin{equation}
\nabla _{a}\sigma _{BB^{\prime }}^{b}=\partial _{a}\sigma _{BB^{\prime
}}^{b}+\Gamma _{ac}{}^{b}\sigma _{BB^{\prime }}^{c}-\gamma _{aB}{}^{C}\sigma
_{CB^{\prime }}^{b}-\gamma _{aB^{\prime }}{}^{C^{\prime }}\sigma
_{BC^{\prime }}^{b}.  \tag{3.35}
\end{equation}%
Differentiating covariantly both sides of either of Eqs. (3.11a) then brings
about the metric condition 
\begin{equation}
\nabla _{a}(\gamma _{BC}\gamma _{B^{\prime }C^{\prime }})=0,  \tag{3.36}
\end{equation}%
and, consequently, also its upper-spinor-index version. It follows that any
Hermitian connecting object for the $\gamma $-formalism bears covariant
constancy, whence we have the somewhat important relation 
\begin{equation}
\func{Re}(\gamma ^{BC}\nabla _{a}\gamma _{BC})=0,  \tag{3.37}
\end{equation}%
together with the one which is obtained from it by interchanging the
spinor-index stairs. Since $\nabla _{a}\delta _{C}{}^{D}=0$ invariantly, we
also obtain 
\begin{equation}
\gamma ^{BD}\nabla _{a}\gamma _{BC}+\gamma _{BC}\nabla _{a}\gamma ^{BD}=0. 
\tag{3.38}
\end{equation}

In both formalisms, Eq. (3.33) ensures a recovery of covariant differential
patterns for world tensors from those for Hermitian spin tensors. It becomes
imperative in any case to regularize the number of spin-affine components so
as to attain a compatible relationship with $\Gamma _{abc}$. The index
configuration of $\vartheta _{aA}{}^{B}$ supplies $32$ real independent
components whence the contracted structure $\vartheta _{aB}{}^{B}$ has to
carry explicitly nowhere-vanishing real and imaginary parts. In the $\gamma $%
-formalism, the real part automatically comes about by expanding the
condition (3.36) and invoking Eq. (3.3) together with its complex conjugate.
We have, in effect, 
\begin{equation}
\nabla _{a}(\gamma _{BC}\gamma _{B^{\prime }C^{\prime }})=(\partial _{a}\log
\mid \gamma \mid ^{2}-2\func{Re}\gamma _{aD}{}^{D})\gamma _{BC}\gamma
_{B^{\prime }C^{\prime }},  \tag{3.39}
\end{equation}%
which immediately produces the constraint 
\begin{equation}
\func{Re}\gamma _{aB}{}^{B}=\partial _{a}\log \mid \gamma \mid .  \tag{3.40}
\end{equation}%
It should be noticed that the individual terms of (3.40) bear world
covariance as $\gamma $ presumably is a world scalar. However, we can not
rewrite it by replacing $\partial _{a}$ with $\nabla _{a}$ because of the
spin-density character of $\gamma $. The original regularization procedure
for the $\gamma $-formalism [5] was carried through by implementing by hand
a make-up constraint for $\gamma _{aB}{}^{B}$ that involves a prescription
of the type 
\begin{equation}
\func{Im}\gamma _{aB}{}^{B}=(-2)\Phi _{a},  \tag{3.41}
\end{equation}%
with $\Phi _{a}$ being a world vector. What should be emphatically observed
in respect of this situation is that covariant differentials in the $\gamma $%
-formalism of any Hermitian $\sigma $-objects, and thence also Eq. (3.39)
itself, remain all unaffected\footnote{%
This applies to the $\varepsilon $-framework as well. The regularization
procedure for the $\varepsilon $-formalism will be entertained later in this
Section.} when purely imaginary world-covariant quantities like $i\iota
_{a}\delta _{B}{}^{C}$ are added to $\gamma _{aB}{}^{C}$. Consequently,
combining (3.40) and (3.41) yields the structure%
\begin{equation}
\gamma _{aB}{}^{B}=-(\theta _{a}+2i\Phi _{a}),  \tag{3.42}
\end{equation}%
with the definition 
\begin{equation}
\theta _{a}\doteqdot \partial _{a}\log (\mid \gamma \mid ^{-1}).  \tag{3.43}
\end{equation}%
When dealing with covariant differentiations in $\mathfrak{M}$, we thus have
to call for the affine relationship 
\begin{equation}
\Gamma _{AA^{\prime }BB^{\prime }CC^{\prime }}+\sigma _{sCC^{\prime
}}\partial _{AA^{\prime }}\sigma _{BB^{\prime }}^{s}=\gamma _{AA^{\prime
}BC}\gamma _{B^{\prime }C^{\prime }}+\gamma _{AA^{\prime }B^{\prime
}C^{\prime }}\gamma _{BC},  \tag{3.44a}
\end{equation}%
along with the splittings [10] 
\begin{equation}
\sigma _{AA^{\prime }}^{a}\sigma _{BB^{\prime }}^{b}\sigma _{CC^{\prime
}}^{c}\Gamma _{a(bc)}=\Gamma _{A(BC)A^{\prime }(B^{\prime }C^{\prime })}+%
\frac{1}{4}\Gamma _{AA^{\prime }}\gamma _{BC}\gamma _{B^{\prime }C^{\prime }}
\tag{3.44b}
\end{equation}%
and 
\begin{equation}
\sigma _{AA^{\prime }}^{a}\sigma _{BB^{\prime }}^{b}\sigma _{CC^{\prime
}}^{c}\Gamma _{a[bc]}=\Theta _{AA^{\prime }BC}\gamma _{B^{\prime }C^{\prime
}}+\Theta _{AA^{\prime }B^{\prime }C^{\prime }}\gamma _{BC},  \tag{3.44c}
\end{equation}%
where 
\begin{equation}
\Gamma _{A(BC)A^{\prime }(B^{\prime }C^{\prime })}=\sigma _{AA^{\prime
}}^{a}\sigma _{BB^{\prime }}^{b}\sigma _{CC^{\prime }}^{c}\hat{s}\Gamma
_{a(bc)},  \tag{3.44d}
\end{equation}%
\begin{equation}
\Gamma _{AA^{\prime }}\doteqdot \sigma _{AA^{\prime }}^{a}\Gamma _{a}=\Gamma
_{AA^{\prime }MM^{\prime }}{}^{MM^{\prime }}  \tag{3.44e}
\end{equation}%
and 
\begin{equation}
2\Theta _{AA^{\prime }BC}{}=\sigma _{AA^{\prime }}^{a}\sigma
_{(B}^{bD^{\prime }}\partial _{C)D^{\prime }}g_{ab}=\Gamma _{A(BC)A^{\prime
}M^{\prime }}{}^{M^{\prime }}=2\Theta _{AA^{\prime }(BC)}{},  \tag{3.44f}
\end{equation}%
with the purely world kernel of (3.44d) being given by the trace-free
relation%
\begin{equation}
\hat{s}\Gamma _{a(bc)}=\Gamma _{a(bc)}-\frac{1}{4}\Gamma _{a}g_{bc}. 
\tag{3.44g}
\end{equation}%
At this point, we can manipulate the index configuration of (3.44a) to
produce the formulae 
\begin{equation}
\Gamma _{A(BC)A^{\prime }(B^{\prime }C^{\prime })}={\small -}\text{ }\sigma
_{AA^{\prime }}^{a}\sigma _{s(B(B^{\prime }}\partial _{\mid a\mid }\sigma
_{C^{\prime })C)}^{s},  \tag{3.45a}
\end{equation}%
\begin{equation}
\gamma _{a(BC)}=\Theta _{aBC}+\frac{1}{2}\sigma _{s(B}^{B^{\prime }}\partial
_{\mid a\mid }\sigma _{C)B^{\prime }}^{s}  \tag{3.45b}
\end{equation}%
and 
\begin{equation}
4\func{Re}\gamma _{aB}{}^{B}=\Gamma _{a}+\sigma _{s}^{BB^{\prime }}\partial
_{a}\sigma _{BB^{\prime }}^{s},  \tag{3.45c}
\end{equation}%
along with the complex conjugate of (3.45b).

For establishing the legitimacy of the splitting (3.44b), it is convenient
to make use of the definition (3.43) to spell out the statement [10] 
\begin{equation}
\sigma _{BB^{\prime }}^{b}\sigma _{CC^{\prime }}^{c}\partial _{AA^{\prime
}}g_{bc}=-[2\theta _{AA^{\prime }}\gamma _{BC}\gamma _{B^{\prime }C^{\prime
}}+g_{bc}\partial _{AA^{\prime }}(\sigma _{BB^{\prime }}^{b}\sigma
_{CC^{\prime }}^{c})],  \tag{3.46a}
\end{equation}%
which amounts to nothing else but a spinor version of the relation (2.9b).
Equations (3.9) imply that the products of $g_{bc}$ with the partial
derivatives of the crossed pieces 
\begin{equation}
(\sigma _{(B[B^{\prime }}^{b}\sigma _{C^{\prime }]C)}^{c},\text{ }\sigma
_{\lbrack B(B^{\prime }}^{b}\sigma _{C^{\prime })C]}^{c}),  \tag{3.46b}
\end{equation}%
both vanish, whereas the product that carries the partial derivative of the
totally symmetric piece is given by 
\begin{equation}
g_{bc}\partial _{AA^{\prime }}(\sigma _{(B(B^{\prime }}^{b}\sigma
_{C^{\prime })C)}^{c})=(-2)\Gamma _{A(BC)A^{\prime }(B^{\prime }C^{\prime
})}.  \tag{3.46c}
\end{equation}%
The contribution that involves the totally antisymmetric piece is expressed
as 
\begin{equation}
g_{bc}\partial _{AA^{\prime }}(\sigma _{\lbrack B[B^{\prime }}^{b}\sigma
_{C^{\prime }]C]}^{c})=\frac{1}{4}\mid \gamma \mid ^{-2}\gamma _{BC}\gamma
_{B^{\prime }C^{\prime }}g_{bc}\partial _{AA^{\prime }}(\mid \gamma \mid
^{2}g^{bc}),  \tag{3.46d}
\end{equation}%
whence fitting pieces together establishes the relevant recovery. We point
out that the torsionlessness condition (2.7) can be expressed as the
configuration 
\begin{equation}
\Gamma _{(ABC)A^{\prime }B^{\prime }C^{\prime }}=\Gamma _{(ABC)(A^{\prime
}B^{\prime })C^{\prime }}=\Gamma _{(ABC)C^{\prime }(A^{\prime }B^{\prime
})}+2\Theta _{(ABC)(A^{\prime }}\gamma _{B^{\prime })C^{\prime }}. 
\tag{3.47a}
\end{equation}%
Since both of the world $\Gamma $-structures of (3.44b) and (3.44c) do not
bear symmetry in $a$, $b$, we can say that Eq. (3.44a) does not generally
lead to the statement\footnote{%
Equation (3.47b) gives rise to typical spin-affine patterns for the class of
conformally flat spacetimes referred to in Section 1.} 
\begin{equation}
\gamma _{A^{\prime }(ABC)}{}=0.  \tag{3.47b}
\end{equation}%
In addition, we can fix up the primed-index symmetry exhibited by the
relation (3.47a) by making use of Eqs. (3.44) and performing the calculation 
\begin{eqnarray*}
\Gamma _{(ABC)A^{\prime }B^{\prime }C^{\prime }} &=&\Gamma _{(ABC)A^{\prime
}(B^{\prime }C^{\prime })}+\Theta _{(ABC)A^{\prime }}\gamma _{B^{\prime
}C^{\prime }} \\
&=&\frac{1}{2}(\Gamma _{(ABC)A^{\prime }B^{\prime }C^{\prime }}+\Gamma
_{(ABC)C^{\prime }(A^{\prime }B^{\prime })}+\frac{1}{2}\Gamma
_{(ABC)C^{\prime }D^{\prime }}{}^{D^{\prime }}\gamma _{A^{\prime }B^{\prime
}}) \\
&&+(\Theta _{(ABC)(A^{\prime }}\gamma _{B^{\prime })C^{\prime }}-\frac{1}{2}%
\Theta _{(ABC)C^{\prime }}\gamma _{A^{\prime }B^{\prime }}) \\
&=&\frac{1}{2}(\Gamma _{(ABC)A^{\prime }B^{\prime }C^{\prime }}+\Gamma
_{(ABC)C^{\prime }(A^{\prime }B^{\prime })})+\Theta _{(ABC)(A^{\prime
}}\gamma _{B^{\prime })C^{\prime }}.
\end{eqnarray*}

The basic $\gamma $-formalism device for computing covariant derivatives of
spin densities is traditionally taken as an affine quantity $\gamma _{a}$
that arises out of the metric prescription [5] 
\begin{equation}
\nabla _{a}\varepsilon _{BC}=0\Leftrightarrow \gamma _{a}-\gamma
_{aB}{}^{B}=0.  \tag{3.48}
\end{equation}%
Consequently, $\gamma _{a}$ behaves under changes of coordinates in $%
\mathfrak{M}$ as a covariant vector. It thus occurs in the formal
configuration%
\begin{equation}
\nabla _{a}\gamma _{BC}=\nabla _{a}(\gamma \varepsilon _{BC})=\varepsilon
_{BC}\nabla _{a}\gamma ,  \tag{3.49a}
\end{equation}%
and likewise enters the expansion%
\begin{equation}
\nabla _{a}\gamma =\partial _{a}\gamma -\gamma \gamma _{a},  \tag{3.49b}
\end{equation}%
which constitutes the prototype in the $\gamma $-formalism for covariant
derivatives of complex spin-scalar densities of weight $+1$. Evidently, the
right-hand side of (3.49b) stands for a covariant expansion for the
independent component of $\gamma _{AB}$. For the density (3.22), we then
have 
\begin{equation}
\nabla _{a}\alpha =\partial _{a}\alpha -\mathfrak{w}\alpha \gamma _{a}. 
\tag{3.50}
\end{equation}%
Needless to say, the computational device that arises from 
\begin{equation}
\nabla _{a}\varepsilon _{B^{\prime }C^{\prime }}=0\Leftrightarrow \bar{\gamma%
}_{a}-\gamma _{aB^{\prime }}{}^{B^{\prime }}=0,  \tag{3.51}
\end{equation}%
appears to be appropriate for the case that involves the complex conjugates
of spin-scalar densities. When differentiating covariantly spin-scalar
densities that bear both weights and antiweights, we must therefore utilize
devices prescribed as suitable linear combinations of $\gamma _{a}$ and $%
\bar{\gamma}_{a}$. For the density (3.23a), for instance, we have 
\begin{equation}
\nabla _{a}\beta =\partial _{a}\beta -\beta (\mathfrak{a}\gamma _{a}+%
\mathfrak{b}\bar{\gamma}_{a}).  \tag{3.52}
\end{equation}%
If $\beta $ bears an absolute weight according to (3.23b), we will get 
\begin{equation}
\nabla _{a}\beta =\partial _{a}\beta -2\mathfrak{a}\beta \func{Re}\gamma
_{a},  \tag{3.53a}
\end{equation}%
that is to say, 
\begin{equation}
\nabla _{a}\beta =\partial _{a}\beta +2\mathfrak{a}\beta \theta _{a}. 
\tag{3.53b}
\end{equation}%
Hence, the combination of the definition (3.43) with the expansion 
\begin{equation}
\nabla _{a}\mid \gamma \mid =\partial _{a}\mid \gamma \mid +\mid \gamma \mid
\theta _{a},  \tag{3.53c}
\end{equation}%
tells us that $\mid \gamma \mid $ is covariantly constant in the $\gamma $%
-formalism. The affine device for the spin-scalar density (3.24) is thus
prescribed as 
\begin{equation}
\nabla _{a}\breve{A}=\partial _{a}\breve{A}-\breve{A}(\mathfrak{a}\gamma
_{a}+\mathfrak{b}\bar{\gamma}_{a}+\mathfrak{c}\func{Re}\gamma _{a}). 
\tag{3.54a}
\end{equation}%
As an interesting example, we have 
\begin{eqnarray}
\nabla _{a}[\exp (i\Phi )] &=&\partial _{a}[\exp (i\Phi )]-\exp (i\Phi
)(\gamma _{a}-\func{Re}\gamma _{a})  \notag \\
&=&i(\partial _{a}\Phi -\func{Im}\gamma _{a})\exp (i\Phi )  \notag \\
&=&i(\partial _{a}\Phi +2\Phi _{a})\exp (i\Phi ).  \TCItag{3.54b}
\end{eqnarray}%
Covariant differential patterns for arbitrary spin-tensor densities can be
specified by invoking the outer-product prescriptions given previously. For
instance, setting 
\begin{equation}
U_{BC...D}\doteqdot \beta T_{BC...D},  \tag{3.55}
\end{equation}%
with $T_{BC...D}$ being some spin tensor, yields the expansion 
\begin{equation}
\nabla _{a}U_{B...}=\partial _{a}U_{B...}-\gamma _{aB}{}^{M}U_{M...}-...-(%
\mathfrak{a}\gamma _{a}+\mathfrak{b}\bar{\gamma}_{a})U_{B...}.  \tag{3.56}
\end{equation}%
The covariant derivative of $\Sigma _{aAA^{\prime }}$, say, is thus written
down as 
\begin{equation}
\nabla _{a}\Sigma _{bBB^{\prime }}=\partial _{a}\Sigma _{bBB^{\prime
}}-\Gamma _{ab}{}^{c}\Sigma _{cBB^{\prime }}-\gamma _{aB}{}^{M}\Sigma
_{bMB^{\prime }}-\gamma _{aB^{\prime }}{}^{M^{\prime }}\Sigma _{bBM^{\prime
}}-\theta _{a}\Sigma _{bBB^{\prime }}.  \tag{3.57}
\end{equation}%
When combined with (3.13a), the property 
\begin{equation}
\nabla _{a}\mid \gamma \mid =0  \tag{3.58}
\end{equation}%
then enables us to state that the derivative (3.57) vanishes. Therefore, the
prescriptions (3.48) and (3.51) imply that all the other $\Sigma $%
-connecting objects must also be thought of as bearing covariant constancy
in the $\gamma $-formalism.

We see from Eqs. (3.30) and (3.32) that the rules for writing covariant
derivatives in both formalisms are symbolically the same, but a
corresponding spin-affine connexion $\Gamma _{aB}{}^{C}$ and its complex
conjugate should effectively take over the computational role within the $%
\varepsilon $-formalism. Thus, for an Hermitian world-spin tensor $\kappa
_{BB^{\prime }}^{b}$, we must have the $\varepsilon $-formalism expansion 
\begin{equation}
\nabla _{a}\kappa _{BB^{\prime }}^{b}=\partial _{a}\kappa _{BB^{\prime
}}^{b}+\Gamma _{ac}{}^{b}\kappa _{BB^{\prime }}^{c}-\Gamma _{aB}{}^{C}\kappa
_{CB^{\prime }}^{b}-\Gamma _{aB^{\prime }}{}^{C^{\prime }}\kappa
_{BC^{\prime }}^{b},  \tag{3.59a}
\end{equation}%
which is manifestly invariant under the world-covariant changes 
\begin{equation}
\Gamma _{aB}{}^{C}\mapsto \Gamma _{aB}{}^{C}+i\iota _{a}\varepsilon
_{B}{}^{C},\text{ }\Gamma _{aB^{\prime }}{}^{C^{\prime }}\mapsto \Gamma
_{aB^{\prime }}{}^{C^{\prime }}-i\iota _{a}\varepsilon _{B^{\prime
}}{}^{C^{\prime }},\text{ }\func{Re}(i\iota _{a})=0.  \tag{3.59b}
\end{equation}%
In Ref. [10], it was observed that a procedure for building up $\Gamma
_{aB}{}^{C}$ could consist in implementing the relationships (3.3) and
taking the limit as $\gamma $ tends to $1$. Putting it into practice would
nevertheless entail the annihilation of $\theta _{a}$, whence the numbers of
independent components of $\Gamma _{ab}{}^{c}{}$ and $\Gamma _{aB}{}^{C}{}$
would have to be regularized from the beginning. Accordingly, we must
necessarily take up the contracted prescription [5] 
\begin{equation}
-\func{Re}\Gamma _{aB}{}^{B}=\Pi _{a},  \tag{3.60}
\end{equation}%
whence the overall expression for $\Gamma _{aB}{}^{B}$ has to be written as 
\begin{equation}
\Gamma _{aB}{}^{B}=-(\Pi _{a}+2i\varphi _{a}),  \tag{3.61}
\end{equation}%
with $\Pi _{a}$ and $\varphi _{a}$ being world vectors. It is well known
[10] that no metric meaning can be assigned to $\func{Re}\Gamma _{aB}{}^{B}$
anyway. When considered together with Eq. (3.43), this fact constitutes one
of the structural differences between the formalisms. The quantities $\Phi
_{a}$ and $\varphi _{a}$ enter the schemes [5, 21] as affine electromagnetic
potentials that fulfill the gauge principle, in addition to satisfying wave
equations having the same form. It was shown in Ref. [11] that, in the
presence of electromagnetic fields, the imaginary part of (3.42) may be
utilized in the limiting case for making up $\Gamma _{aB}{}^{B}$
symbolically. When the limiting procedure is implemented in the absence of
fields, $\Phi _{a}$ turns out to vanish in some gauge frame. We will
describe the limiting process at greater length later upon specifying the
gauge behaviours of typical spin-affine structures.

The right-hand side of the tensor relation (3.21) is also proportional to $%
\tau \varepsilon _{AB}$, with $\tau $ amounting to a complex spin-scalar
density of weight $+1$ given as $\gamma \eta $. Thus, we can write down the
expansion 
\begin{eqnarray}
\nabla _{a}(\tau \varepsilon _{BC}) &=&\partial _{a}(\tau \varepsilon
_{BC})-\tau \Gamma _{aD}{}^{D}\varepsilon _{BC}  \notag \\
&=&(\partial _{a}\tau -\tau \Gamma _{aD}{}^{D})\varepsilon _{BC}=(\nabla
_{a}\tau )\varepsilon _{BC},  \TCItag{3.62}
\end{eqnarray}%
which leads us to stating that the set of affine computational devices for
the $\varepsilon $-formalism can be entirely obtained in any gauge frame
from that for the $\gamma $-formalism just by making the simultaneous
replacements 
\begin{equation}
\theta _{a}\rightarrow \Pi _{a},\text{ }\Phi _{a}\rightarrow \varphi _{a}. 
\tag{3.63}
\end{equation}%
We stress that the prescription (3.60) emerges from 
\begin{equation}
\nabla _{a}(\varepsilon _{BC}\varepsilon _{B^{\prime }C^{\prime }})=0, 
\tag{3.64}
\end{equation}%
whilst Eq. (3.33) appears as the vanishing derivative 
\begin{equation}
\nabla _{a}\Sigma _{BB^{\prime }}^{b}=\partial _{a}\Sigma _{BB^{\prime
}}^{b}+\Gamma _{ac}{}^{b}\Sigma _{BB^{\prime }}^{c}-\Gamma _{aB}{}^{C}\Sigma
_{CB^{\prime }}^{b}-\Gamma _{aB^{\prime }}{}^{C^{\prime }}\Sigma
_{BC^{\prime }}^{b}-\Pi _{a}\Sigma _{BB^{\prime }}^{b}.  \tag{3.65a}
\end{equation}%
It follows that the $\varepsilon $-formalism counterpart of (3.34c) is given
by 
\begin{equation}
\Gamma _{bc}{}^{a}n^{c}=\Sigma _{AA^{\prime }}^{a}(\Gamma _{bB}{}^{A}\zeta
^{B}\zeta ^{A^{\prime }}+\Gamma _{bB^{\prime }}{}^{A^{\prime }}\zeta
^{A}\zeta ^{B^{\prime }})-\zeta ^{A}\zeta ^{A^{\prime }}\partial _{b}\Sigma
_{AA^{\prime }}^{a}+\Pi _{b}n^{a}.  \tag{3.65b}
\end{equation}

The recovery in the $\varepsilon $-formalism of covariant differential
patterns for arbitrary world tensors may be achieved from the requirements 
\begin{equation}
\nabla _{a}u^{b}=\Sigma _{BB^{\prime }}^{b}\nabla _{a}u^{BB^{\prime
}}\Leftrightarrow \nabla _{a}u^{BB^{\prime }}=\Sigma _{b}^{BB^{\prime
}}\nabla _{a}u^{b},  \tag{3.66}
\end{equation}%
where $u^{b}$ amounts to a world vector and $u^{BB^{\prime }}$ is an
Hermitian spin-tensor density of absolute weight $+1$. Some manipulations
involving rearrangements of index configurations then yield the affine
relationship 
\begin{equation}
\Gamma _{AA^{\prime }BB^{\prime }CC^{\prime }}+\Sigma _{sCC^{\prime
}}\partial _{AA^{\prime }}\Sigma _{BB^{\prime }}^{s}=\Gamma _{AA^{\prime
}BC}\varepsilon _{B^{\prime }C^{\prime }}+\Gamma _{AA^{\prime }B^{\prime
}C^{\prime }}\varepsilon _{BC}+\Pi _{AA^{\prime }}\varepsilon
_{BC}\varepsilon _{B^{\prime }C^{\prime }},  \tag{3.67a}
\end{equation}%
where 
\begin{equation}
\Gamma _{AA^{\prime }BB^{\prime }CC^{\prime }}=\Sigma _{AA^{\prime
}}^{a}\Sigma _{BB^{\prime }}^{b}\Sigma _{CC^{\prime }}^{c}\Gamma _{abc}{},%
\text{ }\partial _{AA^{\prime }}=\Sigma _{AA^{\prime }}^{a}\partial _{a}. 
\tag{3.67b}
\end{equation}%
Equations (3.67) exhibit the world covariance of $\Gamma _{aBC}{}$ and its
complex conjugate. The piece $\Gamma _{A(BC)A^{\prime }(B^{\prime }C^{\prime
})}$ and the spinor version of $\Gamma _{a[bc]}$ arising here are both
formally the same as the ones expressed by (3.45). Also, the expression
(3.47a) for the torsionlessness of $\nabla _{a}$ still holds formally, but
the traceful part of $\Gamma _{a(bc)}$ is now subject to 
\begin{equation}
\Gamma _{a}{}+\Sigma _{s}^{BB^{\prime }}\partial _{a}\Sigma _{BB^{\prime
}}^{s}=0.  \tag{3.68}
\end{equation}%
Transvecting (3.67a) with $\varepsilon ^{BC}\varepsilon ^{B^{\prime
}C^{\prime }}$ establishes the appropriateness of the condition (3.68).
Likewise, recalling (3.13a) and (3.43) brings back the $\gamma $-formalism
equality 
\begin{equation}
\begin{array}{l}
\mid \gamma \mid ^{-3}\sigma _{AA^{\prime }}^{a}\sigma _{BB^{\prime
}}^{b}\sigma _{CC^{\prime }}^{c}\Gamma _{abc}{}+\mid \gamma \mid ^{-2}\sigma
_{sCC^{\prime }}\partial _{AA^{\prime }}(\mid \gamma \mid ^{-1}\sigma
_{BB^{\prime }}^{s}) \\ 
=\mid \gamma \mid ^{-3}(\gamma _{AA^{\prime }BC}\gamma _{B^{\prime
}C^{\prime }}+\gamma _{AA^{\prime }B^{\prime }C^{\prime }}\gamma
_{BC}+\theta _{AA^{\prime }}\gamma _{BC}\gamma _{B^{\prime }C^{\prime }}),%
\end{array}
\tag{3.69}
\end{equation}%
provided that 
\begin{equation}
\sigma _{AA^{\prime }}^{a}\sigma _{BB^{\prime }}^{b}\sigma _{CC^{\prime
}}^{c}\Gamma _{abc}{}\text{ }=\mid \gamma \mid ^{3}\Sigma _{AA^{\prime
}}^{a}\Sigma _{BB^{\prime }}^{b}\Sigma _{CC^{\prime }}^{c}\Gamma _{abc}{}. 
\tag{3.70}
\end{equation}

\subsection{Eigenvalue Equations and Metric Expressions}

The covariant constancy of the $\varepsilon $-metric spinors allows the
implementation of the $\gamma $-formalism statement%
\begin{equation}
\nabla _{a}\gamma _{BC}=(\gamma ^{-1}\nabla _{a}\gamma )\gamma _{BC}, 
\tag{3.71}
\end{equation}%
which, when combined with (3.49b), yields the expansion 
\begin{equation}
\nabla _{a}\gamma _{BC}=(\partial _{a}\log \gamma -\gamma _{a})\gamma _{BC}.
\tag{3.72}
\end{equation}%
Equations (3.1) and (3.2) then give the coupled eigenvalue equations [5, 21] 
\begin{equation}
\nabla _{a}\gamma _{BC}=i(\partial _{a}\Phi +2\Phi _{a})\gamma _{BC} 
\tag{3.73a}
\end{equation}%
and 
\begin{equation}
\nabla _{a}\gamma ^{BC}=(-i)(\partial _{a}\Phi +2\Phi _{a})\gamma ^{BC}, 
\tag{3.73b}
\end{equation}%
along with their complex conjugates. By working out the right-hand side of
(3.71), we see that the expansion (3.72) is consistent with Eqs. (3.3),
(3.54) and (3.58), that is to say, 
\begin{eqnarray}
\gamma ^{-1}\nabla _{a}\gamma &=&\frac{1}{2}\gamma ^{BC}\nabla _{a}\gamma
_{BC}=\gamma ^{-1}\nabla _{a}[\mid \gamma \mid \exp (i\Phi )]  \notag \\
&=&\exp (-i\Phi )\nabla _{a}\exp (i\Phi ).  \TCItag{3.74}
\end{eqnarray}%
It should be evident that the occurrence in Eqs. (3.73) of purely imaginary
eigenvalues, is associated to a property of the $\gamma $-formalism which
had been exhibited by the conditions (3.36) and (3.37).

It is observed in Ref. [10] that the partial derivative carried implicitly
by the left-hand side of (3.71) can be isolated by utilizing the
outer-product device 
\begin{equation}
\theta _{a}\gamma _{BC}=(i\partial _{a}\Phi )\gamma _{BC}-\partial
_{a}\gamma _{BC},  \tag{3.75a}
\end{equation}%
which comes from the computational prescription 
\begin{eqnarray}
\theta _{a}\gamma _{BC} &=&\gamma _{BC}\partial _{a}\log [\gamma ^{-1}\exp
(i\Phi )]=\gamma (\partial _{a}\gamma ^{-1})\gamma _{BC}+(i\partial _{a}\Phi
)\gamma _{BC}  \notag \\
&=&\gamma \lbrack \partial _{a}(\gamma ^{-1}\gamma _{BC})-\gamma
^{-1}\partial _{a}\gamma _{BC}]+(i\partial _{a}\Phi )\gamma _{BC}. 
\TCItag{3.75b}
\end{eqnarray}%
Thus, part of the information contained in $\partial _{a}\gamma _{BC}$ gets
annihilated by the information carried by $\theta _{a}\gamma _{BC}$ when we
bring together the individual pieces of $\nabla _{a}\gamma _{BC}$. This
procedure gives rise to the following $\partial $-equations: 
\begin{equation}
\partial _{a}\gamma _{BC}=(i\partial _{a}\Phi -\theta _{a})\gamma _{BC},%
\text{ }\partial _{a}\gamma ^{BC}=(\theta _{a}-i\partial _{a}\Phi )\gamma
^{BC}  \tag{3.76}
\end{equation}%
and 
\begin{equation}
\partial _{a}(\gamma _{BC}\gamma _{B^{\prime }C^{\prime }})=(-2\theta
_{a})\gamma _{BC}\gamma _{B^{\prime }C^{\prime }},\text{ }\partial
_{a}(\gamma ^{BC}\gamma ^{B^{\prime }C^{\prime }})=2\theta _{a}\gamma
^{BC}\gamma ^{B^{\prime }C^{\prime }}.  \tag{3.77}
\end{equation}%
The eigenvalue carried by one of Eqs. (3.76) equals $\partial _{a}\log
\gamma $ whence we can express the parts of the contracted spin-affine
connexion (3.42) as 
\begin{equation}
\theta _{a}=\frac{1}{2}\func{Re}[\gamma ^{BC}(\nabla _{a}-\partial
_{a})\gamma _{BC}]  \tag{3.78a}
\end{equation}%
and\footnote{%
It can be established from Eqs. (3.78) that the world covariance of $\gamma
_{aB}{}^{C}$ rests crucially upon the world invariance of the $\gamma $%
-metric spinors.} 
\begin{equation}
2\Phi _{a}=\frac{1}{2}\func{Im}[\gamma ^{BC}(\nabla _{a}-\partial
_{a})\gamma _{BC}].  \tag{3.78b}
\end{equation}

A system of covariant eigenvalue equations for the non-Hermitian $\sigma $%
-objects arises from Eqs. (3.33) and (3.73). For bringing out the pattern of
a typical eigenvalue, it suffices to derive the equation for any element of
either of the conjugate pairs 
\begin{equation*}
\{(\sigma _{bB}^{A^{\prime }},\Sigma _{bB}^{A^{\prime }}),(\sigma
_{bB^{\prime }}^{A},\Sigma _{bB^{\prime }}^{A})\}.
\end{equation*}%
Thus, taking account of the prescription, say, 
\begin{equation}
\nabla _{a}\sigma _{bB}^{A^{\prime }}=\sigma _{b}^{AA^{\prime }}\nabla
_{a}\gamma _{AB},  \tag{3.79}
\end{equation}%
and employing the expansion 
\begin{equation}
\nabla _{a}\sigma _{bB}^{A^{\prime }}=[\nabla _{a}\exp (i\Phi )]\Sigma
_{bB}^{A^{\prime }}+\exp (i\Phi )\nabla _{a}\Sigma _{bB}^{A^{\prime }}, 
\tag{3.80}
\end{equation}%
yields 
\begin{equation}
\nabla _{a}\sigma _{bB}^{A^{\prime }}=i(\partial _{a}\Phi +2\Phi _{a})\sigma
_{bB}^{A^{\prime }}.  \tag{3.81}
\end{equation}%
It should be noticed that (3.80) and (3.81) imply that 
\begin{equation}
\nabla _{a}\Sigma _{bB}^{A^{\prime }}=0,  \tag{3.82}
\end{equation}%
in agreement with the covariant constancy property of the $\Sigma $-objects.

The information on the spin-affine pieces $\theta _{a}$ and $\Phi _{a}$ is
encoded into Eq. (3.71). In case $\gamma $ is taken as a covariantly
constant quantity in the $\gamma $-formalism, we may recover the expression
(3.43) and achieve a metric specification of $\Phi _{a}$ that enhances the
absence of electromagnetic fields, namely [5] 
\begin{equation}
(-2)\Phi _{a}=\partial _{a}\Phi =\nabla _{a}\Phi .  \tag{3.83}
\end{equation}%
To characterize this situation in an invariant way, one should implement the
condition 
\begin{equation}
\nabla _{a}\gamma _{BC}=0,  \tag{3.84}
\end{equation}%
which evidently produces a commutativity property involving the action of
the metric spinors for the $\gamma $-formalism and the action of the
respective $\nabla $-operator. Equation (3.84) appears as a necessary and
sufficient condition for the non-Hermitian $\sigma $-objects to bear
covariant constancy. A somewhat elegant procedure for illustrating the above
statements [10], amounts to letting $\partial _{a}$ act on the matrix
configuration for $\gamma _{BC}$, while making use of a matrix form of
(3.84). In effect, we have 
\begin{equation}
\left( 
\begin{array}{ll}
0 & \partial _{a}\gamma \\ 
-\partial _{a}\gamma & 0%
\end{array}%
\right) =\left( 
\begin{array}{ll}
0 & \gamma \gamma _{aB}{}^{B}{} \\ 
-\gamma \gamma _{aB}{}^{B}{} & 0%
\end{array}%
\right) ,  \tag{3.85a}
\end{equation}%
whence, in view of (3.49b), we are unambiguously led to 
\begin{equation}
\gamma _{a}{}=\partial _{a}\log \gamma \Leftrightarrow \nabla _{a}\gamma =0.
\tag{3.85b}
\end{equation}%
Equation (3.85b) can be alternatively derived by partially differentiating
both sides of the relations (3.3). We thus obtain the intermediate-stage
configurations 
\begin{equation}
\gamma \partial _{a}\gamma _{BC}-(\partial _{a}\gamma )\gamma _{BC}=\gamma
^{2}\partial _{a}(\gamma ^{-1}\gamma _{BC})=0,  \tag{3.86a}
\end{equation}%
which exhibit the gauge-invariant property\footnote{%
We stress that the operator $\partial _{a}$ is gauge invariant since
arbitrary coordinates on $\mathfrak{M}$ do not bear any spin character at
all.} 
\begin{equation}
\partial _{a}\varepsilon _{BC}=0.  \tag{3.86b}
\end{equation}%
We can accordingly reset Eqs. (3.76) as follows 
\begin{equation}
\partial _{a}\log \gamma =\frac{1}{2}\gamma ^{BC}\partial _{a}\gamma _{BC}. 
\tag{3.86c}
\end{equation}

Apparently, the only procedure for extracting the spacetime-metric
information carried by $\gamma _{aB}{}^{B}$ is associated to the
implementation of the affine prescription (3.45c). With regard to this
observation, the key idea is to introduce the definition 
\begin{equation}
\partial _{a}\log \mu \doteqdot \sigma _{b}^{BB^{\prime }}\partial
_{a}\sigma _{BB^{\prime }}^{b},  \tag{3.87}
\end{equation}%
with $\mu $ thus standing for a mixed real-scalar density of world weight $%
-1 $ and absolute weight $+4$. Hence, recalling (3.43) yields the general
expressions 
\begin{equation}
\mid \gamma \mid ^{4}=\mu (-\mathfrak{g})^{1/2}  \tag{3.88a}
\end{equation}%
and 
\begin{equation}
(-4)\theta _{a}=\partial _{a}\log [\mu (-\mathfrak{g})^{1/2}].  \tag{3.88b}
\end{equation}%
We should observe that the world-spin character of the derivative carried by
the right-hand side of (3.87) can be clearly fixed by contracting with $%
\sigma _{b}^{BB^{\prime }}$ the configuration\footnote{%
Equation (3.87) is compatible with the world-affine transformation laws in $%
\mathfrak{M}$.}%
\begin{equation}
\partial _{a}\sigma _{BB^{\prime }}^{b}=\gamma _{aB}{}^{C}\sigma
_{CB^{\prime }}^{b}+\gamma _{aB^{\prime }}{}^{C^{\prime }}\sigma
_{BC^{\prime }}^{b}-\Gamma _{ac}{}^{b}\sigma _{BB^{\prime }}^{c},  \tag{3.89}
\end{equation}%
which arises from Eq. (3.35) and likewise reinstates the relation (3.45c).
If use is made of both (2.37) and (3.58), we will then conclude that $\mu $
has to satisfy the covariant condition 
\begin{equation}
\nabla _{a}\mu =0.  \tag{3.90}
\end{equation}

If the limit as the pair $(\mid \gamma \mid ,\Phi )$ tends to $(1,0)$ is
carried out, the eigenvalues borne by Eqs. (3.73) will turn out to equal $%
\pm 2i\Phi _{a}$. Consequently, because of the $\nabla $-constancy property
of the $\varepsilon $-metric spinors, the behaviour of the left-hand sides
of those equations can be controlled by the expansion (3.49b). As provided
by the eigenvalue equations (3.76), the description of the limiting process
is based on the gauge-invariant $\partial $-constancy of the $\varepsilon $%
-metric spinors, which implies that the individual pieces of both
eigenvalues tend to zero when the limit is actually implemented. Taking up
covariantly constant $\gamma $-metric spinors would thus make $\Phi _{a}$
into a vanishing gradient, and the outcome of the limit of $\gamma
_{aB}{}^{B}$ would appear as a useless quantity. Therefore, if $\Phi _{a}$
is taken as a gradient, we will have to reconstruct the contracted affine
structures for the $\varepsilon $-formalism apart from the ones for the $%
\gamma $-formalism, but $\varphi _{a}$ will have to carry a gradient
character as well insofar as any shift from one formalism to the other must
not produce any electromagnetic fields. In case the $\gamma $-metric spinors
are taken to have non-vanishing covariant derivatives, the form of the
imaginary part of $\gamma _{aB}{}^{B}$ would be left unchanged. However, we
should still have to take account of (3.60) in order to recover $\Gamma
_{aB}{}^{B}$. We will elaborate further upon this situation in the following
Subsection.

\subsection{Generalized Gauge Transformation Laws}

As the rules for writing covariant derivatives of spin tensors in both
formalisms are symbolically the same, the gauge behaviours of $\gamma
_{aB}{}^{C}$ and $\Gamma _{aB}{}^{C}$ can be specified from one another by
simply replacing kernel letters. The original procedure for establishing
these behaviours [5], amounts in either case to taking up the covariance
requirement 
\begin{equation}
\nabla _{a}^{\prime }\xi _{B}^{\prime }=\Lambda _{B}{}^{C}\nabla _{a}\xi
_{C},  \tag{3.91}
\end{equation}%
with $\xi _{A}$ being an arbitrary spin vector. Hence, by writing out the
expansions of (3.91) explicitly, and using the derivative device 
\begin{equation}
\Lambda _{B}{}^{C}\partial _{a}\xi _{C}=\partial _{a}^{\prime }\xi
_{B}^{\prime }-(\partial _{a}\Lambda _{B}{}^{C})\xi _{C},  \tag{3.92}
\end{equation}%
after invoking the arbitrariness of $\xi _{A}$, we arrive at the
configuration 
\begin{equation}
\vartheta _{aB}^{\prime }{}^{D}\Lambda _{D}{}^{C}=\Lambda
_{B}{}^{D}\vartheta _{aD}{}^{C}+\partial _{a}\Lambda _{B}{}^{C},  \tag{3.93}
\end{equation}%
where the kernel letter $\vartheta $ stands for either $\gamma $ or $\Gamma $%
, as before (see Eq. (3.29)). Obviously, either of the affinities occurring
in (3.93) can be picked out by adequately coupling all the involved
individual pieces with an inverse $\Lambda $-matrix. We have, for instance, 
\begin{equation}
\vartheta _{aB}^{\prime }{}^{C}=\Lambda _{B}{}^{D}\vartheta
_{aD}{}^{M}\Lambda _{M}^{-1}{}^{C}+(\partial _{a}\Lambda _{B}{}^{M})\Lambda
_{M}^{-1}{}^{C}.  \tag{3.94}
\end{equation}%
As remarked in Ref. [10], there is an alternative procedure for deriving the
law (3.94) which appropriately mixes up the unprimed and primed gauge
frames. This consists in applying the Leibniz rule to the requirement
(3.91), likewise supposing that any gauge-matrix components can always be
covariantly differentiated in the same way as ordinary spin tensors. One
thus obtains the correlation 
\begin{equation}
\nabla _{a}^{\prime }\xi _{B}^{\prime }=\nabla _{a}\xi _{B}^{\prime
}-(\nabla _{a}\Lambda _{B}{}^{C})\xi _{C},  \tag{3.95}
\end{equation}%
which immediately yields Eq. (3.93).

The behaviour of any contracted spin-affine structure for either formalism
can be particularly attained by working out the coordinate derivative of the
definition (3.26). For this purpose, we first note that Eqs. (3.76) yield%
\footnote{%
We recall here that the kernel letter $M$ presumably denotes either $\gamma $
or $\varepsilon $.} 
\begin{equation}
\partial _{a}(M{}^{AB}M_{CD})=0,  \tag{3.96}
\end{equation}%
whence it is legitimate to account for the relation 
\begin{equation}
2\partial _{a}\Delta _{{\small \Lambda }}=M{}^{AB}\partial _{a}(\Lambda
_{A}{}^{C}\Lambda _{B}{}^{D})M_{CD}.  \tag{3.97}
\end{equation}%
Additionally, carrying out the $\partial $-expansion borne by the right-hand
side of (3.97) and invoking the prescription (3.93), leads to the value 
\begin{equation}
2\partial _{a}\Delta _{{\small \Lambda }}=U_{a}^{(M)}-V_{a}^{(M)}, 
\tag{3.98a}
\end{equation}%
which carries the contributions 
\begin{equation}
U_{a}^{(M)}=M^{AB}(\vartheta _{aA}^{\prime }{}^{N}\Lambda _{N}{}^{C}\Lambda
_{B}{}^{D}+\vartheta _{aB}^{\prime }{}^{N}\Lambda _{A}{}^{C}\Lambda
_{N}{}^{D})M_{CD}  \tag{3.98b}
\end{equation}%
and 
\begin{eqnarray}
V_{a}^{(M)} &=&M^{AB}(\Lambda _{A}{}^{N}\Lambda _{B}{}^{D}\vartheta
_{aN}{}^{C}+\Lambda _{A}{}^{C}\Lambda _{B}{}^{N}\vartheta _{aN}{}^{D})M_{CD}
\notag \\
&=&M^{AB}(\Lambda _{A}{}^{N}\Lambda _{B}{}^{C}\vartheta _{a[NC]}{}-\Lambda
_{A}{}^{C}\Lambda _{B}{}^{N}\vartheta _{a[NC]}{})  \notag \\
&=&2M^{AB}\Lambda _{A}{}^{C}\Lambda _{B}{}^{D}\vartheta _{a[CD]}{}. 
\TCItag{3.98c}
\end{eqnarray}%
For the $\gamma $-formalism, we use (3.18) to perform the computations 
\begin{eqnarray}
U_{a}^{(\gamma )} &=&\gamma ^{AB}(\gamma _{aA}^{\prime }{}^{M}\gamma
_{MB}^{\prime }+\gamma _{aB}^{\prime }{}^{M}\gamma _{AM}^{\prime })  \notag
\\
&=&2\Delta _{{\small \Lambda }}\gamma ^{\prime AB}\gamma _{a[AB]}^{\prime
}{}=2\Delta _{{\small \Lambda }}\gamma _{aB}^{\prime }{}^{B}  \TCItag{3.99a}
\end{eqnarray}%
and 
\begin{equation}
V_{a}^{(\gamma )}=\gamma ^{AB}\gamma _{aC}{}^{C}\gamma _{AB}^{\prime
}=2\Delta _{{\small \Lambda }}\gamma _{aB}{}^{B}.  \tag{3.99b}
\end{equation}%
In a similar way, for the $\varepsilon $-formalism, we utilize (3.25) to
obtain 
\begin{eqnarray}
U_{a}^{(\varepsilon )} &=&\varepsilon ^{AB}(\Gamma _{aA}^{\prime
}{}^{M}\Lambda _{M}{}^{C}\Lambda _{B}{}^{D}+\Gamma _{aB}^{\prime
}{}^{M}\Lambda _{A}{}^{C}\Lambda _{M}{}^{D})\varepsilon _{CD}  \notag \\
&=&\Delta _{{\small \Lambda }}\varepsilon ^{AB}(\Gamma _{aA}^{\prime
}{}^{M}\varepsilon _{MB}^{\prime }+\Gamma _{aB}^{\prime }{}^{M}\varepsilon
_{AM}^{\prime })  \notag \\
&=&2\Delta _{{\small \Lambda }}\Gamma _{aB}^{\prime }{}^{B}  \TCItag{3.100a}
\end{eqnarray}%
and 
\begin{equation}
V_{a}^{(\varepsilon )}=2\varepsilon ^{AB}\Lambda _{A}{}^{C}\Lambda
_{B}{}^{D}\Gamma _{a[CD]}{}=2\Delta _{{\small \Lambda }}\Gamma _{aB}{}^{B}. 
\tag{3.100b}
\end{equation}%
It follows that 
\begin{equation}
\partial _{a}\Delta _{{\small \Lambda }}=\Delta _{{\small \Lambda }%
}(\vartheta _{aB}^{\prime }{}^{B}-\vartheta _{aB}{}^{B}),  \tag{3.101a}
\end{equation}%
whence Eq. (3.94) can be cast into the form 
\begin{equation}
\vartheta _{aB}^{\prime }{}^{C}=\vartheta _{aB}{}^{C}+\frac{1}{2}(\partial
_{a}\log \Delta _{{\small \Lambda }})\delta _{B}{}^{C}.  \tag{3.101b}
\end{equation}%
Then, making suitable contractions gives rise to the laws 
\begin{equation}
\gamma _{aB}^{\prime }{}^{B}=\gamma _{aB}{}^{B}+\partial _{a}\log \Delta _{%
{\small \Lambda }}  \tag{3.102a}
\end{equation}%
and 
\begin{equation}
\Gamma _{aB}^{\prime }{}^{B}=\Gamma _{aB}{}^{B}+\partial _{a}\log \Delta _{%
{\small \Lambda }},  \tag{3.102b}
\end{equation}%
together with their complex conjugates. We should stress that the metric
prescriptions for lowering and raising spinor indices in both formalisms
must strictly involve quantities defined in the same gauge frames.

From Eqs. (3.102), we see that the gauge behaviours of the individual pieces
of the structures (3.42) and (3.61) have to be specified as 
\begin{equation}
\tau _{a}^{\prime }=\tau _{a}-\partial _{a}\theta ,  \tag{3.103a}
\end{equation}%
\begin{equation}
\theta _{a}^{\prime }=\theta _{a}-\partial _{a}\log \rho  \tag{3.103b}
\end{equation}%
and 
\begin{equation}
\Pi _{a}^{\prime }=\Pi _{a}-\partial _{a}\log \rho ,  \tag{3.104}
\end{equation}%
with the quantity $\tau _{a}$ thus amounting to either $\Phi _{a}$ or $%
\varphi _{a}$. The transformation law for $\mid \gamma \mid $ as given in
Subsection 3.2 can be recovered out of combining (3.43) and (3.103b). By
appealing to (3.76), we can likewise describe the geometric character of $%
\exp (i\Phi )$ from 
\begin{equation}
\partial _{a}^{\prime }\Phi ^{\prime }=\partial _{a}\Phi +2\partial
_{a}\theta .  \tag{3.105}
\end{equation}%
It turns out that the gauge behaviour of the partial derivatives of the $%
\gamma $-metric spinors can be fully described by the law 
\begin{equation}
\partial _{a}^{\prime }\log \gamma ^{\prime }=\partial _{a}\log \gamma
+\partial _{a}\log \Delta _{{\small \Lambda }}.  \tag{3.106}
\end{equation}%
We then conclude that the eigenvalues of Eqs. (3.73) bear gauge invariance,
whence we can establish the invariant character of (3.84) by taking into
consideration the $\gamma $-formalism prescription 
\begin{equation}
\nabla _{a}^{\prime }\gamma _{BC}^{\prime }=\Delta _{{\small \Lambda }%
}\nabla _{a}\gamma _{BC}.  \tag{3.107}
\end{equation}

The establishment of the law (3.103a) characterizes $\Phi _{a}$ and $\varphi
_{a}$ as the electromagnetic potentials of $\gamma _{aB}{}^{B}$ and $\Gamma
_{aB}{}^{B}$, respectively. Equation (3.107) thus shows that if the $\gamma $%
-metric spinors are taken to bear covariant constancy in the unprimed frame,
they will have to be looked upon as covariantly constant entities in the
primed frame as well. Hence, if $\Phi _{a}$ is a gradient in the unprimed
frame, it will also be a gradient in any other frame. Consequently, as had
been observed before, taking the limit as $\gamma $ tends to $1$ would
annihilate both pieces of $\gamma _{aB}{}^{B}$ in the unprimed frame. In
such circumstances, the primed-frame pieces $\Phi _{a}^{\prime }$ and $%
\theta _{a}^{\prime }$ would become proportional to $\partial _{a}\theta $
and $\partial _{a}\log \rho $, whence any contracted affine structures for
the $\varepsilon $-formalism would have indeed to be entirely reconstructed
in accordance with the prescriptions (3.61) and (3.104). It should be clear
that the gauge behaviours of $\partial \gamma $-equations like (3.76) and
(3.77) may be controlled in any case by Eq. (3.106). Therefore, one can
state a metric principle that describes in a gauge-invariant fashion the
geometric structure of the $\gamma $-formalism as regards the presence or
absence of electromagnetic fields.

We can covariantly keep track of gauge behaviours by assuming that any $%
\nabla $-derivative of some spin tensor or density can be carried out in any
frame regardless of whether the kernel letter of the object to be
differentiated is primed or unprimed. Let us, in effect, consider the $%
\gamma $-formalism expansion 
\begin{equation}
\nabla _{a}\gamma _{BC}^{\prime }=\partial _{a}\gamma _{BC}^{\prime }-\gamma
_{aM}{}^{M}\gamma _{BC}^{\prime }.  \tag{3.108}
\end{equation}%
Interchanging the roles of the frames and making use of (3.102a) yields 
\begin{equation}
\nabla _{a}^{\prime }\gamma _{BC}=\nabla _{a}\gamma _{BC}-(\partial _{a}\log
\Delta _{{\small \Lambda }})\gamma _{BC},  \tag{3.109}
\end{equation}%
whence the covariant derivative carried by (3.108) obeys the relation 
\begin{equation}
\nabla _{a}\gamma _{BC}^{\prime }=\nabla _{a}^{\prime }\gamma _{BC}^{\prime
}+(\partial _{a}\log \Delta _{{\small \Lambda }})\gamma _{BC}^{\prime }. 
\tag{3.110}
\end{equation}%
As a consequence of Eq. (3.110), we can account for the contracted
derivatives 
\begin{equation}
\gamma ^{BC}\nabla _{a}^{\prime }\gamma _{BC}=\gamma ^{BC}\nabla _{a}\gamma
_{BC}-\partial _{a}\log (\Delta _{{\small \Lambda }})^{2}  \tag{3.111a}
\end{equation}%
and 
\begin{equation}
\gamma ^{\prime BC}\nabla _{a}\gamma _{BC}^{\prime }=\gamma ^{\prime
BC}\nabla _{a}^{\prime }\gamma _{BC}^{\prime }+\partial _{a}\log (\Delta _{%
{\small \Lambda }})^{2},  \tag{3.111b}
\end{equation}%
which clearly reflect the interchange of frames implemented above. We can
see that if either of Eqs. (3.111) had been considered alone, then the
gauge-frame prescription for the other could be obtained by effecting the
substitution 
\begin{equation}
\Delta _{{\small \Lambda }}\mapsto \delta _{{\small \Lambda }}.  \tag{3.112}
\end{equation}%
By taking account of (3.109), we write down the expansions 
\begin{eqnarray}
\nabla _{a}^{\prime }\gamma _{BC}^{\prime } &=&\nabla _{a}^{\prime }(\Delta
_{{\small \Lambda }}\gamma _{BC})=\Delta _{{\small \Lambda }}\nabla
_{a}^{\prime }\gamma _{BC}+(\nabla _{a}^{\prime }\Delta _{{\small \Lambda }%
})\gamma _{BC}  \notag \\
&=&\Delta _{{\small \Lambda }}\nabla _{a}\gamma _{BC}+(\nabla _{a}^{\prime
}\Delta _{{\small \Lambda }}-\partial _{a}^{\prime }\Delta _{{\small \Lambda 
}})\gamma _{BC},  \TCItag{3.113}
\end{eqnarray}%
which suggest ascribing a gauge-scalar character to $\Delta _{\Lambda }$,
namely\footnote{%
Equation (3.114) enables one to say that the functions $\rho $ and $\theta $
carried by the definition (3.14a) are world-spin scalars.} 
\begin{equation}
\nabla _{a}^{\prime }\Delta _{{\small \Lambda }}=\partial _{a}^{\prime
}\Delta _{{\small \Lambda }}=\partial _{a}\Delta _{{\small \Lambda }}=\nabla
_{a}\Delta _{{\small \Lambda }}.  \tag{3.114}
\end{equation}%
From Eq. (3.113), it also follows that 
\begin{equation}
\gamma ^{\prime BC}\nabla _{a}^{\prime }\gamma _{BC}^{\prime }=\gamma
^{BC}\nabla _{a}\gamma _{BC},  \tag{3.115}
\end{equation}%
whence the condition (3.36) is subject to the homogeneous law 
\begin{equation}
\nabla _{a}^{\prime }(\gamma _{BC}^{\prime }\gamma _{B^{\prime }C^{\prime
}}^{\prime })=\mid \Delta _{{\small \Lambda }}\mid ^{2}\nabla _{a}(\gamma
_{BC}\gamma _{B^{\prime }C^{\prime }}).  \tag{3.116}
\end{equation}

A covariant mixed-frame property arises when we work out covariant
derivatives of the unprimed-index $\gamma $-metric spinors for the primed
frame. For instance, taking (3.114) into account leads to 
\begin{equation}
\nabla _{a}\gamma _{BC}^{\prime }=\nabla _{a}(\Delta _{{\small \Lambda }%
}\gamma _{BC})=\Delta _{{\small \Lambda }}\nabla _{a}\gamma _{BC}+(\partial
_{a}\Delta _{{\small \Lambda }})\gamma _{BC},  \tag{3.117a}
\end{equation}%
whence, because of Eqs. (3.108)-(3.110), we can write 
\begin{equation}
\nabla _{a}^{\prime }(\Delta _{{\small \Lambda }}\gamma _{BC})+(\partial
_{a}\Delta _{{\small \Lambda }})\gamma _{BC}=\Delta _{{\small \Lambda }%
}\nabla _{a}^{\prime }\gamma _{BC}+(2\partial _{a}\Delta _{{\small \Lambda }%
})\gamma _{BC}.  \tag{3.117b}
\end{equation}%
Equation (3.109) then yields the prescription 
\begin{equation}
\delta _{{\small \Lambda }}\nabla _{a}\gamma _{BC}^{\prime }=\nabla
_{a}^{\prime }\gamma _{BC}+2(\partial _{a}\log \Delta _{{\small \Lambda }%
})\gamma _{BC},  \tag{3.118}
\end{equation}%
which upon transvection with $\gamma ^{\prime BC}$ gives 
\begin{equation}
\gamma ^{BC}\nabla _{a}^{\prime }\gamma _{BC}=\gamma ^{\prime BC}\nabla
_{a}\gamma _{BC}^{\prime }-\partial _{a}\log (\Delta _{{\small \Lambda }%
})^{4}.  \tag{3.119}
\end{equation}%
Therefore, the sum of contracted $\nabla $-derivatives having the same
gauge-frame mixing is maintained when we interchange the frames, namely 
\begin{equation}
\gamma ^{\prime BC}\nabla _{a}\gamma _{BC}^{\prime }+\gamma _{BC}^{\prime
}\nabla _{a}\gamma ^{\prime BC}=\gamma ^{BC}\nabla _{a}^{\prime }\gamma
_{BC}+\gamma _{BC}\nabla _{a}^{\prime }\gamma ^{BC}.  \tag{3.120}
\end{equation}

An important property of the covariant derivative prescriptions we have
exhibited is that they can be used as a metric tool for looking into the
structure of the transformation laws for the contracted spin affinities of
the $\gamma $-formalism [10]. The best way of describing this situation is
to observe that a requirement of the form of Eq. (3.91) comes out when we
insert into the relation (3.110) the expansion 
\begin{equation}
\nabla _{a}\gamma _{BC}^{\prime }=\Lambda _{B}{}^{L}\Lambda _{C}{}^{M}\nabla
_{a}\gamma _{LM}+\nabla _{a}(\Lambda _{B}{}^{L}\Lambda _{C}{}^{M})\gamma
_{LM}.  \tag{3.121}
\end{equation}%
Hence, implementing (3.114) in the form 
\begin{equation}
\nabla _{a}(\Lambda _{B}{}^{L}\Lambda _{C}{}^{M})\gamma _{LM}=(\partial
_{a}\Delta _{{\small \Lambda }})\gamma _{BC},  \tag{3.122}
\end{equation}%
produces the statement%
\begin{equation}
\nabla _{a}^{\prime }\gamma _{BC}^{\prime }=\Lambda _{B}{}^{L}\Lambda
_{C}{}^{M}\nabla _{a}\gamma _{LM},  \tag{3.123}
\end{equation}%
which effectively recovers the laws (3.102a) and (3.107). In both gauge
frames, there occurs annihilation of part of the information carried by the
covariant derivatives of $\Lambda _{B}{}^{L}\Lambda _{C}{}^{M}$ when the
overall differential expansions are appropriately contracted with $\gamma
_{LM}$ or $\gamma _{LM}^{\prime }$. The amount of information annihilated in
each frame is not gauge invariant, and can be calculated by performing the
relevant expansion. What results is, in effect, that the pieces 
\begin{equation}
(\Delta _{\Lambda }\gamma _{aM}{}^{M}\gamma _{BC},\text{ }\Delta _{\Lambda
}\gamma _{aM}^{\prime }{}^{M}\gamma _{BC}^{\prime }),  \tag{3.124}
\end{equation}%
cancel out when the contracted derivatives are individually built up. To
establish this statement, we rewrite (3.110) as 
\begin{equation}
\nabla _{a}^{\prime }\gamma _{BC}^{\prime }=\nabla _{a}\gamma _{BC}^{\prime
}-\nabla _{a}(\Lambda _{B}{}^{L}\Lambda _{C}{}^{M})\gamma _{LM}, 
\tag{3.125a}
\end{equation}%
or, more explicitly, as\footnote{%
We notice that Eqs. (3.125) recover the relation (3.122).} 
\begin{equation}
\nabla _{a}(\Lambda _{B}{}^{L}\Lambda _{C}{}^{M})\gamma _{LM}=\partial
_{a}(\Lambda _{B}{}^{L}\Lambda _{C}{}^{M})\gamma _{LM}.  \tag{3.125b}
\end{equation}%
Particularly, the pieces occurring in the configuration 
\begin{equation}
\gamma ^{BC}\nabla _{a}(\Lambda _{B}{}^{L}\Lambda _{C}{}^{M})\gamma
_{LM}=\gamma ^{BC}\partial _{a}(\Lambda _{B}{}^{L}\Lambda _{C}{}^{M})\gamma
_{LM},  \tag{3.126}
\end{equation}%
carry only gauge-invariant information.

At this stage, it is expedient to reexpress (3.94) as%
\begin{equation}
\gamma _{aBC}^{\prime }{}=\Lambda _{B}{}^{L}\Lambda _{C}{}^{M}\gamma
_{aLM}{}+(\partial _{a}\Lambda _{B}{}^{L})\Lambda _{C}{}^{M}\gamma _{LM}. 
\tag{3.127}
\end{equation}%
Because of the pattern of Eq. (3.14c), we can also write out the relation%
\begin{equation}
(\partial _{a}\Delta _{{\small \Lambda }})\gamma _{BC}=2(\partial
_{a}\Lambda _{B}{}^{L})\Lambda _{C}{}^{M}\gamma _{LM},  \tag{3.128}
\end{equation}%
whence 
\begin{equation}
\gamma _{aBC}^{\prime }{}=\Lambda _{B}{}^{L}\Lambda _{C}{}^{M}\gamma
_{aLM}{}+\frac{1}{2}(\partial _{a}\Delta _{{\small \Lambda }})\gamma _{BC}, 
\tag{3.129}
\end{equation}%
which recovers the law (3.101b). Now, multiplying Eq. (3.129) by $\gamma
^{\prime BC}$ reinstates the law (3.102a), since 
\begin{equation}
\gamma ^{\prime BC}\Lambda _{B}{}^{L}\Lambda _{C}{}^{M}\gamma
_{aLM}{}=\delta _{{\small \Lambda }}\gamma ^{BC}\Lambda _{B}{}^{L}\Lambda
_{C}{}^{M}\gamma _{a[LM]}{}=\gamma _{aB}{}^{B}  \tag{3.130a}
\end{equation}%
and 
\begin{equation}
\frac{1}{2}\gamma ^{\prime BC}(\partial _{a}\Delta _{{\small \Lambda }%
})\gamma _{BC}=\delta _{{\small \Lambda }}\partial _{a}\Delta _{{\small %
\Lambda }}=\partial _{a}\log \Delta _{{\small \Lambda }}.  \tag{3.130b}
\end{equation}%
Hence, if we implement the splittings 
\begin{equation}
\gamma _{aBC}^{\prime }{}=\gamma _{a(BC)}^{\prime }{}+\frac{1}{2}\gamma
_{aM}^{\prime }{}^{M}\gamma _{BC}^{\prime }  \tag{3.131a}
\end{equation}%
and 
\begin{equation}
\Lambda _{B}{}^{L}\Lambda _{C}{}^{M}\gamma _{aLM}{}=\Lambda
_{B}{}^{L}\Lambda _{C}{}^{M}\gamma _{a(LM)}{}+\frac{1}{2}\Delta _{{\small %
\Lambda }}\gamma _{aM}{}^{M}\gamma _{BC},  \tag{3.131b}
\end{equation}%
we will obtain the spin-tensor prescription 
\begin{equation}
\gamma _{a(BC)}^{\prime }{}=\Lambda _{B}{}^{L}\Lambda _{C}{}^{M}\gamma
_{a(LM)}{}=\Delta _{{\small \Lambda }}\gamma _{a(BC)}{},  \tag{3.132}
\end{equation}%
along with the law 
\begin{equation}
\gamma _{aBC}^{\prime }{}=\Lambda _{B}{}^{L}\Lambda _{C}{}^{M}\gamma
_{a(LM)}{}+\frac{1}{2}\Delta _{{\small \Lambda }}(\gamma
_{aM}{}^{M}+\partial _{a}\log \Delta _{{\small \Lambda }})\gamma _{BC}. 
\tag{3.133}
\end{equation}

Upon proceeding to the derivation of the transformation laws for the $%
\varepsilon $-formalism, we must recall the structure (3.94) and work out
the primed-frame configuration 
\begin{equation}
\Gamma _{aBC}^{\prime }{}=\Gamma _{aB}^{\prime }{}^{M}\varepsilon
_{MC}^{\prime }.  \tag{3.134}
\end{equation}%
The relations (3.114) and (3.128) are still valid as they stand there since
both formalisms involve one and the same gauge group, but the law (3.133)
has to be replaced with 
\begin{equation}
\Gamma _{aBC}^{\prime }{}=(\Delta _{{\small \Lambda }})^{-1}\Lambda
_{B}{}^{L}\Lambda _{C}{}^{M}\Gamma _{a(LM)}{}+\frac{1}{2}(\Gamma
_{aM}{}^{M}+\partial _{a}\log \Delta _{{\small \Lambda }})\varepsilon _{BC}.
\tag{3.135}
\end{equation}%
Equations (3.101b) and (3.102b) are consequently recovered, and we can write
the prescription 
\begin{equation}
\Gamma _{a(BC)}^{\prime }{}=(\Delta _{{\small \Lambda }})^{-1}\Lambda
_{B}{}^{L}\Lambda _{C}{}^{M}\Gamma _{a(LM)}{}=\Gamma _{a(BC)}{},  \tag{3.136}
\end{equation}%
whence $\Gamma _{a(BC)}{}$ is an invariant spin-tensor density of weight $-1$%
. It can therefore be said that the symmetric parts of any spin-affine
connexions for both formalisms carry a gauge-covariant character. By making
use of Eqs. (3.133) and (3.135) along with the trivial equality 
\begin{equation}
\rho \partial _{a}\rho =\func{Re}(\overline{\Delta }_{\Lambda }\partial
_{a}\Delta _{\Lambda }),  \tag{3.137}
\end{equation}%
we also establish that the relationships (3.44a) and (3.67a) behave
covariantly.

One of the most remarkable analogies between world and spin configurations
is reflected by the fact that covariant differentials in both formalisms of
any typical geometric objects carry the same gauge characters as the
differentiated objects themselves. This property exhibits the existence of a
formal analogy between covariant derivatives of world and spin quantities in 
$\mathfrak{M}$. It just comes from the combination of the outer-product
extension of the requirement (3.91) with the prescriptions for building up
arbitrary spin-tensor densities. For example, the gauge behaviour of the
expansion (3.56) is specified by 
\begin{equation}
\nabla _{a}^{\prime }U_{BC...D}^{\prime }=(\Delta _{{\small \Lambda }})^{%
\mathfrak{a}}(\bar{\Delta}_{{\small \Lambda }})^{\mathfrak{b}}\Lambda
_{B}{}^{L}\Lambda _{C}{}^{M}...\Lambda _{D}{}^{N}\nabla _{a}U_{LM...N}. 
\tag{3.138}
\end{equation}%
The prescription (3.54b) thus undergoes the transformation 
\begin{equation}
\nabla _{a}^{\prime }\exp (i\Phi ^{\prime })=\Delta _{{\small \Lambda }}\mid
\Delta _{{\small \Lambda }}\mid ^{-1}\nabla _{a}\exp (i\Phi ),  \tag{3.139}
\end{equation}%
while $\nabla _{a}S_{AA^{\prime }}^{b}$ behaves as 
\begin{equation}
\nabla _{a}^{\prime }\sigma _{AA^{\prime }}^{\prime b}=\mid \Delta _{{\small %
\Lambda }}\mid \nabla _{a}\sigma _{AA^{\prime }}^{b},\text{ }\nabla
_{a}^{\prime }\Sigma _{AA^{\prime }}^{\prime b}=\nabla _{a}\Sigma
_{AA^{\prime }}^{b}.  \tag{3.140}
\end{equation}%
Equation (3.140) may establish the gauge invariance of the $\nabla $%
-constancy property of the elements of the set (3.8).

\section{SPIN CURVATURE AND WAVE EQUATIONS}

We shall now describe systematically the curvature spinors of $\gamma
_{aB}{}^{C}$ and $\Gamma _{aB}{}^{C}$. The pertinent computational devices
carry the definition of a set of spinor differential operators that
constitute the bivector configuration for $\nabla _{\lbrack a}\nabla _{b]}$.
A rough form of such operators was first utilized in Ref. [12] for deriving
a system of wave equations for some classical spinning fields. Upon working
out the procedures that yield the wave equations for gravitons, we will have
necessarily to implement a version of the gravitational Bianchi identity
which amounts to an extension of the one borne by the spinor classification
schemes mentioned earlier. As before, we will bring out the geometric
quantities for the $\gamma $-formalism without leaving out their $%
\varepsilon $-formalism counterparts.

A particularly remarkable feature of the $\gamma \varepsilon $-framework is
that whereas any curvature spinors for the $\gamma $-formalism are subject
to tensorial gauge transformation laws, the corresponding ones for the $%
\varepsilon $-formalism carry a gauge-invariant density character. In both
formalisms, any conjugate gravitational and electromagnetic wave functions
supply dynamical states for gravitons and photons of opposite handednesses.
The gravitational pieces of the curvature splittings for both formalisms may
likewise give rise to a common gauge-invariant expression for the
cosmological constant. It turns out indeed that a system of gauge-covariant
field and wave equations bearing prescribed index configurations is what
controls the propagation of gravitons and photons in $\mathfrak{M}$.

Obviously, all the main procedures shall be completed in the presence of
electromagnetic fields. In Subsection 4.1, the relevant commutator
structures along with the curvature spinors are constructed. The
electromagnetic field and wave equations are exhibited in Subsection 4.2. We
will exhibit the gravitational statements subsequently in Subsection 4.3. In
respect of the formalisms themselves, any wave functions shall be taken as
classical fields from the physical point of view. Thus, there will not be
henceforth any attempt to provide a quantum description of gravitons and
photons. The inclusion of the description of Dirac fields in $\mathfrak{M}$
is made in Subsection 4.4. Either of the potentials of Eq. (3.103a) will be
denoted as $\Phi _{a}$.

\subsection{Commutators and Curvature Spinors}

The information on the curvature splittings that arise in both formalisms is
carried by the covariant commutator [5] 
\begin{equation}
\lbrack \nabla _{a},\nabla _{b}]S^{cDD^{\prime }}\doteqdot 2\nabla _{\lbrack
a}(\nabla _{b]}S^{cDD^{\prime }})=0{\large ,}  \tag{4.1}
\end{equation}%
where $S^{cDD^{\prime }}$ is one of the entries of the set (3.8). Expanding
the middle configuration of (4.1) and invoking the covariant differential
prescriptions of Subsection 3.3, yields the relation 
\begin{equation}
S^{cAB^{\prime }}W_{abA}{}^{B}+S^{cBA^{\prime }}W_{abA^{\prime
}}{}^{B^{\prime }}+S^{hBB^{\prime }}R_{abh}{}^{c}=0,  \tag{4.2}
\end{equation}%
with 
\begin{equation}
W_{abA}{}^{B}=2\partial _{\lbrack a}\vartheta _{b]A}{}^{B}-(\vartheta
_{aA}{}^{C}\vartheta _{bC}{}^{B}-\vartheta _{bA}{}^{C}\vartheta
_{aC}{}^{B})=W_{[ab]A}{}^{B}  \tag{4.3}
\end{equation}%
being the defining expression for a typical Infeld-van der Waerden mixed
curvature object for either formalism. The explicit expansion for the $%
\varepsilon $-formalism version of (4.1) carries a term proportional to $%
\partial _{\lbrack a}\Pi _{b]}$ which may be taken to vanish [10]. This
point will be touched upon again in Section 5. Hence, transvecting (4.2)
with $S_{cDB^{\prime }}$ gives 
\begin{equation}
2W_{abA}{}^{B}+\delta _{A}{}^{B}W_{abA^{\prime }}{}^{A^{\prime
}}=S_{AB^{\prime }}^{c}S^{dBB^{\prime }}R_{abcd}{},  \tag{4.4}
\end{equation}%
whence we can write down the contracted statement 
\begin{equation}
2\func{Re}W_{abA}{}^{A}=R_{abh}{}^{h}\equiv 0.  \tag{4.5}
\end{equation}%
Evidently, the procedure that yields Eq. (4.5) brings about annihilation of
the information carried by $R_{abc}{}^{d}$, whence the trace $W_{abA}{}^{A}$
appears as a purely imaginary quantity in either formalism. The simplest
manner of deriving the spin-affine expressions for the conjugate $W$-traces
of both formalisms is to contract the free spinor indices of (4.3),
verifying thereafter that the contracted pattern for the involved quadratic $%
\vartheta $-piece vanishes identically. We thus obtain the electromagnetic
contribution%
\begin{equation}
W_{abA}{}^{A}=2\partial _{\lbrack a}\vartheta _{b]A}{}^{A}=(-4i)\partial
_{\lbrack a}\Phi _{b]}.  \tag{4.6}
\end{equation}%
It is observed in Refs. [10, 11] that the $W$-objects for both formalisms
can be alternatively obtained from%
\begin{equation}
\lbrack \nabla _{a},\nabla _{b}]\zeta ^{C}=W_{abM}{}^{C}\zeta ^{M}, 
\tag{4.7}
\end{equation}%
where $\zeta ^{C}$ is some spin vector. Furthermore, we can recover the
expression (4.3) from (4.7) by replacing $\zeta ^{C}$ with a spin quantity
defined as the outer product of a gauge-invariant world vector with a
suitable Hermitian $S$-matrix. The gravitational contribution to the
curvature structure of either formalism amounts to the piece 
\begin{equation}
W_{ab(AB)}{}=\frac{1}{2}S_{AB^{\prime }}^{c}S_{B}^{dB^{\prime }}R_{abcd}{}, 
\tag{4.8}
\end{equation}%
which really bears the symmetries exhibited by Eqs. (3.9). Then, combining
(4.6) and (4.8) leads to the splitting 
\begin{equation}
W_{abAB}{}=\frac{1}{2}S_{AB^{\prime }}^{c}{}S_{B}^{dB^{\prime
}}R_{abcd}{}-iF_{ab}M_{AB},  \tag{4.9}
\end{equation}%
with $F_{ab}$ being the Maxwell tensor 
\begin{equation}
F_{ab}\doteqdot 2\partial _{\lbrack a}\Phi _{b]}=2\nabla _{\lbrack a}\Phi
_{b]}.  \tag{4.10}
\end{equation}%
A symmetrization over the indices $A$ and $B$ of Eq. (4.9) obviously causes
annihilation of the electromagnetic information carried by $W_{abAB}{}$.

In the $\gamma $-formalism, we have the covariant prescription 
\begin{equation}
W_{abAB}^{\prime }={}\Lambda _{A}{}^{C}\Lambda _{B}{}^{D}W_{abCD}=\Delta _{%
{\small \Lambda }}W_{abAB}.  \tag{4.11}
\end{equation}%
The symmetric pieces $W_{ab(AB)}{}$ and $W_{ab(A^{\prime }B^{\prime })}{}$
for the $\varepsilon $-formalism behave, respectively, as invariant
spin-tensor densities of weight $-1$ and antiweight $-1$, whence we have the
law 
\begin{equation}
W_{abAB}^{\prime }{}=(\Delta _{{\small \Lambda }})^{-1}\Lambda
_{A}{}^{C}\Lambda _{B}{}^{D}(W_{ab(CD)}+\frac{1}{2}W_{abM}{}^{M}\varepsilon
_{CD})=W_{abAB},  \tag{4.12}
\end{equation}%
along with the complex conjugates of (4.11) and (4.12). It should be pointed
out that $W_{abA}{}^{B}$ thus amounts to a gauge-invariant world-spin 
\textit{tensor} in both formalisms. The overall curvature spinors of either $%
\gamma _{aB}{}^{C}$ or $\Gamma _{aB}{}^{C}$ arise from the bivector
configuration borne by (4.9). We have, in effect, 
\begin{equation}
S_{AA^{\prime }}^{a}S_{BB^{\prime }}^{b}W_{abCD}=M_{A^{\prime }B^{\prime
}}\omega _{ABCD}+M_{AB}\omega _{A^{\prime }B^{\prime }CD},  \tag{4.13}
\end{equation}%
where 
\begin{equation}
\omega _{ABCD}=\omega _{(AB)CD}\doteqdot \frac{1}{2}S_{AA^{\prime
}}^{a}S_{B}^{bA^{\prime }}W_{abCD}  \tag{4.14a}
\end{equation}%
and 
\begin{equation}
\omega _{A^{\prime }B^{\prime }CD}=\omega _{(A^{\prime }B^{\prime
})CD}\doteqdot \frac{1}{2}S_{AA^{\prime }}^{a}S_{B^{\prime }}^{bA}W_{abCD}. 
\tag{4.14b}
\end{equation}%
Owing to the gauge characters of the $W$-objects, the curvature spinors for
the $\gamma $-formalism are subject to the tensor laws 
\begin{equation}
\omega _{ABCD}^{\prime }=\Lambda _{A}{}^{L}\Lambda _{B}{}^{M}\Lambda
_{C}{}^{R}\Lambda _{D}{}^{S}\omega _{LMRS}=(\Delta _{{\small \Lambda }%
})^{2}\omega _{ABCD}  \tag{4.15a}
\end{equation}%
and 
\begin{equation}
\omega _{A^{\prime }B^{\prime }CD}^{\prime }=\Lambda _{A^{\prime
}}{}^{L^{\prime }}\Lambda _{B^{\prime }}{}^{M^{\prime }}\Lambda
_{C}{}^{R}\Lambda _{D}{}^{S}\omega _{L^{\prime }M^{\prime }RS}=\mid \Delta _{%
{\small \Lambda }}\mid ^{2}\omega _{A^{\prime }B^{\prime }CD},  \tag{4.15b}
\end{equation}%
whereas the ones for the $\varepsilon $-formalism are invariant spin-tensor
densities prescribed by 
\begin{equation}
\omega _{ABCD}^{\prime }=(\Delta _{{\small \Lambda }})^{-2}\Lambda
_{A}{}^{L}\Lambda _{B}{}^{M}\Lambda _{C}{}^{R}\Lambda _{D}{}^{S}\omega
_{LMRS}=\omega _{ABCD}  \tag{4.16a}
\end{equation}%
and 
\begin{equation}
\omega _{A^{\prime }B^{\prime }CD}^{\prime }=\mid \Delta _{{\small \Lambda }%
}\mid ^{-2}\Lambda _{A^{\prime }}{}^{L^{\prime }}\Lambda _{B^{\prime
}}{}^{M^{\prime }}\Lambda _{C}{}^{R}\Lambda _{D}{}^{S}\omega _{L^{\prime
}M^{\prime }RS}=\omega _{A^{\prime }B^{\prime }CD}.  \tag{4.16b}
\end{equation}

It is demonstrated in Ref. [10] that the Riemann-Christoffel curvature
structure of $\mathfrak{M}$ can be completely recovered from the pair 
\begin{equation}
\mathbf{G}=(\omega _{AB(CD)},\text{ }\omega _{A^{\prime }B^{\prime }(CD)}). 
\tag{4.17}
\end{equation}%
The elements of the pair for each formalism thus enter the corresponding
spinor expression for $R_{abcd}$ according to the gauge-covariant Hermitian
prescription 
\begin{equation}
R_{AA^{\prime }BB^{\prime }CC^{\prime }DD^{\prime }}=\hspace{-1pt}%
(M_{A^{\prime }B^{\prime }}M_{C^{\prime }D^{\prime }}\omega _{AB(CD)}\hspace{%
-1pt}+M_{AB}M_{C^{\prime }D^{\prime }}\omega _{A^{\prime }B^{\prime }(CD)})+%
\text{c.c.},  \tag{4.18}
\end{equation}%
with the symbol "c.c."\ denoting an overall complex-conjugate piece. This
property was established by utilizing the expansion (4.18) along with some
metric formulae and the expression 
\begin{equation}
R_{abcd}\hspace{-2pt}=\hspace{-1pt}S_{a}^{AA^{\prime }}S_{b}^{BB^{\prime
}}S_{c}^{CC^{\prime }}S_{d}^{DD^{\prime }}R_{AA^{\prime }BB^{\prime
}CC^{\prime }DD^{\prime }},  \tag{4.19}
\end{equation}%
to rewrite the right-hand side of (4.8) as 
\begin{equation}
\frac{1}{2}S_{CA^{\prime }}^{c}S_{D}^{dA^{\prime }}R_{abcd}{}=S_{A^{\prime
}[a}^{A}S_{b]}^{BA^{\prime }}\omega _{AB(CD)}+S_{A[a}^{A^{\prime
}}S_{b]}^{B^{\prime }A}\omega _{A^{\prime }B^{\prime }(CD)}.  \tag{4.20}
\end{equation}%
The above-mentioned procedure recovers the symmetries borne by (4.14). It
really annihilates the entire complex-conjugate piece of (4.18), and
likewise allows one to pick up the elements of the $\mathbf{G}$-pair from $%
R_{abcd}$. Hence, the gravitational curvature spinors of either formalism
are defined as the entries of the pair defined as Eq. (4.17). The symmetries
exhibited by the configuration (4.20) correspond to the skew symmetry in the
indices of the pairs $ab$ and $cd$ borne by $R_{abcd}$, in accordance with
(2.13a). In view of the spacetime symmetry (2.13c), we have also to demand
the index-pair symmetries 
\begin{equation}
\omega _{AB(CD)}=\omega _{(CD)AB},\text{ }\omega _{A^{\prime }B^{\prime
}(CD)}=\omega _{(CD)A^{\prime }B^{\prime }}.  \tag{4.21}
\end{equation}%
Whence the second entry of the $\mathbf{G}$-pair has to be regarded as an
Hermitian entity in both formalisms. There is no fixed prescription for
ordering its indices since unprimed and primed spinor indices have been
taking algebraically independent values. The spinor $\omega _{A^{\prime
}B^{\prime }(CD)}$ thus possesses nine real independent components while $%
\omega _{AB(CD)}$ possesses eleven, with the number of independent
components of $R_{abcd}$ being thereupon recovered in both formalisms. This
component prescription was given originally in Ref. [15].

To attain a cosmological interpretation of the gravitational spinors, it is
convenient to reset (4.18) as 
\begin{equation}
R_{AA^{\prime }BB^{\prime }CC^{\prime }DD^{\prime }}=(M_{A^{\prime
}B^{\prime }}M_{C^{\prime }D^{\prime }}\text{X}_{ABCD}+M_{AB}M_{C^{\prime
}D^{\prime }}\Xi _{CA^{\prime }DB^{\prime }})+\text{c.c.},  \tag{4.22}
\end{equation}%
with the X$\Xi $-spinors being defined by 
\begin{equation}
\text{X}_{ABCD}\doteqdot \frac{1}{4}M^{A^{\prime }B^{\prime }}M^{C^{\prime
}D^{\prime }}R_{AA^{\prime }BB^{\prime }CC^{\prime }DD^{\prime }}=\omega
_{AB(CD)}  \tag{4.23a}
\end{equation}%
and 
\begin{equation}
\Xi _{CA^{\prime }DB^{\prime }}\doteqdot \frac{1}{4}M^{AB}M^{C^{\prime
}D^{\prime }}R_{AA^{\prime }BB^{\prime }CC^{\prime }DD^{\prime }}=\omega
_{A^{\prime }B^{\prime }(CD)}.  \tag{4.23b}
\end{equation}%
In fact, the developments leading to this insight [12] had supported a
spinor translation of Einstein's equations. Thus, we initially note that the
first of Eqs. (4.21) yields the statement 
\begin{equation}
M^{AD}\text{X}_{A(BC)D}=0\Leftrightarrow M^{BC}\text{X}_{(A\mid BC\mid D)}=0,
\tag{4.24}
\end{equation}%
which right away produces the relations 
\begin{equation}
M^{AD}\text{X}_{ABCD}=\varpi M_{BC}\Leftrightarrow M^{BC}\text{X}%
_{ABCD}=\varpi M_{AD}  \tag{4.25a}
\end{equation}%
and 
\begin{equation}
\text{X}_{AB}{}^{AB}=2\varpi ,  \tag{4.25b}
\end{equation}%
with $\varpi $ obviously standing for a world-spin invariant in both
formalisms.\footnote{%
The quantity $\varpi $ is the same in both formalisms. This fact will be
considered further in Section 5.} Hence, by taking account of the first-left
dual pattern%
\begin{equation}
^{\ast }R_{AA^{\prime }BB^{\prime }CC^{\prime }DD^{\prime
}}=[(-i)(M_{A^{\prime }B^{\prime }}M_{C^{\prime }D^{\prime }}\text{X}%
_{ABCD}-M_{AB}M_{C^{\prime }D^{\prime }}\Xi _{CA^{\prime }DB^{\prime }})]+%
\text{c.c.},  \tag{4.26}
\end{equation}%
which comes directly from the combination of (2.28a), (3.12a) and (4.22),
and invoking one of the properties (2.29), we deduce the reality statement%
\begin{equation}
M_{A^{\prime }D^{\prime }}M^{BC}\text{X}_{ABCD}=M_{AD}M^{B^{\prime
}C^{\prime }}\text{X}_{A^{\prime }B^{\prime }C^{\prime }D^{\prime }}, 
\tag{4.27}
\end{equation}%
whence $\func{Im}\varpi =0$. Either of the $\gamma \varepsilon $-expressions
for the Ricci tensor of $\mathfrak{M}$ then appears as 
\begin{equation}
R_{AA^{\prime }BB^{\prime }}=2(\varpi M_{AB}M_{A^{\prime }B^{\prime }}-\Xi
_{AA^{\prime }BB^{\prime }}).  \tag{4.28}
\end{equation}%
Consequently, from (2.46), we can conclude that the $\Xi $-spinor of either
formalism is associated to $\Xi _{ab}$, that is to say,%
\begin{equation}
\Xi _{ab}=S_{a}^{AA^{\prime }}S_{b}^{BB^{\prime }}\Xi _{AA^{\prime
}BB^{\prime }}.  \tag{4.29}
\end{equation}%
For the Ricci scalar, we thus have%
\begin{equation}
R=8\varpi ,  \tag{4.30}
\end{equation}%
whereas the spinor version of the field equations (2.54a) is simply written
as%
\begin{equation}
2\Xi _{AA^{\prime }BB^{\prime }}=\kappa (T_{AA^{\prime }BB^{\prime }}-\frac{1%
}{4}TM_{AB}M_{A^{\prime }B^{\prime }}).  \tag{4.31}
\end{equation}%
We emphasize that the quantity $\Lambda $ defined in Ref. [12] always obeys
the relations $\Lambda =\varkappa $ and $\varpi =3\Lambda $, whilst the
equality $\lambda =2\varpi $ holds only when $T=0$. It follows that, when
only traceless sources are present, the spinor expression for the Einstein
tensor appears as [11]%
\begin{equation}
G_{AA^{\prime }BB^{\prime }}=-2\Xi _{AA^{\prime }BB^{\prime }}-\lambda
M_{AB}M_{A^{\prime }B^{\prime }}.  \tag{4.32}
\end{equation}

The symmetries of X$_{ABCD}$ as given by (4.21) and (4.24) considerably
simplify the four-index reduction formula [12]%
\begin{eqnarray}
\hspace{-0.1cm}\hspace{-0.01cm}\text{X}_{ABCD}\hspace{-0.07cm} &=&\hspace{%
-0.07cm}\text{X}_{(ABCD)}-\frac{1}{4}(M_{AB}\text{X}^{M}{}_{(MCD)}+M_{AC}%
\text{X}^{M}{}_{(MBD)}+M_{AD}\text{X}^{M}{}_{(MBC)})  \notag \\
&&\hspace{-0.07cm}-\frac{1}{3}(M_{BC}\text{X}^{M}{}_{A(MD)}+M_{BD}\text{X}%
^{M}{}_{A(MC)})-\frac{1}{2}M_{CD}\text{X}_{AB}{}^{M}{}_{M}.  \TCItag{4.33}
\end{eqnarray}%
This property affords us the expansion%
\begin{equation}
\text{X}_{ABCD}=\text{X}_{(ABCD)}-\frac{2}{3}\varpi M_{A(C}M_{D)B}, 
\tag{4.34}
\end{equation}%
along with 
\begin{equation}
\text{X}_{(ABCD)}=\text{X}_{A(BCD)}=\text{X}_{(ABC)D}.  \tag{4.35}
\end{equation}%
Additionally, we stress that the Hermitian configuration 
\begin{eqnarray}
&&(M_{A(C}M_{D)B}M_{A^{\prime }B^{\prime }}M_{C^{\prime }D^{\prime }})+\text{%
c.c.}  \notag \\
&=&M_{AD}M_{BC}M_{A^{\prime }D^{\prime }}M_{B^{\prime }C^{\prime
}}-M_{AC}M_{BD}M_{A^{\prime }C^{\prime }}M_{B^{\prime }D^{\prime }}, 
\TCItag{4.36}
\end{eqnarray}%
gives rise to the splitting 
\begin{eqnarray}
&&M_{A^{\prime }B^{\prime }}M_{C^{\prime }D^{\prime }}(\text{X}_{(ABCD)}-%
\text{X}_{ABCD})+\text{c.c.}  \notag \\
&=&\frac{2}{3}\varpi (M_{AD}M_{BC}M_{A^{\prime }D^{\prime }}M_{B^{\prime
}C^{\prime }}-M_{AC}M_{BD}M_{A^{\prime }C^{\prime }}M_{B^{\prime }D^{\prime
}}).  \TCItag{4.37}
\end{eqnarray}

The electromagnetic contribution to the curvature spinors for either
formalism amounts to the pair of contracted pieces [10] 
\begin{equation}
\mathbf{E}=(\omega _{ABC}{}^{C},\text{ }\omega _{A^{\prime }B^{\prime
}C}{}^{C}),  \tag{4.38a}
\end{equation}%
which enter the bivector decomposition 
\begin{equation}
S_{AA^{\prime }}^{a}S_{BB^{\prime }}^{b}F_{ab}=\frac{i}{2}(M_{A^{\prime
}B^{\prime }}\omega _{ABC}{}^{C}+M_{AB}\omega _{A^{\prime }B^{\prime
}C}{}^{C}).  \tag{4.38b}
\end{equation}%
These electromagnetic spinors obey the conjugacy relations 
\begin{equation}
\omega _{ABC}{}^{C}=-\hspace{1pt}\omega _{ABC^{\prime }}{}^{C^{\prime
}},\omega _{A^{\prime }B^{\prime }C}{}^{C}=-\hspace{1pt}\omega _{A^{\prime
}B^{\prime }C^{\prime }}{}^{C^{\prime }}.  \tag{4.38c}
\end{equation}%
From Eq. (4.10), we get the relationships 
\begin{equation}
\omega _{ABC}{}^{C}=2i\nabla _{(A}^{C^{\prime }}\Phi _{B)C^{\prime }},\text{ 
}\omega _{A^{\prime }B^{\prime }C}{}^{C}=2i\nabla _{(A^{\prime }}^{C}\Phi
_{B^{\prime })C},  \tag{4.39}
\end{equation}%
whence we are led to the general spinor splittings 
\begin{equation}
\omega _{ABCD}=\omega _{(AB)(CD)}+\frac{1}{2}\omega _{(AB)L}{}^{L}M_{CD} 
\tag{4.40a}
\end{equation}%
and 
\begin{equation}
\omega _{A^{\prime }B^{\prime }CD}=\omega _{(A^{\prime }B^{\prime })(CD)}+%
\frac{1}{2}\omega _{(A^{\prime }B^{\prime })L}{}^{L}M_{CD},  \tag{4.40b}
\end{equation}%
together with their complex conjugates. Whereas the electromagnetic pieces
of Eqs. (4.40) behave in the $\gamma $-formalism as spin tensors, they occur
in the $\varepsilon $-formalism as invariant spin-tensor densities subject
to the laws 
\begin{equation}
\omega _{ABC}^{\prime }{}^{C}=(\Delta _{{\small \Lambda }})^{-1}\Lambda
_{A}{}^{L}\Lambda _{B}{}^{M}\omega _{LMC}{}^{C}=\omega _{ABC}{}^{C} 
\tag{4.41a}
\end{equation}%
and 
\begin{equation}
\omega _{A^{\prime }B^{\prime }C}^{\prime }{}^{C}=(\bar{\Delta}_{{\small %
\Lambda }})^{-1}\Lambda _{A^{\prime }}{}^{L^{\prime }}\Lambda _{B^{\prime
}}{}^{M^{\prime }}\omega _{L^{\prime }M^{\prime }C}{}^{C}=\omega _{A^{\prime
}B^{\prime }C}{}^{C}.  \tag{4.41b}
\end{equation}

As regards the computations that produce the derivation of the wave
equations for both formalisms [14], the key covariant derivative pattern is
written out as 
\begin{equation}
\lbrack \nabla _{AA^{\prime }},\nabla _{BB^{\prime }}]=M_{A^{\prime
}B^{\prime }}\Delta _{AB}+M_{AB}\Delta _{A^{\prime }B^{\prime }}.  \tag{4.42}
\end{equation}%
The $\Delta $-kernels involved on the right-hand side of (4.42) are both
symmetric second-order differential operators which bear linearity as well
as the Leibniz-rule property. In the $\gamma $-formalism, they behave
formally under gauge transformations as covariant spin tensors, with the
respective defining expressions being written as%
\begin{equation}
\Delta _{AB}=\nabla _{C^{\prime }(A}\nabla _{B)}^{C^{\prime }}-i\beta
_{C^{\prime }(A}\nabla _{B)}^{C^{\prime }}=-\nabla _{(A}^{C^{\prime }}\nabla
_{B)C^{\prime }}  \tag{4.43}
\end{equation}%
and 
\begin{equation}
\Delta _{A^{\prime }B^{\prime }}=\nabla _{C(A^{\prime }}\nabla _{B^{\prime
})}^{C}+i\beta _{C(A^{\prime }}\nabla _{B^{\prime })}^{C}=-\nabla
_{(A^{\prime }}^{C}\nabla _{B^{\prime })C},  \tag{4.44}
\end{equation}%
where $i\beta _{a}$ amounts to the eigenvalue carried by Eq. (3.73a). For
the $\varepsilon $-formalism, we have 
\begin{equation}
\Delta _{AB}=\nabla _{C^{\prime }(A}\nabla _{B)}^{C^{\prime }},\text{ }%
\Delta _{A^{\prime }B^{\prime }}=\nabla _{C(A^{\prime }}\nabla _{B^{\prime
})}^{C},  \tag{4.45}
\end{equation}%
with $\Delta _{AB}$ and $\Delta _{A^{\prime }B^{\prime }}$ thus behaving as
invariant spin-tensor densities of weight $-1$ and antiweight $-1$,
respectively. It is useful to remark that the covariant constancy of $%
M^{AB}M^{A^{\prime }B^{\prime }}$ enables one to define the contravariant
form of any $\Delta $-operator. In particular, the $\gamma $-formalism
version of $\Delta ^{AB}$, for instance, appears as 
\begin{equation}
\Delta ^{AB}=-(\nabla ^{C^{\prime }(A}\nabla _{C^{\prime }}^{B)}+i\beta
^{C^{\prime }(A}\nabla _{C^{\prime }}^{B)}),  \tag{4.46a}
\end{equation}%
or, equivalently, as 
\begin{equation}
\Delta ^{AB}=\nabla _{C^{\prime }}^{(A}\nabla ^{B)C^{\prime }},  \tag{4.46b}
\end{equation}%
with the relevant defining structure being in either formalism set as%
\footnote{%
Because of the symmetry of the $\Delta $-operators, there is no need for
staggering their indices.} 
\begin{equation}
\Delta ^{AB}\doteqdot M^{AC}M^{BD}\Delta _{CD}=M^{A(C}M^{D)B}\nabla
_{C}^{M^{\prime }}\nabla _{DM^{\prime }}.  \tag{4.47}
\end{equation}

One of the implications of the eventual presence of electromagnetic pieces
in curvature splittings is that an appropriate number of additional
contributions carrying terms of the same type as the entries of (4.38a) must
be incorporated into any $\Delta $-derivatives of arbitrary outer-product
configurations. Equations (4.7) and (4.42) suggest that some of the most
elementary derivatives should be prescribed in either formalism as 
\begin{equation}
\Delta _{AB}\zeta ^{C}=\omega _{ABM}{}^{C}\zeta ^{M}=\text{X}%
_{ABM}{}^{C}\zeta ^{M}+\frac{1}{2}\omega _{ABM}{}^{M}\zeta ^{C}  \tag{4.48a}
\end{equation}%
and 
\begin{equation}
\Delta _{A^{\prime }B^{\prime }}\zeta ^{C}=\omega _{A^{\prime }B^{\prime
}M}{}^{C}\zeta ^{M}=\Xi _{A^{\prime }B^{\prime }M}{}^{C}\zeta ^{M}+\frac{1}{2%
}\omega _{A^{\prime }B^{\prime }M}{}^{M}\zeta ^{C}.  \tag{4.48b}
\end{equation}%
The basic prescriptions for computing $\Delta $-derivatives of a covariant
spin vector $\xi _{A}$ can be obtained from (4.48) by carrying out Leibniz
expansions of the product $\zeta ^{C}\xi _{C}$. We then have\footnote{%
When acting on a world-spin scalar $h$, the $\Delta $-operators recover the
torsionlessness of $\nabla _{a}$ as $\Delta _{AB}h=0$ and $\Delta
_{A^{\prime }B^{\prime }}h=0$.} 
\begin{equation}
\Delta _{AB}\xi _{C}=-\hspace{1pt}\omega _{ABC}{}^{M}\xi _{M}=-\hspace{1pt}(%
\text{X}_{ABC}{}^{M}\xi _{M}+\frac{1}{2}\omega _{ABM}{}^{M}\xi _{C}) 
\tag{4.49a}
\end{equation}%
and 
\begin{equation}
\Delta _{A^{\prime }B^{\prime }}\xi _{C}=-\hspace{1pt}\omega _{A^{\prime
}B^{\prime }C}{}^{M}{}\xi _{M}=-\hspace{1pt}(\Xi _{A^{\prime }B^{\prime
}C}{}^{M}{}\xi _{M}+\frac{1}{2}\omega _{A^{\prime }B^{\prime }M}{}^{M}\xi
_{C}),  \tag{4.49b}
\end{equation}%
along with the complex conjugates of (4.48) and (4.49). For the complex
spin-scalar density defined by (3.22), we can write the derivatives 
\begin{equation}
\Delta _{AB}\alpha =-\mathfrak{w}\alpha \omega _{ABC}{}^{C}  \tag{4.50a}
\end{equation}%
and 
\begin{equation}
\Delta _{A^{\prime }B^{\prime }}\alpha =-\mathfrak{w}\alpha \omega
_{A^{\prime }B^{\prime }C}{}^{C},  \tag{4.50b}
\end{equation}%
which are usually thought of as coming from the integrability condition [40] 
\begin{equation}
\lbrack \nabla _{a},\nabla _{b}]\alpha =2\alpha \nabla _{\lbrack a}(\alpha
^{-1}\nabla _{b]}\alpha )=(-2\mathfrak{w}\alpha )\partial _{\lbrack
a}\vartheta _{b]}=2i\mathfrak{w}\alpha F_{ab},  \tag{4.51}
\end{equation}%
with $\vartheta _{a}$ standing for either of the affine devices $\gamma _{a}$
and $\Gamma _{aB}{}^{B}$. It is obvious that the right-hand sides of (4.50)
and (4.51) will turn out to vanish when gradient potentials are allowed for.
Because of the presupposition that both $\partial _{\lbrack a}\theta _{b]}$
and $\partial _{\lbrack a}\Pi _{b]}$ should vanish, any real spin-scalar
densities must behave in either formalism as numerical constants with
respect to the action of the $\Delta $-operators. The patterns of $\Delta $%
-derivatives of some spin-tensor density can certainly be specified from
Leibniz expansions like 
\begin{equation}
\Delta _{AB}(\alpha B_{C...D})=(\Delta _{AB}\alpha )B_{C...D}+\alpha \Delta
_{AB}B_{C...D},  \tag{4.52}
\end{equation}%
with $B_{C...D}$ being a spin tensor. It follows that if we invoke once more
the outer-product extension of the requirement (3.91), observing that Eq.
(3.114) entails the constancy of $\Delta _{\Lambda }$ with respect to the
action of $\nabla _{\lbrack a}\nabla _{b]}$, we shall conclude that the
gauge behaviours of any $\Delta $-derivatives bear both homogeneity and
linearity in either formalism. For example, we have the $\gamma $-formalism
law 
\begin{equation}
\Delta _{AB}^{\prime }(\alpha ^{\prime }B_{C...D}^{\prime })=(\Delta _{%
{\small \Lambda }})^{\mathfrak{w}}\Lambda _{A}{}^{G}\Lambda
_{B}{}^{H}\Lambda _{C}{}^{L}...\Lambda _{D}{}^{M}\Delta _{GH}(\alpha
B_{L...M}).  \tag{4.53}
\end{equation}

There are some situations of practical interest wherein the calculation of $%
\Delta $-derivatives may be carried out as if electromagnetic pieces were
absent from curvature splittings [10]. The first point concerning this
observation is related to the fact that there occurs a cancellation of those
pieces whenever $\Delta $-derivatives of arbitrary Hermitian quantities are
explicitly computed in either formalism, independently of which allowable
index configurations for the $\Delta $-operators are implemented. Such a
cancellation likewise happens when we let $\Delta $-operators act freely
upon spin tensors of valences $\{a,a;0,0\}$ and $\{0,0;c,c\}$. For $%
\mathfrak{w}<0$, it still occurs in the expansion (4.52) when the valence of 
$B_{C...D}$ equals $\{0,-2\mathfrak{w};0,0\}$ and $\func{Im}\alpha \neq 0$
everywhere. A similar property also holds for cases that involve outer
products between contravariant spin tensors and complex spin-scalar
densities having suitable positive weights.

\subsection{Wave Equations for Photons}

In both formalisms, the wave functions for photons in $\mathfrak{M}$
constitute the bivector decomposition given by Eqs. (4.38). The relevant
definitions are expressed as 
\begin{equation}
\phi _{AB}\doteqdot \frac{i}{2}\omega _{ABC}{}^{C},\text{ }\phi _{A^{\prime
}B^{\prime }}\doteqdot \frac{i}{2}\omega _{A^{\prime }B^{\prime }C}{}^{C}, 
\tag{4.54}
\end{equation}%
together with the field-potential relationships 
\begin{equation}
\phi _{AB}=-\nabla _{(A}^{C^{\prime }}\Phi _{B)C^{\prime }},\text{ }\phi
_{A^{\prime }B^{\prime }}=-\nabla _{(A^{\prime }}^{C}\Phi _{B^{\prime })C} 
\tag{4.55a}
\end{equation}%
and 
\begin{equation}
\phi {}^{AB}=\bigtriangledown _{C^{\prime }}^{(A}\Phi ^{B)C^{\prime }},\text{
}\phi {}^{A^{\prime }B^{\prime }}=\bigtriangledown _{C}^{(A^{\prime }}\Phi
^{B^{\prime })C}.  \tag{4.55b}
\end{equation}%
These wave functions are inextricably rooted into the curvature structure of 
$\mathfrak{M}$, being locally considered as massless uncharged fields of
spin $\pm 1$. At each point of $\mathfrak{M}$, they represent the six
geometric degrees of freedom of $W_{abC}{}^{C}$, in accordance with the
expansion 
\begin{equation}
S_{AA^{\prime }}^{a}S_{BB^{\prime }}^{b}F_{ab}=M_{A^{\prime }B^{\prime
}}\phi _{AB}+M_{AB}\phi _{A^{\prime }B^{\prime }}  \tag{4.56}
\end{equation}%
and its dual 
\begin{equation}
S_{AA^{\prime }}^{a}S_{BB^{\prime }}^{b}F_{ab}^{\ast }=i(M_{AB}\phi
_{A^{\prime }B^{\prime }}-M_{A^{\prime }B^{\prime }}\phi _{AB}).  \tag{4.57}
\end{equation}%
In the $\varepsilon $-formalism, $\phi _{AB}$ and $\phi _{A^{\prime
}B^{\prime }}$ bear gauge invariance, with any rearrangements of the indices
carried by (4.54) likewise leading to gauge-invariant fields. On the other
hand, the only index configurations that yield invariant fields in the $%
\gamma $-formalism are supplied by $\phi _{A}{}^{B}$ and $\phi _{A^{\prime
}}{}^{B^{\prime }}$, which visibly carry an invariant spin-tensor character
in the $\varepsilon $-formalism as well. The corresponding field equations
may arise from the coupled conjugate statements 
\begin{equation}
\nabla ^{AA^{\prime }}(S_{AA^{\prime }}^{a}S_{BB^{\prime
}}^{b}F_{ab}+iS_{AA^{\prime }}^{a}S_{BB^{\prime }}^{b}F_{ab}^{\ast })=0 
\tag{4.58a}
\end{equation}%
and 
\begin{equation}
\nabla ^{AA^{\prime }}(S_{AA^{\prime }}^{a}S_{BB^{\prime
}}^{b}F_{ab}-iS_{AA^{\prime }}^{a}S_{BB^{\prime }}^{b}F_{ab}^{\ast })=0. 
\tag{4.58b}
\end{equation}%
We then have the Maxwell equations 
\begin{equation}
\nabla ^{AA^{\prime }}(M_{A^{\prime }B^{\prime }}\phi _{AB})=0,\text{ }%
\nabla ^{AA^{\prime }}(M_{AB}\phi _{A^{\prime }B^{\prime }})=0.  \tag{4.59}
\end{equation}

In the $\gamma $-formalism, the statements (4.59) amount to the eigenvalue
equations 
\begin{equation}
\nabla ^{AB^{\prime }}\phi _{AB}=i\beta ^{AB^{\prime }}\phi
_{AB}\Leftrightarrow \nabla _{AB^{\prime }}\phi ^{AB}=(-i)\beta _{AB^{\prime
}}\phi ^{AB}  \tag{4.60a}
\end{equation}%
and 
\begin{equation}
\nabla ^{BA^{\prime }}\phi _{A^{\prime }B^{\prime }}=(-i)\beta ^{BA^{\prime
}}\phi _{A^{\prime }B^{\prime }}\Leftrightarrow \nabla _{BA^{\prime }}\phi
^{A^{\prime }B^{\prime }}=i\beta _{BA^{\prime }}\phi ^{A^{\prime }B^{\prime
}},  \tag{4.60b}
\end{equation}%
with the $\beta $-spinor being the same as the one carried by the
definitions (4.43). The specification of the gauge behaviours of Eqs. (4.60)
can be attained from the law 
\begin{equation}
(\nabla ^{\prime AB^{\prime }}-i\beta ^{\prime AB^{\prime }})\phi
_{AB}^{\prime }=\exp (2i\theta )(\nabla ^{AB^{\prime }}-i\beta ^{AB^{\prime
}})\phi _{AB},  \tag{4.61}
\end{equation}%
whence the gauge invariance of Maxwell's equations turns out to be exhibited
by either 
\begin{equation}
\nabla ^{\prime AB^{\prime }}\phi _{A}^{\prime }{}^{B}=\rho ^{-1}\nabla
^{AB^{\prime }}\phi _{A}{}^{B}=0  \tag{4.62}
\end{equation}%
or the complex conjugate of (4.62). Clearly, this result appears to be
compatible with the gauge invariance of the vacuum equations 
\begin{equation}
\nabla ^{a}F_{ab}=0,\text{ }\nabla ^{a}F_{ab}^{\ast }=0,  \tag{4.63}
\end{equation}%
with the second of which standing for the electromagnetic Bianchi identity.
In the $\varepsilon $-formalism, Eqs. (4.59) are reduced to the
gauge-invariant massless-free-field equations 
\begin{equation}
\nabla ^{AB^{\prime }}\phi _{AB}{}=0,\text{ }\nabla ^{BA^{\prime }}\phi
_{A^{\prime }B^{\prime }}{}=0.  \tag{4.64}
\end{equation}%
The gauge invariance of (4.64) is independent of any choices of index
configurations because of the $\nabla $-constancy of the $\varepsilon $%
-metric spinors.

In either formalism, the basic procedure for obtaining the wave equation
that controls the propagation of $\phi _{A}{}^{B}$, amounts to operating on
it with the $\nabla $-splitting 
\begin{equation}
\nabla _{A^{\prime }}^{C}\nabla ^{AA^{\prime }}=\Delta ^{AC}-\frac{1}{2}%
M^{AC}\square ,  \tag{4.65a}
\end{equation}%
and working out the resulting structure. For completing the calculational
steps in a systematic fashion, it is necessary to take account of the
algebraic rules 
\begin{equation}
2\nabla _{\lbrack C}^{A^{\prime }}\nabla _{A]A^{\prime }}=M_{AC}\square
=\nabla _{D}^{A^{\prime }}(M_{CA}\nabla _{A^{\prime }}^{D})  \tag{4.65b}
\end{equation}%
and 
\begin{equation}
2\nabla _{A^{\prime }}^{[C}\nabla ^{A]A^{\prime }}=M^{CA}\square =\nabla
_{A^{\prime }}^{D}(M^{AC}\nabla _{D}^{A^{\prime }}),  \tag{4.65c}
\end{equation}%
along with their complex conjugates and the gauge-invariant definition 
\begin{equation}
\square \doteqdot S_{MM^{\prime }}^{a}S^{bMM^{\prime }}\nabla _{a}\nabla
_{b}=\nabla _{MM^{\prime }}\nabla ^{MM^{\prime }}.  \tag{4.65d}
\end{equation}%
In the $\gamma $-formalism, we thus have 
\begin{equation}
\nabla _{A^{\prime }}^{C}\nabla ^{AA^{\prime }}\phi _{A}{}^{B}=\Delta
^{AC}\phi _{A}{}^{B}-\frac{1}{2}\gamma ^{AC}\square \phi _{A}{}^{B}=0. 
\tag{4.66}
\end{equation}%
Because of the valence pattern of $\phi _{A}{}^{B}$, the $\Delta $-expansion
of (4.66) just carries the X-spinor, namely 
\begin{equation}
\Delta ^{AC}\phi _{A}{}^{B}=\text{X}^{AC}{}_{M}{}^{B}\phi _{A}{}^{M}-\text{X}%
^{AC}{}_{A}{}^{M}\phi _{M}{}^{B}=\Delta ^{A(B}\phi _{A}{}^{C)}.  \tag{4.67}
\end{equation}%
Explicit calculations [10] show that the symmetry in $B$ and $C$ brought out
by (4.67) can be established by allowing for the result%
\begin{equation}
\Delta ^{A[C}\phi _{A}{}^{B]}=0.  \tag{4.68}
\end{equation}%
Hence, by rearranging the indices of the middle configuration of the
expansion (4.67) and invoking (4.34), we get the contribution 
\begin{equation}
\Delta ^{AB}\phi _{A}{}^{C}=\frac{4}{3}\varpi \phi ^{BC}-\omega
^{(ABCD)}\phi _{AD},  \tag{4.69}
\end{equation}%
which leads us to the gauge-invariant equation 
\begin{equation}
(\square +\frac{R}{3})\phi _{A}{}^{B}=-2\Psi _{AD}{}{}^{BC}\phi _{C}{}^{D}, 
\tag{4.70a}
\end{equation}%
with the definition 
\begin{equation}
\Psi _{ABCD}\doteqdot \omega _{(ABCD)}=\text{X}_{(ABCD)}.  \tag{4.70b}
\end{equation}

Since $\phi _{A}{}^{B}$ bears a tensor character in both formalisms, the $%
\varepsilon $-formalism expansion for $\Delta ^{AC}\phi _{A}{}^{B}$ is
formally the same as (4.67), whence the corresponding wave equation is an
invariant tensor statement of the same form as (4.70a). The $\varepsilon $%
-formalism wave equation for $\phi _{AB}$ may of course be readily written
as 
\begin{equation}
(\square +\frac{R}{3})\phi _{AB}{}{}=2\Psi _{AB}{}{}^{CD}\phi _{CD}{}. 
\tag{4.71}
\end{equation}%
This result agrees with the fact that the wave function $\phi _{AB}$ for the 
$\varepsilon $-formalism is a two-index spin-tensor density of weight $-1$.
Consequently, one might still implement the purely gravitational pattern of
(4.67) upon expanding $\Delta ^{AB}\phi _{AC}{}$. The $\gamma $-formalism
version of Eq. (4.71) emerges from working out the configuration 
\begin{equation}
2\Delta ^{AC}\phi _{AB}{}-\gamma ^{AC}\square \phi _{AB}{}=\nabla
_{A^{\prime }}^{C}(2i\beta ^{AA^{\prime }}\phi _{AB}),  \tag{4.72a}
\end{equation}%
with the pertinent equation amounting, in effect, to the spin-tensor
statement 
\begin{equation}
(\square -2i\beta ^{h}\nabla _{h}-\Upsilon _{(\mathcal{P})}+\frac{R}{3})\phi
_{AB}{}{}=2\Psi _{AB}{}{}^{CD}\phi _{CD}{}  \tag{4.72b}
\end{equation}%
and 
\begin{equation}
\Upsilon _{(\mathcal{P})}\doteqdot \beta ^{h}\beta _{h}+i(\square \Phi
+2\nabla _{h}\Phi ^{h}).  \tag{4.72c}
\end{equation}%
It was shown in Ref. [10] that the right-hand side of (4.72a) is essentially
constituted by the Leibniz contributions 
\begin{equation}
\beta ^{AA^{\prime }}\nabla _{CA^{\prime }}\phi _{AB}=(\beta ^{h}\nabla _{h}-%
\frac{1}{2}i\beta ^{h}\beta _{h})\phi _{BC}{}  \tag{4.72d}
\end{equation}%
and 
\begin{equation}
(\nabla _{CA^{\prime }}\beta ^{AA^{\prime }})\phi _{AB}=(\frac{1}{2}\square
\Phi +\nabla _{h}\Phi ^{h})\phi _{BC}{}+2\phi _{C}{}{}^{A}{}{}\phi
_{AB}{}{}{}.  \tag{4.72e}
\end{equation}%
By combining pieces together, we can see that the (skew) non-linear term $%
4i\phi _{C}{}{}^{A}{}{}\phi _{AB}$ cancels out because of the expansion 
\begin{equation}
2\Delta ^{AC}\phi _{AB}{}=\frac{R}{3}\phi _{B}{}{}^{C}{}{}-2\Psi
_{B}{}{}^{CMN}\phi _{MN}{}-2\omega ^{AC}{}_{M}{}^{M}\phi _{AB}{}. 
\tag{4.72f}
\end{equation}

In either formalism, the wave equation for $\Phi _{AA^{\prime }}$ can be
derived by working out any of the relationships (4.55). For instance, 
\begin{equation}
(-2)\phi _{A}{}^{B}=\nabla ^{BB^{\prime }}\Phi _{AB^{\prime }}+M^{BC}\nabla
_{A}^{B^{\prime }}\Phi _{CB^{\prime }},  \tag{4.73a}
\end{equation}%
whence 
\begin{equation}
\nabla ^{AA^{\prime }}\nabla ^{BB^{\prime }}\Phi _{AB^{\prime }}+\nabla
^{AA^{\prime }}(M^{BC}\nabla _{A}^{B^{\prime }}\Phi _{CB^{\prime }})=0. 
\tag{4.73b}
\end{equation}%
For the first piece of (4.73b), we may utilize the operator splitting 
\begin{equation}
\nabla ^{AA^{\prime }}\nabla ^{BB^{\prime }}=\nabla ^{BA^{\prime }}\nabla
^{AB^{\prime }}+M^{AB}(\frac{1}{2}M^{A^{\prime }B^{\prime }}\square +\nabla
_{C}^{(A^{\prime }}\nabla ^{B^{\prime })C}),  \tag{4.74}
\end{equation}%
to obtain the expression 
\begin{equation}
\nabla ^{AA^{\prime }}\nabla ^{BB^{\prime }}\Phi _{AB^{\prime }}=M^{AB}(%
\frac{1}{2}M^{A^{\prime }B^{\prime }}\square +\nabla _{C}^{(A^{\prime
}}\nabla ^{B^{\prime })C})\Phi _{AB^{\prime }}+\nabla ^{BA^{\prime }}\Theta ,
\tag{4.75a}
\end{equation}%
where $\Theta $ is the Lorentz world scalar\footnote{%
We emphasize that the quantity $\Theta $ transforms under the action of the
gauge group as $\Theta ^{\prime }=\Theta -\square \theta $.} 
\begin{equation}
\Theta \doteqdot S_{MM^{\prime }}^{a}S^{bMM^{\prime }}\nabla _{a}\Phi
_{b}=\nabla _{MM^{\prime }}\Phi ^{MM^{\prime }}.  \tag{4.75b}
\end{equation}%
For the other piece of (4.73b), we have the calculation 
\begin{eqnarray}
\nabla ^{AA^{\prime }}(M^{BC}\nabla _{A}^{B^{\prime }}\Phi _{CB^{\prime }})
&=&\nabla ^{AA^{\prime }}(M^{BC}\nabla _{(A}^{B^{\prime }}\Phi _{C)B^{\prime
}}+\frac{1}{2}M^{BC}M_{CA}\Theta )  \notag \\
&=&(-\frac{1}{2})\nabla ^{BA^{\prime }}\Theta ,  \TCItag{4.76}
\end{eqnarray}%
with the field equation (4.62) having been employed.

The complex conjugates of Eqs. (4.46) supply the $\gamma $-formalism
configuration 
\begin{equation}
\nabla ^{AA^{\prime }}\nabla ^{BB^{\prime }}\Phi _{AB^{\prime }}=\gamma
^{AB}(\frac{1}{2}\gamma ^{A^{\prime }B^{\prime }}\square \Phi _{AB^{\prime
}}+\Delta ^{A^{\prime }B^{\prime }}\Phi _{AB^{\prime }})+\nabla ^{BA^{\prime
}}\Theta ,  \tag{4.77}
\end{equation}%
whence adding together (4.76) and (4.77) produces the structure 
\begin{equation}
\gamma ^{AB}(\gamma ^{A^{\prime }B^{\prime }}\square \Phi _{AB^{\prime
}}+2\Delta ^{A^{\prime }B^{\prime }}\Phi _{AB^{\prime }})+\nabla
^{BA^{\prime }}\Theta =0.  \tag{4.78}
\end{equation}%
By virtue of the Hermiticity of $\Phi _{AB^{\prime }}$, the $\Delta $%
-expansion of (4.78) as prescribed by Eqs. (4.49) carries only the
gravitational contributions borne by 
\begin{equation}
\Delta ^{A^{\prime }B^{\prime }}\Phi _{AB^{\prime }}=\frac{1}{2}%
R{}_{A}{}^{A^{\prime }BB^{\prime }}\Phi _{BB^{\prime }},  \tag{4.79}
\end{equation}%
with $R{}_{AA^{\prime }BB^{\prime }}{}$ being given by the expression
(4.28). Some trivial manipulations then yield the statement%
\begin{equation}
\square \Phi _{AA^{\prime }}+R{}_{AA^{\prime }}{}^{BB^{\prime }}\Phi
_{BB^{\prime }}-\nabla _{AA^{\prime }}\Theta =0.  \tag{4.80}
\end{equation}%
Under the cosmological circumstances of Eqs. (2.50), we may reinstate (4.80)
as%
\begin{equation}
(\square +\lambda )\Phi _{AA^{\prime }}-\nabla _{AA^{\prime }}\Theta =0. 
\tag{4.81}
\end{equation}

It has become obvious that the $\varepsilon $-formalism version of $\Delta
^{A^{\prime }B^{\prime }}\Phi _{AB^{\prime }}$ bears the same form as the
structure (4.79). Combining (4.75) and (4.76) thus leads to a wave equation
bearing the same form as the statement (4.80). Since the action of either $%
\square $-operator on any appropriate Hermitian $S$-matrix produces a
vanishing outcome, we can establish that electromagnetic potentials for both
formalisms must coincide with each other when electromagnetic fields are
present. If instead of (4.73a) we had used the configuration for either $%
\phi _{AB}$ or $\phi ^{AB}$, we would have derived the same wave equation
for $\Phi _{AA^{\prime }}$ as the ones exhibited above. In either formalism,
the pattern of the traditional spacetime wave equation for $\Phi _{a}$ could
therefore be recovered from (4.80) just by invoking the requirement (3.33).
In accordance with Ref. [10], we stress that the main point regarding the
situation at issue is associated to a commonness feature of the Maxwell
bivectors carried by the formalisms. Apparently, it gets strengthened when
one carries out the world computation 
\begin{eqnarray}
\nabla ^{b}F_{ba} &=&\nabla ^{b}(\nabla _{b}\Phi _{a}-\nabla _{a}\Phi
_{b})=\square \Phi _{a}-g^{bh}([\nabla _{h},\nabla _{a}]+\nabla _{a}\nabla
_{h})\Phi _{b}  \notag \\
&=&\square \Phi _{a}-[\nabla _{b},\nabla _{a}]\Phi ^{b}-\nabla _{a}\Theta
=\square \Phi _{a}+R_{a}{}^{b}\Phi _{b}-\nabla _{a}\Theta .  \TCItag{4.82}
\end{eqnarray}

\subsection{Wave Equations for Gravitons}

The totally symmetric curvature piece defined by Eq. (4.70b) is one of the
Weyl spinor fields. In both formalisms, such objects enter together with
their complex conjugates into the spinor expression for the Weyl tensor $%
C_{abcd}$ of $\mathfrak{M}$, according to the scheme [12, 13]%
\begin{equation}
S_{AA^{\prime }}^{a}S_{BB^{\prime }}^{b}S_{CC^{\prime }}^{c}S_{DD^{\prime
}}^{d}C_{abcd}=M_{A^{\prime }B^{\prime }}M_{C^{\prime }D^{\prime }}\Psi
_{ABCD}+\text{c.c.}.  \tag{4.83}
\end{equation}%
At each point of $\mathfrak{M}$, the conjugate $\Psi $-fields for either
formalism are taken to represent the ten independent degrees of freedom of $%
g_{ab}$. Physically, they are massless uncharged wave functions carrying
spin $\pm 2$, which lie deeply in the gravitational structure of $\mathfrak{M%
}$. The derivation of the relevant field equations usually employs the
expression (4.26) along with the second of Eqs. (2.29), to work out the
coupled conjugate relations [10] 
\begin{equation}
M^{C^{\prime }D^{\prime }}\nabla ^{AA^{\prime }}{}^{\ast }R_{AA^{\prime
}BB^{\prime }CC^{\prime }DD^{\prime }}=0  \tag{4.84a}
\end{equation}%
and 
\begin{equation}
M^{CD}\nabla ^{AA^{\prime }}{}{}^{\ast }R_{AA^{\prime }BB^{\prime
}CC^{\prime }DD^{\prime }}=0,  \tag{4.84b}
\end{equation}%
which constitute the spinor version of the gravitational Bianchi identity.

In the $\gamma $-formalism, Eq. (4.84a) takes the explicit form 
\begin{equation}
\nabla _{B^{\prime }}^{A}\text{X}_{ABCD}-2i\beta _{B^{\prime }}^{A}\text{X}%
_{ABCD}=\nabla _{B}^{A^{\prime }}\Xi _{A^{\prime }B^{\prime }CD},  \tag{4.85}
\end{equation}%
which can be rewritten as%
\begin{equation}
\nabla ^{AA^{\prime }}(\text{X}_{ABC}{}^{D}\gamma _{A^{\prime }B^{\prime
}})=\nabla ^{AA^{\prime }}(\Xi _{A^{\prime }B^{\prime }C}{}^{D}\gamma _{AB}).
\tag{4.86}
\end{equation}%
Hence, performing a symmetrization over the indices $B$, $C$ and $D$ of
(4.85), and recalling the property (4.35), yields the statement 
\begin{equation}
\nabla _{B^{\prime }}^{A}\Psi _{ABCD}-2i\beta _{B^{\prime }}^{A}\Psi
_{ABCD}=\nabla _{(B}^{A^{\prime }}\Xi _{CD)A^{\prime }B^{\prime }}. 
\tag{4.87}
\end{equation}%
As emphasized in Ref. [11], the skew parts in $B$ and $C$ of the terms
involved in (4.86) produce a differential gravitational relationship which
does not depend upon whether electromagnetic fields are present or absent.
We have, in effect, 
\begin{equation}
\nabla _{B^{\prime }}^{A}\text{X}_{A[BC]D}-2i\beta _{B^{\prime }}^{A}\text{X}%
_{A[BC]D}=\nabla _{\lbrack B}^{A^{\prime }}\Xi _{C]DA^{\prime }B^{\prime }},
\tag{4.88}
\end{equation}%
whence, after performing some calculations, we obtain 
\begin{equation}
(-8)\nabla ^{AA^{\prime }}\Xi _{AA^{\prime }BB^{\prime }}=\nabla
_{BB^{\prime }}R.  \tag{4.89}
\end{equation}%
The procedure that leads to the statement (4.87) annihilates the information
carried by the $\varpi $-piece of (4.34). In vacuum, we can then write down
the gauge-covariant eigenvalue equations 
\begin{equation}
\nabla ^{AB^{\prime }}\Psi _{ABCD}=2i\beta ^{AB^{\prime }}\Psi
_{ABCD}\Leftrightarrow \nabla _{AB^{\prime }}\Psi ^{ABCD}=(-2i)\beta
_{AB^{\prime }}\Psi ^{ABCD},  \tag{4.90}
\end{equation}%
which can be rewritten as the invariant massless-free-field equation 
\begin{equation}
\nabla ^{AA^{\prime }}\Psi _{AB}{}^{CD}=0.  \tag{4.91}
\end{equation}%
From the transformation law (4.15a), we see that the $\varepsilon $%
-formalism version of $\Psi _{AB}{}^{CD}$ amounts to an invariant
spin-tensor wave function, whence the corresponding field equation is
formally the same as the statement (4.91).

For the purpose of deriving the wave equations for gravitons in both
formalisms, we may follow up the same starting procedure as that for the
electromagnetic situation. In the $\gamma $-formalism, we thus allow for the
splitting 
\begin{equation}
\nabla _{A^{\prime }}^{E}\nabla ^{AA^{\prime }}\Psi _{AB}{}^{CD}=\Delta
^{AE}\Psi _{AB}{}^{CD}-\frac{1}{2}\gamma ^{AE}\square \Psi _{AB}{}^{CD}=0, 
\tag{4.92}
\end{equation}%
and account for Eq. (4.34) to get the calculational result [10, 11]%
\begin{equation}
\Delta ^{AE}\Psi _{AB}{}^{CD}=\frac{R}{4}\Psi
{}^{CDE}{}_{B}-3Q^{(CDEL)}{}\gamma _{LB},  \tag{4.93a}
\end{equation}%
along with the definition%
\begin{equation}
Q^{CDEL}{}\doteqdot \Psi _{MN}{}^{CD}\Psi {}^{ELMN}{}  \tag{4.93b}
\end{equation}%
and the expansion%
\begin{equation}
4Q^{(CDEL)}{}=Q^{(CDE)L}{}+Q^{(CDL)E}{}+Q^{(CEL)D}{}+Q^{(DEL)C}{}=4Q^{(CDE)L}{}.
\tag{4.93c}
\end{equation}%
Consequently, one is led to the gauge-invariant vacuum equation 
\begin{equation}
(\square +\frac{R}{2})\Psi {}_{AB}{}^{CD}=6\Psi _{MN}{}^{(CD}\Psi
{}^{EL)MN}\gamma _{EA}\gamma _{LB}.  \tag{4.94}
\end{equation}%
The $\varepsilon $-formalism version of the splitting (4.92) reads 
\begin{equation}
\nabla _{A^{\prime }}^{E}\nabla ^{AA^{\prime }}\Psi _{AB}{}^{CD}=\Delta
^{AE}\Psi _{AB}{}^{CD}-\frac{1}{2}\varepsilon ^{AE}\square \Psi
_{AB}{}^{CD}=0.  \tag{4.95}
\end{equation}%
As the index configuration of $\Psi _{AB}{}^{CD}$ yields a spin-tensor
character in both formalisms, we can say that the computation of the $\Delta 
$-derivative of (4.95) bears the same form as that implemented above as Eqs.
(4.93). It is also clear that any $\Delta $-derivatives of $\Psi {}_{ABCD}{}$
within the $\varepsilon $-formalism carry only gravitational contributions%
\footnote{%
This observation is evidently similar to that made previously concerning the 
$\varepsilon $-formalism version of $\phi _{AB}$.} since we are supposedly
dealing with a four-index spin-tensor density of weight $-2$. Hence, we can
write the $\varepsilon $-formalism statement 
\begin{equation}
(\square +\frac{R}{2})\Psi {}_{ABCD}{}=6\Psi _{MN(AB}{}\Psi {}_{CD)}{}^{MN}.
\tag{4.96}
\end{equation}%
The $\gamma $-formalism pattern carrying $\Delta ^{AE}\Psi {}_{ABCD}{}$
appears as 
\begin{equation}
(2\Delta {}^{AE}+2i\beta ^{EB^{\prime }}\nabla _{B^{\prime }}^{A}-\gamma
^{AE}\square )\Psi {}_{ABCD}{}=(-4i)\nabla ^{EB^{\prime }}(\beta _{B^{\prime
}}^{A}\Psi {}_{ABCD}{}).  \tag{4.97}
\end{equation}%
It may be seen [10] that some of the pieces of (4.97) can be manipulated so
as to give the contributions 
\begin{equation}
2\Delta _{E}{}^{A}\Psi _{ABCD}=\frac{R}{2}\Psi _{BCDE}{}-6Q_{(BCDE)}+8i\phi
_{E}{}^{A}\Psi _{ABCD}{},  \tag{4.98a}
\end{equation}%
\begin{equation}
2i\beta _{E}^{B^{\prime }}\nabla _{B^{\prime }}^{A}\Psi _{ABCD}{}=2(\beta
^{h}\beta _{h})\Psi _{BCDE}{}  \tag{4.98b}
\end{equation}%
and 
\begin{equation}
(-4i)\nabla _{E}^{B^{\prime }}(\beta _{B^{\prime }}^{A}\Psi
_{ABCD}{})=(2\beta ^{h}\beta _{h}+4i\beta ^{h}\nabla _{h}+\Upsilon _{(%
\mathcal{G})})\Psi _{BCDE}{}+8i\phi _{E}{}^{A}\Psi _{ABCD}{},  \tag{4.98c}
\end{equation}%
with 
\begin{equation}
\Upsilon _{(\mathcal{G})}\doteqdot 2(\beta ^{h}\beta _{h}+\Upsilon _{(%
\mathcal{P})}),  \tag{4.99}
\end{equation}%
and $\Upsilon _{(\mathcal{P})}$ being given by (4.72c). The resulting wave
equation is then written as 
\begin{equation}
(\square -4i\beta ^{h}\nabla _{h}-\Upsilon _{(\mathcal{G})}+\frac{R}{2})\Psi
{}_{ABCD}{}=6\Psi _{MN(AB}{}\Psi {}_{CD)}{}^{MN}.  \tag{4.100}
\end{equation}

Equations (4.94) and (4.100) can be derived from one another by taking
account of the differential prescriptions 
\begin{equation}
\square \Psi _{ABCD}=\square (\Psi _{AB}{}^{LM}\gamma _{LC}\gamma _{MD}),%
\text{ }\square (\gamma _{LC}\gamma _{MD})=(-\hspace{2pt}\overline{\Upsilon }%
_{(\mathcal{G})})\gamma _{LC}\gamma _{MD}  \tag{4.101a}
\end{equation}%
and 
\begin{equation}
2(\nabla _{a}\Psi _{AB}{}^{LM})\nabla ^{a}(\gamma _{LC}\gamma
_{MD})=4(2\beta ^{h}\beta _{h}+i\beta ^{h}\nabla _{h})\Psi _{ABCD}. 
\tag{4.101b}
\end{equation}%
By following up this procedure, we can deduce Eq. (4.100) without having to
perform the somewhat lengthy calculations that yield the contributions
(4.98). It becomes obvious that the $\gamma $-formalism vacuum wave equation
for $\Psi ^{ABCD}$ might also be derived by making use of a similar
procedure which takes up the configurations 
\begin{equation}
\square \Psi ^{ABCD}=\square (\gamma ^{AL}\gamma ^{BM}\Psi _{LM}{}^{CD}),%
\text{ }\square (\gamma ^{AL}\gamma ^{BM})=(-\hspace{2pt}\Upsilon _{(%
\mathcal{G})})\gamma ^{AL}\gamma ^{BM}  \tag{4.102a}
\end{equation}%
and 
\begin{equation}
2\nabla ^{a}(\gamma ^{AL}\gamma ^{BM})(\nabla _{a}\Psi
_{LM}{}^{CD})=4(2\beta ^{h}\beta _{h}-i\beta ^{h}\nabla _{h})\Psi ^{ABCD}. 
\tag{4.102b}
\end{equation}%
We have, in effect, 
\begin{equation}
(\square +4i\beta ^{h}\bigtriangledown _{h}-\overline{\Upsilon }_{(\mathcal{G%
})}+\frac{R}{2})\Psi ^{ABCD}=6\Psi _{MN}{}^{(AB}\Psi ^{CD)MN}.  \tag{4.103}
\end{equation}%
As had been established in Ref. [10], the $\gamma $-formalism wave equations
satisfied by any fields of valences $\{a,0;0,0\}$ and $\{0,a;0,0\}$, as well
as their complex-conjugate versions, can be obtained from each other by
invoking the interchange rule\footnote{%
This rule gives the equation $(\square +2i\beta ^{h}\bigtriangledown _{h}-%
\overline{\Upsilon }_{(\mathcal{P})}+\frac{R}{3})\phi {}^{AB}{}=2\Psi
^{AB}{}_{CD}\phi {}^{CD}{}$ straightaway from (4.72b). We should notice that
both $\Upsilon _{(\mathcal{P})}$ and $\Upsilon _{(\mathcal{G})}$ bear gauge
invariance.} 
\begin{equation}
i\beta ^{h}\nabla _{h}\leftrightarrow (-i)\beta ^{h}\nabla _{h},\text{ }%
(\Upsilon _{(\mathcal{P})},\Upsilon _{(\mathcal{G})})\leftrightarrow (%
\overline{\Upsilon }_{(\mathcal{P})},\overline{\Upsilon }_{(\mathcal{G})}). 
\tag{4.104}
\end{equation}

\subsection{Wave Equations for Dirac Fields}

As in the case of world-spin curvature objects, the Infeld-van der Waerden
treatment of Dirac fields [5] entirely left out the decompositions that
occur in operator bivector expansions for covariant differential
commutators. The achievement of the spinor computational techniques utilized
in the previous Subsections has also afforded [33] a complete description of
the interaction couplings carried by the wave equations for Dirac fields in $%
\mathfrak{M}$. A notable feature of these configurations is that they are
strictly exhibited by the patterns of the $\gamma $-formalism equations
which control the propagation of the fields. Only couplings of Dirac
particles with underlying photons are brought about by the relevant
derivation procedures, there actually occurring \textit{no} couplings that
involve wave functions for gravitons. In fact, the interaction pieces turn
out all to be cancelled when we set up the wave equations for the $%
\varepsilon $-formalism.

The issue concerning the description of the fundamental couplings between
Dirac fields and Infeld-van der Waerden photons is now entertained. Of
course, the curvature splittings of $\mathfrak{M}$ will once again be
assumed to carry nowhere-vanishing electromagnetic contributions. Like the
situation of the original formulation [5], any Dirac field will be
physically thought of as a classical wave function. However, no specific
energy character will throughout what follows be attributed to it. The $%
\Delta $-operator prescriptions of Subsection (4.1) will be used so many
times here that we shall not refer to them explicitly upon deriving our wave
equations.

A Dirac system in $\mathfrak{M}$ can be defined in either formalism as the
conjugate field pairs borne by the set 
\begin{equation}
\mathbf{D}=\{\{\psi ^{A},\chi _{A^{\prime }}\},\{\chi _{A},\psi ^{A^{\prime
}}\}\}.  \tag{4.105}
\end{equation}%
All fields of this set are usually taken to possess the same rest mass $m$.
The entries of each pair have the opposite helicity values $+1/2$ and $-1/2$%
, but such values get reversed when we pass from one pair to the other. In
addition, each of the pairs carries the same electric charge, with the
charge of one pair being opposite to the charge of the other pair. In the $%
\gamma $-formalism, any element of the set (4.105) behaves as a spin vector
under the action of the gauge group. The unprimed and primed elements of the
former pair appear in the $\varepsilon $-formalism as spin-vector densities
of weight $+1/2$ and antiweight $-1/2$, respectively. It is clear that the
weights of the $\varepsilon $-formalism version of the conjugate fields turn
out to be the other way about.

In both formalisms, the theory of Dirac fields was originally taken [5] as
the combination of the statements 
\begin{equation}
\nabla _{AA^{\prime }}\psi ^{A}=(-i)\mu \chi _{A^{\prime }},\text{ }\nabla
^{AA^{\prime }}\chi _{A^{\prime }}=(-i)\mu \psi ^{A}  \tag{4.106}
\end{equation}%
with their complex conjugates.\footnote{%
The coupling constant borne by (4.106) carries the normalized rest mass $\mu
=m/\sqrt{2}$.} In the $\gamma $-formalism, the field equations (4.106) are
equivalent to 
\begin{equation}
\nabla ^{AA^{\prime }}\psi _{A}=i(\mu \chi ^{A^{\prime }}+\beta ^{AA^{\prime
}}\psi _{A}),\text{ }\nabla _{AA^{\prime }}\chi ^{A^{\prime }}=i(\mu \psi
_{A}+\beta _{AA^{\prime }}\chi ^{A^{\prime }}).  \tag{4.107}
\end{equation}%
The $\varepsilon $-formalism version of (4.107) is given by 
\begin{equation}
\nabla ^{AA^{\prime }}\psi _{A}=i\mu \chi ^{A^{\prime }},\text{ }\nabla
_{AA^{\prime }}\chi ^{A^{\prime }}=i\mu \psi _{A},  \tag{4.108}
\end{equation}%
which evidently can be recast into the form of (4.106), with the wave
functions $\{\psi _{A},$ $\chi ^{A^{\prime }}\}$ showing up as spin-vector
densities of weight $-1/2$ and antiweight $+1/2$. Hence, if we operate with $%
\nabla _{B}^{A^{\prime }}$ on the first of Eqs. (4.106), likewise
implementing the field equation for $\chi _{A^{\prime }}$, we will arrive at
the $\gamma $-formalism statement 
\begin{equation}
(\gamma _{AB}\square -2\Delta _{AB})\psi ^{A}=(-2)\mu ^{2}\psi _{B}, 
\tag{4.109}
\end{equation}%
which amounts to the wave equation 
\begin{equation}
(\square +\frac{R}{4}+m^{2})\psi ^{A}=(-2i)\phi ^{A}{}_{B}\psi ^{B}. 
\tag{4.110}
\end{equation}%
A similar procedure yields the wave equation for $\chi _{A^{\prime }}$%
\begin{equation}
(\square +\frac{R}{4}+m^{2})\chi _{A^{\prime }}=2i\phi {}_{A^{\prime
}}{}^{B^{\prime }}\chi _{B^{\prime }},  \tag{4.111}
\end{equation}%
which accordingly comes from the configuration 
\begin{equation}
(2\Delta ^{A^{\prime }B^{\prime }}-\gamma ^{A^{\prime }B^{\prime }}\square
)\chi _{A^{\prime }}=(-2)\mu ^{2}\chi ^{B^{\prime }}.  \tag{4.112}
\end{equation}%
The $\varepsilon $-formalism counterparts of Eqs. (4.109) and (4.112)
involve the derivatives 
\begin{equation}
\Delta _{AB}\psi ^{A}=-\frac{R}{8}\psi _{B},\text{ }\Delta ^{A^{\prime
}B^{\prime }}\chi _{A^{\prime }}=\frac{R}{8}\chi ^{B^{\prime }},  \tag{4.113}
\end{equation}%
whence the corresponding wave equations are written as\footnote{%
In the $\varepsilon $-formalism, we also have $(\square +\frac{R}{4}%
+m^{2})\psi _{A}=0$ and $(\square +\frac{R}{4}+m^{2})\chi ^{A^{\prime }}=0$.}
\begin{equation}
(\square +\frac{R}{4}+m^{2})\psi ^{A}=0,\text{ }(\square +\frac{R}{4}%
+m^{2})\chi _{A^{\prime }}=0.  \tag{4.114}
\end{equation}%
It becomes evident that the reason for the non-occurrence of Maxwell-Dirac
interactions within the $\varepsilon $-formalism is related to the
spin-density character of the respective Dirac wave functions.

A particular procedure for deriving the $\gamma $-formalism wave equations
for the fields of the pair $\{\psi _{A},\chi ^{A^{\prime }}\}$ consists in
allowing suitably indexed $\nabla $-operators to act through Eqs. (4.107),
taking up thereafter either the contravariant differential configuration
(4.65c) or its complex conjugate. For $\psi _{A}$, for instance, we thus
have the differential relation 
\begin{equation}
\Delta ^{AB}\psi _{A}-\frac{1}{2}\gamma ^{AB}\square \psi _{A}=i\nabla
_{A^{\prime }}^{B}(\mu \chi ^{A^{\prime }}+\beta ^{AA^{\prime }}\psi _{A}). 
\tag{4.115}
\end{equation}%
Some calculations similar to those for photons performed anteriorly, supply
the following contributions to the right-hand side of Eq. (4.115): 
\begin{equation}
i\beta ^{AA^{\prime }}\nabla _{A^{\prime }}^{B}\psi _{A}=\frac{1}{2}(\beta
^{h}\beta _{h})\psi ^{B}-i\gamma ^{AB}(\beta ^{h}\nabla _{h}\psi _{A})-\mu
\beta ^{BA^{\prime }}\chi _{A^{\prime }}  \tag{4.116}
\end{equation}%
and 
\begin{equation}
i(\nabla _{A^{\prime }}^{B}\beta ^{AA^{\prime }})\psi _{A}=\frac{i}{2}%
(\nabla _{h}\beta ^{h})\psi ^{B}+2i\phi ^{AB}\psi _{A}.  \tag{4.117}
\end{equation}%
It should be noticed that the computation which produces the right-hand side
of (4.117) absorbs one of the relations (4.55). Then, implementing the
expression 
\begin{equation}
\Delta ^{AB}\psi _{A}=\frac{R}{8}\psi ^{B}+i\phi ^{AB}\psi _{A},  \tag{4.118}
\end{equation}%
along with the second of Eqs. (4.107), yields 
\begin{equation}
(\square +\frac{R}{4}+m^{2}-2i\beta ^{h}\nabla _{h}-\Upsilon _{(\mathcal{P}%
)})\psi _{A}=2i\phi {}_{A}{}^{B}\psi _{B},  \tag{4.119}
\end{equation}%
with $\Upsilon _{(\mathcal{P})}$ being given by the definition (4.72c). For $%
\chi ^{A^{\prime }}$, we likewise obtain the formulae 
\begin{equation}
i\beta _{AA^{\prime }}\nabla _{B^{\prime }}^{A}\chi ^{A^{\prime }}=\frac{1}{2%
}(\beta ^{h}\beta _{h})\chi _{B^{\prime }}+i\gamma _{A^{\prime }B^{\prime
}}(\beta ^{h}\nabla _{h}\chi ^{A^{\prime }})-\mu \beta _{AB^{\prime }}\psi
^{A},  \tag{4.120}
\end{equation}%
\begin{equation}
i(\nabla _{B^{\prime }}^{A}\beta _{AA^{\prime }})\chi ^{A^{\prime }}=\frac{i%
}{2}(\nabla _{h}\beta ^{h})\chi _{B^{\prime }}-2i\phi _{A^{\prime }B^{\prime
}}\chi ^{A^{\prime }}  \tag{4.121}
\end{equation}%
and 
\begin{equation}
\Delta _{A^{\prime }B^{\prime }}\chi ^{A^{\prime }}=i\phi _{A^{\prime
}B^{\prime }}\chi ^{A^{\prime }}-\frac{R}{8}\chi _{B^{\prime }},  \tag{4.122}
\end{equation}%
which lead us to the equation 
\begin{equation}
(\square +\frac{R}{4}+m^{2}-2i\beta ^{h}\nabla _{h}-\Upsilon _{(\mathcal{P}%
)})\chi ^{A^{\prime }}=(-2i)\phi ^{A^{\prime }}{}_{B^{\prime }}\chi
^{B^{\prime }}.  \tag{4.123}
\end{equation}

The consistency between the $\gamma $-formalism wave equations we have
exhibited can be verified by taking into account the prescriptions 
\begin{equation}
\square \gamma ^{BC}=(-\Upsilon _{(\mathcal{P})})\gamma ^{BC},\text{ }%
\square \gamma _{BC}=(-\overline{\Upsilon }_{(\mathcal{P})})\gamma _{BC} 
\tag{4.124a}
\end{equation}%
and%
\begin{equation}
\square (\psi ^{A})=\gamma ^{AB}\square \psi _{B}+(\square \gamma ^{AB})\psi
_{B}+2(\nabla ^{h}\gamma ^{AB})\nabla _{h}\psi _{B},  \tag{4.124b}
\end{equation}%
along with 
\begin{equation}
\Delta _{AB}\psi _{C}-(\Delta _{AB}\psi ^{M})\gamma _{MC}=2i\phi _{AB}\psi
_{C}  \tag{4.125a}
\end{equation}%
and%
\begin{equation}
\Delta _{A^{\prime }B^{\prime }}\chi _{C^{\prime }}-(\Delta _{A^{\prime
}B^{\prime }}\chi ^{M^{\prime }})\gamma _{M^{\prime }C^{\prime }}=(-2i)\phi
_{A^{\prime }B^{\prime }}\chi _{C^{\prime }}.  \tag{4.125b}
\end{equation}%
We can then state that the right-hand sides of such wave equations amount to
the only structures which carry the interaction patterns produced by the
propagation in $\mathfrak{M}$ of the fields borne by the pairs $\{\psi
^{A},\chi _{A^{\prime }}\}$ and $\{\psi _{A},\chi ^{A^{\prime }}\}$.
Remarkably enough, these coupling configurations are not affected by the
implementation of any devices for changing valences like the ones of Eqs.
(4.124) and (4.125).

\section{CONCLUDING REMARKS}

The only spacetime-metric character of the $\varepsilon $-formalism is
carried by Eqs. (3.7b) and (3.68), which effectively yield the expressions 
\begin{equation*}
\mathfrak{e}=K(-\mathfrak{g})^{-1/2},\text{ }\Sigma _{h}^{BB^{\prime
}}\partial _{a}\Sigma _{BB^{\prime }}^{h}=\partial _{a}\log \mathfrak{e},
\end{equation*}%
where $K$ stands for a constant positive-definite world-spin invariant. An $%
\varepsilon $-formalism counterpart of the condition (3.90) can therefore be
brought into the overall metric picture, according to the requirement 
\begin{equation*}
\nabla _{a}\mathfrak{e}=0.
\end{equation*}%
The transformation law (3.104) suggests the implementation of a prescription
of the type 
\begin{equation*}
\Pi _{a}=\partial _{a}\log E^{-1}\Rightarrow \partial _{\lbrack a}\Pi
_{b]}=0,
\end{equation*}%
with $E$ amounting to a covariantly constant world-invariant spin-scalar
density of absolute weight $+1$ that carries no specific metric meaning.
This prescription can be considered as a formal physically meaningless
counterpart of Eq. (3.43), which is associated to the spin-displacement
configuration 
\begin{equation*}
\Pi _{a}dx^{a}=-E^{-1}dE.
\end{equation*}%
It also guarantees the genuineness of the $\varepsilon $-formalism version
of Eqs. (4.1) and (4.3) since 
\begin{equation*}
\nabla _{\lbrack a}(\Pi _{b]}\Sigma ^{cDD^{\prime }})=0.
\end{equation*}

It has become manifest that the strongest way of characterizing $\Phi _{a}$
and $\varphi _{a}$ as affine electromagnetic potentials is afforded by the
commutators that yield the curvature spinors of $\gamma _{aB}{}^{C}$ and $%
\Gamma _{aB}{}^{C}$. As we had mentioned in Section 1, one of the
traditional properties of the $\gamma $-formalism is that the presence or
absence of electromagnetic fields can be controlled by means of any of the
metric devices provided by Eqs. (3.73) and (3.81). The derivatives (4.48)
and (4.49) supply alternative "electromagnetic switches"\ of the form 
\begin{equation*}
\Delta _{AB}\gamma _{CD}=(\Delta _{AB}\gamma )\varepsilon _{CD}=2i\phi
_{AB}\gamma _{CD}.
\end{equation*}%
Then, whenever $\Phi _{a}$ is taken as a gradient, we may allow for the
relationship 
\begin{equation*}
(-2)\phi _{AB}=\Delta _{AB}\Phi =0,
\end{equation*}%
which obviously brings out the Christoffel property of $\Gamma _{abc}$ as
expressed by 
\begin{equation*}
\lbrack \nabla _{a},\nabla _{b}]\Phi =0.
\end{equation*}%
Another noteworthy difference between the formalisms is related to the
non-availability of any $\varepsilon $-counterparts of such electromagnetic
devices.

A gauge-covariant form of the limiting process gets clearly exhibited when
we call for the $\gamma \varepsilon $-formulae 
\begin{equation*}
\Theta _{aBC}^{(\gamma )}=\gamma \Theta _{aBC}^{(\varepsilon )},\text{ }%
\Gamma _{A(BC)A^{\prime }(B^{\prime }C^{\prime })}^{(\gamma )}=\mid \gamma
\mid ^{3}\Gamma _{A(BC)A^{\prime }(B^{\prime }C^{\prime })}^{(\varepsilon )}
\end{equation*}%
and 
\begin{equation*}
\sigma _{h(B}^{D^{\prime }}\partial _{\mid a\mid }\sigma _{C)D^{\prime
}}^{h}=\gamma \Sigma _{h(B}^{D^{\prime }}\partial _{\mid a\mid }\Sigma
_{C)D^{\prime }}^{h},\text{ }\gamma _{a(BC)}=\gamma \Gamma _{a(BC)}.
\end{equation*}%
Consequently, we can write down the affine configuration 
\begin{equation*}
\vartheta _{aBC}=\frac{1}{2}(S_{(B}^{bD^{\prime }}\partial _{C)D^{\prime
}}g_{ab}+S_{b(B}^{D^{\prime }}\partial _{\mid a\mid }S_{C)D^{\prime
}}^{b}+\vartheta _{aD}{}^{D}M_{BC}),
\end{equation*}%
together with Eq. (3.61) and the explicit $\gamma $-formalism expression 
\begin{equation*}
\gamma _{aB}{}^{B}=\frac{1}{4}(\Gamma _{a}+\sigma _{s}^{BB^{\prime
}}\partial _{a}\sigma _{BB^{\prime }}^{s})-2i\Phi _{a}.
\end{equation*}%
We have thus been able to build up a metric expression for $\gamma _{aBC}$
and likewise to construct out of utilizing the limiting procedure the
corresponding configuration for $\Gamma _{aBC}$.

The implementation of the relation $\partial _{\lbrack a}\Pi _{b]}=0$
particularly ensures that the pattern (4.3) for the $\gamma $-formalism
equals its $\varepsilon $-formalism counterpart, that is to say, 
\begin{equation*}
W_{abA}^{(\gamma )}{}^{B}=W_{abA}^{(\varepsilon )}{}^{B}\Leftrightarrow
W_{abAB}^{(\gamma )}{}=\gamma W_{abAB}^{(\varepsilon )}{}.
\end{equation*}%
Indeed, the $W$-objects for both formalisms may also arise from the
combination of Eq. (4.1) with either of the commutators 
\begin{equation*}
\lbrack \nabla _{a},\nabla _{b}]u^{CC^{\prime }}=S_{c}^{CC^{\prime
}}R_{abh}{}^{c}u^{h},\text{ }[\nabla _{a},\nabla _{b}]u_{CC^{\prime
}}=-S_{CC^{\prime }}^{c}\hspace{1pt}R_{abc}{}^{h}u_{h}.
\end{equation*}%
Suitably contracted versions of these structures lead to purely
gravitational configurations like 
\begin{equation*}
\Delta _{AB}u^{BC^{\prime }}=\Xi _{ABD^{\prime }}{}^{C^{\prime
}}u^{BD^{\prime }}-\frac{R}{8}u_{A}{}^{C^{\prime }},
\end{equation*}%
whence, in either formalism, we may account for the Hermitian expansions 
\begin{equation*}
\lbrack \nabla _{a},\nabla _{b}]u^{CC^{\prime }}=W_{abD}{}^{C}u^{DC^{\prime
}}+W_{abD^{\prime }}{}^{C^{\prime }}u^{CD^{\prime }}
\end{equation*}%
and 
\begin{equation*}
\lbrack \nabla _{a},\nabla _{b}]u_{CC^{\prime }}=-\hspace{2pt}%
(W_{abC}{}^{D}u_{DC^{\prime }}+W_{abC^{\prime }}{}^{D^{\prime
}}u_{CD^{\prime }}).
\end{equation*}%
The combination of these results with the relations 
\begin{equation*}
R_{AA^{\prime }BB^{\prime }CC^{\prime }DD^{\prime }}^{(\gamma )}=\mid \gamma
\mid ^{4}R_{AA^{\prime }BB^{\prime }CC^{\prime }DD^{\prime }}^{(\varepsilon
)}
\end{equation*}%
and 
\begin{equation*}
\omega _{ABCD}^{(\gamma )}=\gamma ^{2}\omega _{ABCD}^{(\varepsilon )},\text{ 
}\omega _{A^{\prime }B^{\prime }CD}^{(\gamma )}=\mid \gamma \mid ^{2}\omega
_{A^{\prime }B^{\prime }CD}^{(\varepsilon )},
\end{equation*}%
establishes the $\gamma \varepsilon $-commonness of the gravitational
quantities $\varkappa $ and $\varpi $, and additionally enhances the
correspondence principle involved in the limiting process. While the
torsionlessness property of $\nabla _{a}$ may be expressed in terms of
spin-affinity pieces such as in Eq. (3.47a), the spin-curvature version of
it may be exhibited by the statements (4.27) and (4.85).

One can attain a confirmation of the result regarding the tensor behaviour
of the $\gamma $-formalism wave equations for gravitons and photons in $%
\mathfrak{M}$ by invoking the gauge invariance of $\beta _{a}$ along with
the transformation law for $\Theta $ and the homogeneous pattern 
\begin{equation*}
\square ^{\prime }(\breve{A}^{\prime }T_{BC...D}^{\prime })=(\Delta _{%
{\small \Lambda }})^{\mathfrak{a}}(\bar{\Delta}_{{\small \Lambda }})^{%
\mathfrak{b}}\mid \Delta _{{\small \Lambda }}\mid ^{\mathfrak{c}}\Lambda
_{B}{}^{L}\Lambda _{C}{}^{M}...\Lambda _{D}{}^{N}\square (\breve{A}%
T_{LM...N}).
\end{equation*}%
This procedure takes up implicitly the gauge invariance of the $\Upsilon $%
-functions defined by (4.72c) and (4.99). The sourceful version of Eq.
(4.100) amounts to 
\begin{equation*}
(\square -4i\beta ^{h}\nabla _{h}-\Upsilon _{(\mathcal{G})}+\frac{R}{2})\Psi
{}_{ABCD}{}-6\Psi _{MN(AB}{}\Psi {}_{CD)}{}^{MN}=-\kappa s_{ABCD},
\end{equation*}%
with%
\begin{equation*}
s_{ABCD}=\gamma _{L(A}\nabla _{B}^{A^{\prime }}\nabla ^{B^{\prime
}L}T_{CD)A^{\prime }B^{\prime }},
\end{equation*}%
whence the rule (4.104) still holds for it, namely%
\begin{equation*}
(\square +4i\beta ^{h}\bigtriangledown _{h}-\overline{\Upsilon }_{(\mathcal{G%
})}+\frac{R}{2})\Psi ^{ABCD}-6\Psi _{MN}{}^{(AB}\Psi ^{CD)MN}=-\kappa
s^{ABCD}.
\end{equation*}%
In the $\varepsilon $-formalism, we correspondingly obtain, for instance,%
\begin{equation*}
(\square +\frac{R}{2})\Psi {}_{ABCD}{}-6\Psi _{MN(AB}{}\Psi
{}_{CD)}{}^{MN}=-\kappa \nabla _{(A}^{A^{\prime }}\nabla _{B}^{B^{\prime
}}T_{CD)A^{\prime }B^{\prime }}.
\end{equation*}

The $\gamma \varepsilon $-formalisms have afforded us an elementary
description of generally relativistic spacetime geometry. We emphasize
further that their inner structure may suggest looking upon them as an
intrinsic part of general relativity. Thus, the occurrence of
electromagnetic configurations in spacetime curvatures could lead us to
thinking of Infeld-van der Waerden photons as generally relativistic
objects. Consequently, we could expect that some theoretical insights may
eventually be gained into the situation which deals with the physical
properties of the radiation background of the universe.

\end{document}